\documentclass{jfm}

\usepackage{graphicx}
\usepackage{natbib}

\ifCUPmtlplainloaded \else
  \checkfont{eurm10}
  \iffontfound
    \IfFileExists{upmath.sty}
      {\typeout{^^JFound AMS Euler Roman fonts on the system,
                   using the 'upmath' package.^^J}%
       \usepackage{upmath}}
      {\typeout{^^JFound AMS Euler Roman fonts on the system, but you
                   dont seem to have the}%
       \typeout{'upmath' package installed. JFM.cls can take advantage
                 of these fonts,^^Jif you use 'upmath' package.^^J}%
      }
  \else
  \fi
\fi

\ifCUPmtlplainloaded \else
  \checkfont{msam10}
  \iffontfound
    \IfFileExists{amssymb.sty}
      {\typeout{^^JFound AMS Symbol fonts on the system, using the
                'amssymb' package.^^J}%
       \usepackage{amssymb}%
         \let\leq=\leqslant
         
      }{}
  \fi
\fi

\ifCUPmtlplainloaded \else
  \IfFileExists{amsbsy.sty}
    {\typeout{^^JFound the 'amsbsy' package on the system, using it.^^J}%
     \usepackage{amsbsy}}
    {}
\fi

\newsavebox{\astrutbox}
\sbox{\astrutbox}{\rule[-5pt]{0pt}{20pt}}

\usepackage{amsmath,bm}
\usepackage{subfigure}

\title{Effects of shock topology on temperature field in compressible turbulence}
\author[Qionglin Ni]%
{Q\ls I\ls O\ls N\ls G\ls L\ls I\ls N\ns N\ls I\ls$^{1,2}$
\thanks{Email address for correspondence: niql.pku@gmail.com}, and\ns
S\ls H\ls I\ls Y\ls I\ns C\ls H\ls E\ls N\ls$^1$
\thanks{Email address for correspondence: syc@pku.edu.cn}}
\affiliation{$^1$State Key Laboratory for Turbulence and Complex Systems, College of Engineering,
Peking University, Beijing 100871, People's Republic of China \\
$^2$Department of Physics, University of Rome "Tor Vergata", Via della Ricerca Scientifica 1, 00133, Rome, Italy}

\date{?; revised ?; accepted ?.}

\begin{document}
\maketitle

\begin{abstract}
Effects of two types of shock topology, namely, small-scale shocklet and large-scale shock wave, on the statistics of
temperature in compressible turbulence were investigated by numerical simulations. The shocklet and shock wave are
caused by the purely solenoidal and primarily compressive modes of large-scale random forces, respectively.
Hereafter, the corresponding two flows are abbreviated as SFT (solenoidal forced turbulence) and CFT (compressive
forced turbulence), respectively. It shows that in SFT the temperature spectrum follows the $k^{-5/3}$ power law,
and the temperature field has the "ramp-cliff" structures. By contrast, in CFT the temperature spectrum defers to
the $k^{-2}$ power law, and the temperature field is dominated by the large-scale rarefaction and compression.
The power-law exponents for the probability distribution function (p.d.f.) of large negative dilatation are $-2.5$
in SFT and $-3.5$ in CFT, very close to those computed from a theoretical model. In CFT, the collapse of the p.d.f.
for temperature increment to the same distribution indicates the saturation of scaling exponent at high order numbers.
For the isentropic assumption of thermodynamic variables, it shows that the derivation in SFT grows with the
turbulent Mach number ($M_t$), and for same $M_t$, the variables in CFT are more anisentropic. The angle statistics of
CFT shows that the temperature gradient is preferentially perpendicular to the anisotropic strain rate tensor. In detail,
it tends to be parallel with the first eigenvector and be orthogonal with the second and third eigenvectors. By
employing a "coarse-graining" approach, we investigated the cascade of temperature. It shows that the temperature variance
at large scales is increased by the viscous dissipation at small scales and the pressure-dilatation at moderate
scales, but is decreased by the subgrid-scale (SGS) temperature flux, which preferentially transfers in the orientation
where the temperature gradient anti-aligns with the SGS temperature-velocity coupling. The distributions of pressure-dilatation
and its two components prove the fact that the negligible contribution of pressure-dilatation at small scales is due to the
cancelations of high values between rarefaction and compression regions. The strongly positive skewness of the p.d.f.s of
pressure-dilatation implies that the conversion from kinetic to internal energy through compression is more intense than
the opposite process through rarefaction. Furthermore, it shows that in SFT the fluctuations of pressure-dilatation
approximately follow the Zeman model \citep{Zeman91}.
\end{abstract}


\section{Introduction}

Since the earlier seminal work of Corrsin \citep{Corrsin51}, much effort has been devoted to studying the statistics of
temperature fluctuations in turbulence, including astrophysics \citep{Cattaneo03}, geophysics \citep{Stevens05} and
engineering \citep{Hill76}. For incompressible turbulence, the statistical interplay between temperature and velocity is
decoupled, and thus, the temperature is regarded as a passive field \citep{Sreenivasan91,Shraiman00,Warhaft00}. Nevertheless,
for convective or compressible turbulence, because of the substantial impact to velocity through buoyancy or pressure, the
temperature always behaves as an active field. Therefore, the properties of temperature statistics are the central issues.

\citet{Belmonte96} experimentally studied the small-scale features of temperature in a thermal convection. They found that
the temperature acts actively when the skewness product of temperature and its temporal derivative is positive, otherwise,
it acts passively. The comparative experiments on the turbulent Rayleigh-Bernard convection performed by \citet{Zhou08}
showed that when the dynamical timescale is above (below) the Balgiano timescale, the temperature behaves as an active
(passive) field. A fascinating feature of passive field is that the scaling exponent of structure function
saturates for high order numbers, which is believed to be related to the so-called "ramp-cliff" structures.
Therefore, it is natural to ask whether a similar saturation appears for temperature. In fact, the findings in both
experiments \citep{Zhou02} and simulations \citep{Celani01,Zhou13} showed that the temperature is more intermittent
than a passive field, and possesses a saturated exponent of $0.8$, even smaller than the Burgers saturated exponent of $1.0$
\citep{Mitra05}.

In terms of temperature in compressible turbulence, although it does not satisfy the standard definition of an active
field, the nonlinear coupling to velocity makes it act actively. However, so far there are very few
studies addressed this topic. In his theoretical analysis, \citet{Canuto97} developed a model for handling compressible
convection in the presence of large-scale flows, and obtained the dynamic equations for the mean and variance of temperature.
\citet{Ni12} numerically investigated the statistics of temperature in one-dimensional compressible turbulence. They found that
the temperature undergoes downscale cascade and follows the Kolmogorov picture. Moreover, the scaling exponent of temperature
structure function is close to the Burgers scaling, indicating saturation at high order numbers. Recently, \citet{Donzis13}
carried out compressible turbulence simulations spanning the range of $M_t = 0.09\sim 0.61$. Their results were that:
(1) the temperature spectrum defers to the $k^{-5/3}$ power law; (2) temperature fluctuations are less correlated to other
thermodynamic variables, and the covariance between density and temperature contributes to the scaling of the mean and variance
of pressure; and (3) the p.d.f. of temperature fluctuations is basically independent of turbulent Mach number. In detail,
The p.d.f. tails for the positive component of temperature fluctuations are log-normal, while those for the negative component
of temperature fluctuations retain a Mach-number dependence.

Previous simulations in compressible turbulence \citep{Wang11,Wang12a,Wang12b,Wang13a,Wang13b} showed that there
exists a strong connection between the shock topology and the forcing scheme. In particular, the flows driven by the
solenoidal and compressive forces generate the small-scale shocklets and large-scale shock waves, respectively.
This in turn, greatly influences the statistics of fields such as density, velocity and temperature.

In this paper, we use the two groups data from the numerical simulations driven by the large-scale, random,
solenoidal and compressive forcings to study the temperature in compressible turbulence, focusing on the effects
of shock topology on the small-scale statistics of temperature. The flows are computed on a $1024^3$ grid by adopting
a high-precision hybrid method \citep{Wang10}, and the stationary values of the turbulent Mach number $M_t$ and
Taylor microscale Reynolds number $Re_{\lambda}$ are $(1.03, 255)$ for SFT and $(0.62, 164)$ for CFT, respectively.
A systemic investigation on the fundamental statistics of temperature including the spectrum and field structure
is actualized. Then, we report the statistics of dilatation, the application of isentropic assumption to the
thermodynamic variables, and the angle statistics of temperature gradient on local flow structures. By employing
a "coarse-graining" approach to the temperature variance budget, we analyze the cascade of temperature, in particular,
the crucial role of pressure-dilatation in the transport of temperature fluctuations. This paper is part of
a series of investigations on scalar transport in compressible turbulence.
In three companion papers \citep{Ni15a,Ni15b,Ni15c},
we have examined carefully the statistical differences between active and passive scalars, the effects of Mach
and Schmidt numbers on scalar mixing. We hope that this comprehensive study will advance our understanding of the
small-scale statistics of temperature in compressible turbulence.

The reminder of this paper is organized as follows. The governing equations and system parameters, as well as the
details of simulation method, are described in Section 2. The basic statistics of the simulated flows is reported
in Section 3. In the following two sections, we discuss the isentropic approximation of thermodynamic variables,
and the statistical properties of temperature gradient. The analysis of the cascade of temperature is presented
in Section 6. In Section 7, we present the summary and conclusions.

\section{Governing Equations and System Parameters}

We simulate the statistically stationary compressible turbulence driven by the large-scale velocity forcing. Besides,
a cooling function is added at large scales for removing accumulated internal energy at small scales. Similar to
\citet{Ni15a}, here we use the reference length $L$, density $\rho$, velocity $U$ and temperature $T_0$ to normalize
the compressible flow. Then we obtain the reference Mach number $M\equiv U/c_0$, where $c_0=\sqrt{\gamma RT_0}$ is
the reference sound speed, $R=C_p-C_v$ is the specific gas constant, and $\gamma=C_p/C_v$ is the specific heat ratio,
with $C_p$ and $C_v$ representing the two specific heats at constant pressure and volume, respectively. By adding
the reference dynamical viscosity $\mu_0$ and thermal conductivity $\kappa_0$, we obtain another two basic
parameters: the reference Reynolds number $Re\equiv UL/\mu_0$ and the reference Prandtl number
$Pr\equiv\mu_0C_p/\kappa_0$. In our simulations, the values of $\gamma$ and $Pr$ are set as $1.4$ and $0.7$,
respectively.

Based on the above procedure, the governing equations of the simulated flows, in dimensionless form,
are written as
\begin{eqnarray}
&& \frac{\partial}{\partial t}\rho + \frac{\partial}{\partial x_j}\big(\rho u_j
\big)=0,
\label{density}\\
&& \frac{\partial}{\partial t}\big(\rho u_i\big) + \frac{\partial}{\partial x_j}
\big[\rho u_iu_j + p\delta_{ij}/\gamma M^2\big]
=\frac{1}{Re}\frac{\partial}{\partial x_j}\sigma _{ij} + \rho{\cal F}_i,
\label{momentum} \\
&& \frac{\partial}{\partial t}{\cal E} + \frac{\partial}{\partial x_j}
\big[({\cal E}+ p/\gamma M^2)u_j\big]=\frac{1}{\alpha}\frac{\partial}{\partial x_j}
\big(\kappa\frac{\partial T}{\partial x_j}\big) + \frac{1}{Re}\frac{\partial}
{\partial x_j}\big(\sigma _{ij}u_i\big) - \Lambda + \rho{\cal F}_j u_j,
\label{energyeqn} \\
&& p=\rho T,
\label{state}
\end{eqnarray}
where $\alpha\equiv PrRe(\gamma-1)M^2$. The primary variables are density $\rho$, velocity $u_i$, temperature $T$
and pressure $p$. $F_j$ is the dimensionless large-scale velocity forcing
\begin{equation}
F_j = \sum_{l=1}^{2}\hat{F_j}(\textbf{k}_l)\exp(i\textbf{k}_l\textbf{x}) + c.c..
\end{equation}
Here $i\equiv\sqrt{-1}$, and $\hat{F_j}$ is the Fourier amplitude, which has only a solenoidal component perpendicular to
$\textbf{k}_l$ for SFT \citep{Ni2013,Ni15a}, but has another compressive component parallel to $\textbf{k}_l$
for CFT \citep{Ni15b}. The viscous stress $\sigma_{ij}$ and total energy per unit volume ${\cal E}$ are defined by
\begin{equation}
\sigma _{ij} \equiv \mu \big(\frac{\partial u_i}{\partial
x_j}+\frac{\partial u_j}{\partial x_i}\big)-\frac{2}{3}\mu\theta\delta_{ij},
\end{equation}
\begin{equation}
{\cal E} \equiv \frac{p}{(\gamma-1)\gamma M^2}+\frac{1}{2}\rho\big(u_ju_j\big),
\end{equation}
where $\theta=\partial u_k/\partial x_k$ is the velocity divergence or dilatation, a variable that quantifies the local
rate of rarefaction ($\theta>0$) or compression ($\theta<0$). We now give the expressions of the dimensionless dynamical
viscosity and thermal conductivity \citep{Sutherland92}, to complete the system
\begin{equation}
\mu, \kappa = \frac{1.4042T^{1.5}}{T+0.4042}.
\end{equation}
The large-scale velocity forcing presented in Equation (2.5) injects the same amount of energy into
the two lowest spherical wavenumber shells. In particular, the energy injection in SFT is only perpendicular to
the wavenumber vector, while that in CFT is both parallel and perpendicular to the wavenumber vector, and the
ratio is $1:20$. In addition, to keep temperature staying in a statistically stationary state, the velocity
forcing and cooling function should satisfy the relation: $\langle\Lambda\rangle =\langle{\rho\cal F}_ju_j\rangle$,
where $\langle\cdot\rangle$ indicates ensemble average.

\begin{table*}
\caption{Simulated parameters and resulting flow statistics.}
\begin{center}
\small
\begin{tabular*}{0.95\textwidth}{@{\extracolsep{\fill}}rccccccccccccc}
\hline\hline
\emph{Flow} &\emph{Grid} &$Re_{\lambda}$ &$M_t$ &$k_{max}\eta$ &$\tau$ &$L_f$ &$L_{Tf}$ &$E_K$ &$E_T$ &$\theta'/\omega'$  \\
\hline
SFT &$1024^3$ &$255$ &$1.03$ &$3.34$ &$1.11$ &$1.44$ &$0.80$ &$2.13$ &$0.51$ &$0.34$   \\
CFT &$1024^3$ &$162$ &$0.62$ &$3.18$ &$1.23$ &$1.56$ &$1.12$ &$2.67$ &$0.53$ &$1.36$   \\
\hline
\end{tabular*}
\begin{tabular*}{0.95\textwidth}{@{\extracolsep{\fill}}rccccccccccccc}
$S_3$ &$S_{3m}$ &$S_{3T}$ &$u'$ &$T'$ &$\rho'$ &$u'_c/u'_s$ &$\langle\epsilon\rangle$
&$\langle\epsilon_s\rangle/\langle\epsilon\rangle$ &$\langle\epsilon_c\rangle/\langle\epsilon\rangle$ &$\langle\epsilon_m\rangle/\langle\epsilon\rangle$ &$\langle\epsilon_T\rangle$   \\
\hline
$-2.2$ &$-2.0$ &$0.18$ &$2.22$ &$0.12$ &$0.29$ &$0.24$ &$0.54$ &$88\%$ &$14\%$ &$-2\%$ &$0.05$   \\
$-16.9$ &$-15.7$ &$0.76$ &$2.19$ &$0.23$ &$0.52$ &$1.56$ &$0.79$ &$32\%$ &$68\%$ &$0\%$ &$0.16$   \\
\hline\hline
\end{tabular*}
\normalsize \label{table1}
\end{center}
\end{table*}

The system is solved numerically by adopting a hybrid method in a cubic-box grid resolution of $1024^3$ with
periodic boundary conditions. This method applies a seventh-order weighted essentially non-oscillatory (WENO)
scheme \citep{Balsara00} for shock region and an eighth-order compact central finite difference (CCFD) scheme \citep{Lele92}
for smooth region outside shock. A flux-based conservative formulation is implemented to optimize
the handling at the interface between the shock and smooth regions. In addition, a new numerical hyperviscosity
formulation is used to improve the numerical instability without compromising the accuracy of the hybrid method \citep{Wang10}.
To obtain statistical averages of interested variables, a total of ten stationary flow fields
were used. The values of large-eddy turnover time $\tau$ are $1.11$ in SFT and $1.23$ in CFT. Furthermore, in
this paper, we always use the turbulent Mach number $M_t$ and Taylor microscale Reynolds number $Re_\lambda$,
rather than the reference Mach number $M$ and Reynolds number $Re$, to measure the compressible flows,
which are defined as follows
\begin{equation}
M_t\equiv M\frac{u'}{\langle{\sqrt{T}}\rangle},\quad Re_{\lambda}\equiv Re\frac{u'\lambda\langle\rho\rangle}
{\sqrt{3}\langle\mu\rangle},
\end{equation}
where $u'\equiv\sqrt{\langle u_1^2+u_2^2+u_3^2\rangle}$ is the r.m.s. velocity magnitude, and $\lambda$ is the Taylor microscale,
defined as $\lambda\equiv u'/\sqrt{\langle(\partial u_1/\partial x_1)^2+(\partial u_2/\partial x_2)^2
+(\partial u_3/\partial x_3)^2\rangle}$. In our simulations, the stationary average values of ($M_t$, $Re_{\lambda}$)
are ($1.03$, $255$) for SFT and ($0.62$, $164$) for CFT.

\section{Fundamental Statistics of Compressible Flows}

In this section, we first describe the simulated parameters of the compressible flows, then report
the spectrum and field structure. The section is ended up with the discussion on the statistics of
dilatation.

\subsection{\emph{Simulated Parameters, Spectra and Structure Fields}}

Table~\ref{table1} summarizes some overall statistics of the compressible flows. The resolution parameter
$k_{max}\eta$ are $3.34$ in SFT and $3.18$ in CFT, where $\eta=[\langle\mu/(Re\rho)\rangle^3/<\epsilon/\rho>]^{1/4}$
is the Kolmogorov scale, and $k_{max}=N/2=512$ is the maximum wavenumber in our simulations. It means that
for both flows, the fine-scale structures in smooth regions are well resolved by the hybrid method.
Here we point out that although the thickness of shock is comparable to the grid length and is not directly
resolved by the WENO scheme, the total amount of dissipation across shock is independent of numerical viscosity.

The integral scales for velocity and temperature are computed by
\begin{equation}
L_f=\frac{3\pi}{2u'^2}\int\limits^{\infty}_0\frac{E(k)}{k}dk, \quad
L_{Tf}=\frac{\pi}{T'^2}\int\limits^{\infty}_0\frac{E_T(k)}{k}dk,
\end{equation}
where $E(k)$ and $E_T(k)$ are the spectra of kinetic energy and temperature per unit mass, respectively
\begin{equation}
\int\limits^{\infty}_0E(k)dk=\frac{1}{2}u'^2, \quad \int\limits^{\infty}_0E_T(k)dk=\frac{1}{2}T'^2.
\end{equation}
The ratio of the r.m.s. magnitudes of dilatation to vorticity, $\theta'/\omega'$, are $0.34$ in SFT and $1.36$ in CFT,
showing that at small scales, the effect of compressibility in is overwhelming in the compressive forced flow.
Here $\theta'=\sqrt{\langle\theta^2\rangle}$ and $\omega'=\sqrt{\langle\omega_1^2+\omega_2^2+\omega_3^2\rangle}$
are the r.m.s. magnitudes of dilatation and vorticity, respectively.

The compressible character can be further demonstrated by ensemble averages of the skewness of velocity derivative
and the mixed skewness of velocity-temperature derivatives
\begin{equation}
S_3=\frac{\big[\langle(\frac{\partial u_1}{\partial x_1})^3+(\frac{\partial u_2}{\partial x_2})^3
+(\frac{\partial u_3}{\partial x_3})^3\rangle\big]/3}
{\big[\langle(\frac{\partial u_1}{\partial x_1})^2+(\frac{\partial u_2}{\partial x_2})^2
+(\frac{\partial u_3}{\partial x_3})^2\rangle/3\big]^{3/2}},
\end{equation}
\begin{equation}
S_{3m}=\frac{\big[\langle\frac{\partial u_1}{\partial x_1}(\frac{\partial T}{\partial x_1})^2
+\frac{\partial u_2}{\partial x_2}(\frac{\partial T}{\partial x_2})^2
+\frac{\partial u_3}{\partial x_3}(\frac{\partial T}{\partial x_3})^2\rangle\big]/3}
{\big[\langle(\frac{\partial u_1}{\partial x_1})^2+(\frac{\partial u_2}{\partial x_2})^2
+(\frac{\partial u_3}{\partial x_3})^2\rangle/3\big]^{1/2}
\big[\langle(\frac{\partial T}{\partial x_1})^2+(\frac{\partial T}{\partial x_2})^2
+(\frac{\partial T}{\partial x_3})^2\rangle/3\big]}.
\end{equation}
In SFT $S_3$ and $S_{m3}$ are $-2.2$ and $-2.0$, respectively; however, in CFT their magnitudes increase
to $S_3=-16.9$ and $S_{3m}=-15.7$. This indicates that the presence of large-scale shock waves leads the velocity
field of CFT to be anisotropic. By contrast, for ensemble average of the skewness of temperature derivative
\begin{equation}
S_{3T}=\frac{\big[\langle(\frac{\partial T}{\partial x_1})^3+(\frac{\partial T}{\partial x_2})^3
+(\frac{\partial T}{\partial x_3})^3\rangle\big]/3}
{\big[\langle(\frac{\partial T}{\partial x_1})^2+(\frac{\partial T}{\partial x_2})^2
+(\frac{\partial T}{\partial x_3})^2\rangle/3\big]^{3/2}}.
\end{equation}
The values are $0.18$ for SFT and $0.76$ for CFT, implying that the temperature field is approximately isotropic
in compressible turbulence, espcially in the solenoidal forced flow. In addition, the r.m.s. magnitude of temperature
fluctuations is $0.12$ in SFT and $0.23$ in CFT, while that of density fluctuations reaches as high as $0.29$ in
SFT and $0.52$ in CFT. This reveals that compared to temperature, the compressibility makes density vary more intense.

The application of Helmholtz decomposition \citep{Samtaney01} to velocity field gives that
\begin{equation}
u_i = u_{is} + u_{ic},
\end{equation}
where $u_{is}$ is the solenoidal component satisfying $\partial u_{is}/\partial x_i = 0$, and $u_{ic}$ is the compressive
component satisfying $\varepsilon_{ijk}\partial u_{kc}/\partial x_j = 0$, with $\varepsilon_{ijk}$ representing the Levi-Civita
symbol. The ratio for the r.m.s. magnitudes of the two components, $u'_c/u'_s$, is $0.24$ in SFT and $1.56$ in CFT.
Furthermore, \citet{Andreopoulos00} showed that the viscous dissipation rate $\epsilon=\sigma_{ij}S_{ij}/Re$
can be divided into three parts: the solenoidal part $\epsilon_s=(\mu/Re)\omega_i\omega_i$, the dilatation part
$\epsilon_c=(4/3)(\mu/Re)\theta^2$, and the residual part $\epsilon_m=(2\mu/Re)[(\partial u_i/\partial x_j)
(\partial u_j/\partial x_i)-\theta^2]$.
In our simulations, the percentages of the first two parts are ($88\%$, $14\%$) for SFT and ($32\%$, $68\%$)
for CFT, meaning that in SFT most kinetic energy is dissipated through vortices stretching, while in CFT the
dissipation of kinetic energy is dominated by rarefaction and compression.The temperature dissipation rate is
defined as
\begin{equation}
\epsilon_T \equiv \kappa\big(\partial T/\partial x_j\big)^2.
\end{equation}
Table~\ref{table1} shows that the ensemble-average value, $\langle\epsilon_T\rangle$, is $0.05$ in SFT
and $0.16$ in CFT, indicating that compared to small-scale shocklets, there are more temperature fluctuations
depleted by large-scale shock waves.

\begin{figure}
\centerline{\includegraphics[width=8cm]{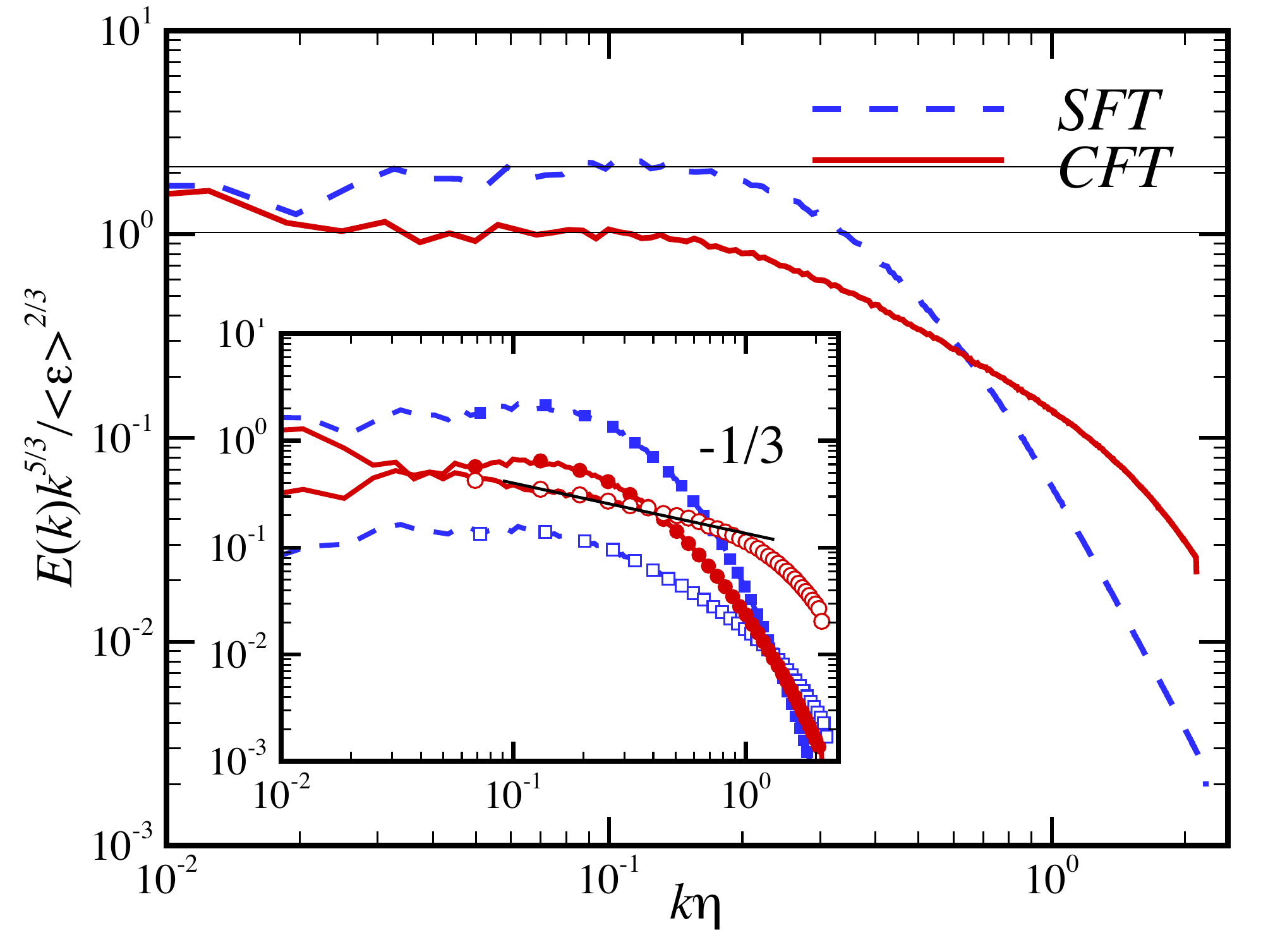}}
\caption{Compensated spectrum of kinetic energy, where the dashed and solid lines are for SFT and CFT, respectively.
Inset: solenoidal (solid) and compressive (open) components, where the symbols of squares and circles are for SFT
and CFT, respectively.}
\label{fig:fig1}
\end{figure}
\begin{figure}
\centerline{\includegraphics[width=8cm]{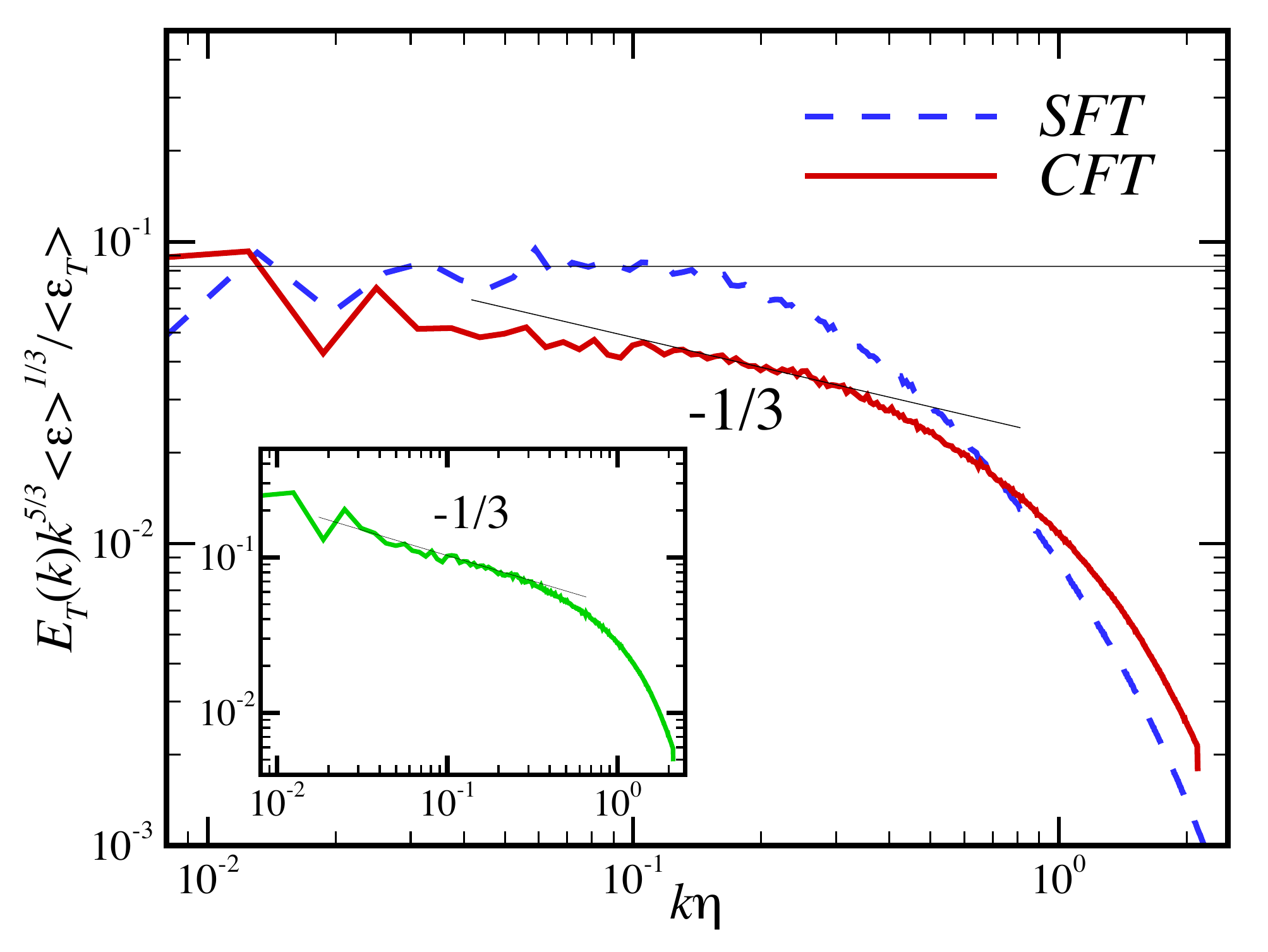}}
\caption{Compensated spectrum of temperature, where the dashed and solid lines are for SFT and CFT, respectively.
The dotted line is for the compensated spectrum of density-weighted temperature in CFT.}
\label{fig:fig2}
\end{figure}

Figure~\ref{fig:fig1} presents the compensated kinetic energy spectra from the two simulated flows.
Here an operational definition of the inertial range is identified for the overall kinetic energy
spectrum, namely, $C_K = E(k)k^{5/3}/\langle\epsilon\rangle^{2/3}$, where $C_K$ is the Kolmogorov
constant. In SFT, $C_K$ is found to be $2.17$ for the range of $0.03\leq k\eta \leq0.1$, a bit higher
than the typical values observed in incompressible turbulence \citep{Wang96}. By contrast, for the same
range is CFT, $C_K$ has a much smaller value of $1.05$. In the inset we plot the solenoidal and compressive
components of the kinetic energy spectra. For the solenoidal components, the appearance of spectral
bumps at high wavenumber reveals the nonlocal feature for the transfer of the solenoidal component of
velocity fluctuations in the crossover region between inertial and dissipative ranges. Moreover, it shows
that the compressive component in SFT follows the $k^{-5/3}$ power law, while that in CFT defers to the
$k^{-2}$ power law over than one decade wavenumber range. As one known, previous studies on Burgers
turbulence \citep{Bec07} showed that it is the existence of large-scale shock wave gives rise to
the $k^{-2}$ scaling.

In Figure~\ref{fig:fig2} we plot the compensated spectra of temperature. Similar to that in compressible flow, the
temperature spectrum in SFT displays the $k^{-5/3}$ power law, and its scaling constant can be operational defined
as $C_T=E_T(k)k^{5/3}\langle\epsilon\rangle^{1/3}/\langle\epsilon_T\rangle$. It shows that in the range of
$0.06\leq k\eta\leq0.12$, the value of $C_T$ is about $0.08$, much smaller than the typical values observed from
passive scalar. By contrast, the spectrum of temperature in CFT has the $k^{-2}$ power law, indicating the dominant
motion of large-scale shock wave. In the inset we present the spectrum of the density-weighted temperature,
$\rho^{1/3}T$, from CFT. The result is that the spectrum still follows the $k^{-2}$ power law, which is different
from the $k^{-5/3}$ power law for the spectrum of the density-weighted velocity, $\rho^{1/3}\textbf{u}$, reported
by \citet{Kritsuk07}. It implies that in compressible turbulence the density and temperature do not have the
direct statistical coupling, they only connect with each other through velocity.

\begin{figure}
\centerline{\includegraphics[width=10cm,angle=-90]{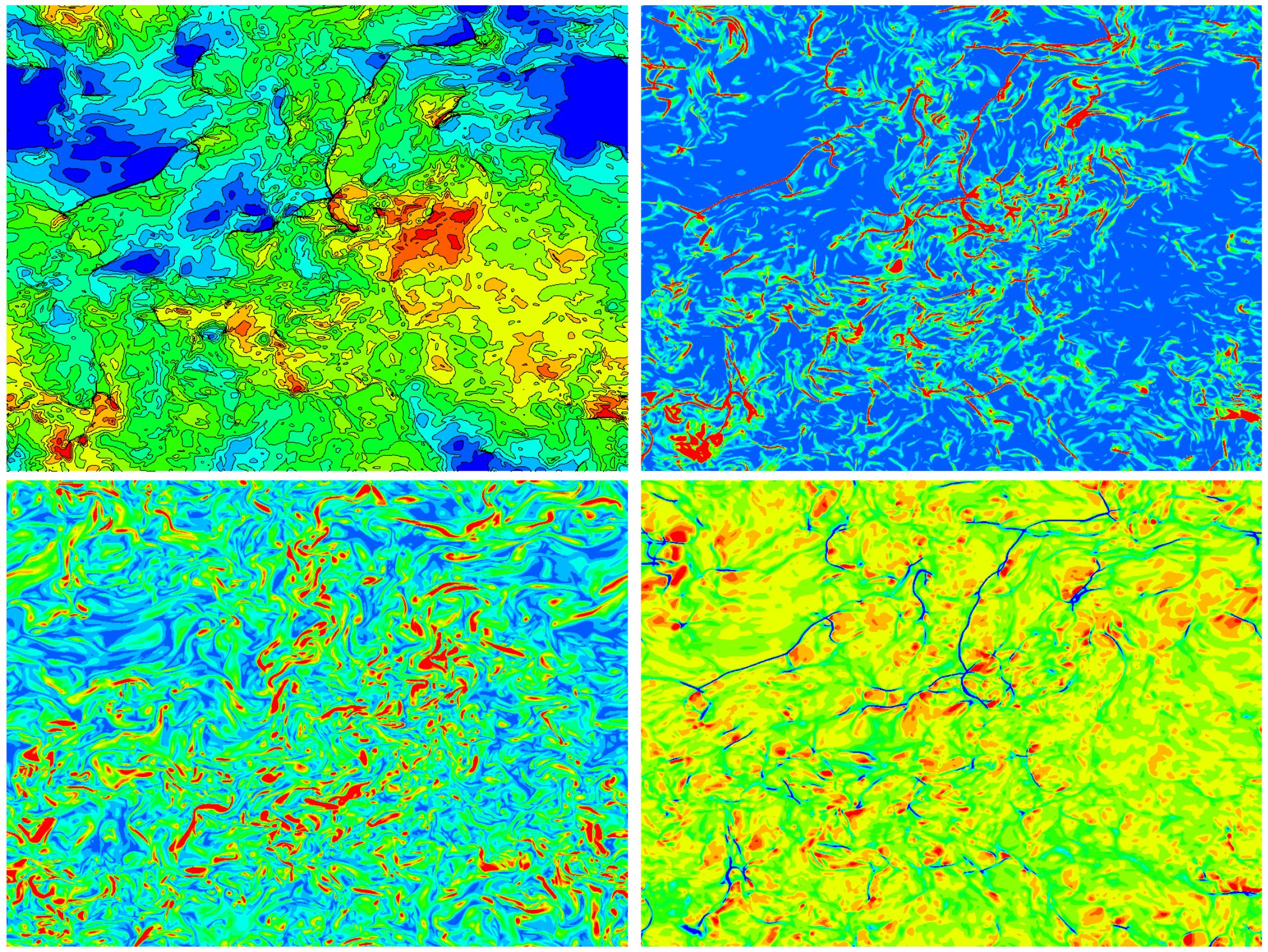}}
\caption{Two-dimensional contours of temperature (top left), temperature dissipation rate
(top right), vorticity magnitude (bottom left), and dilatation (bottom right) in SFT.}
\label{fig:fig3}
\end{figure}
\begin{figure}
\centerline{\includegraphics[width=10cm,angle=-90]{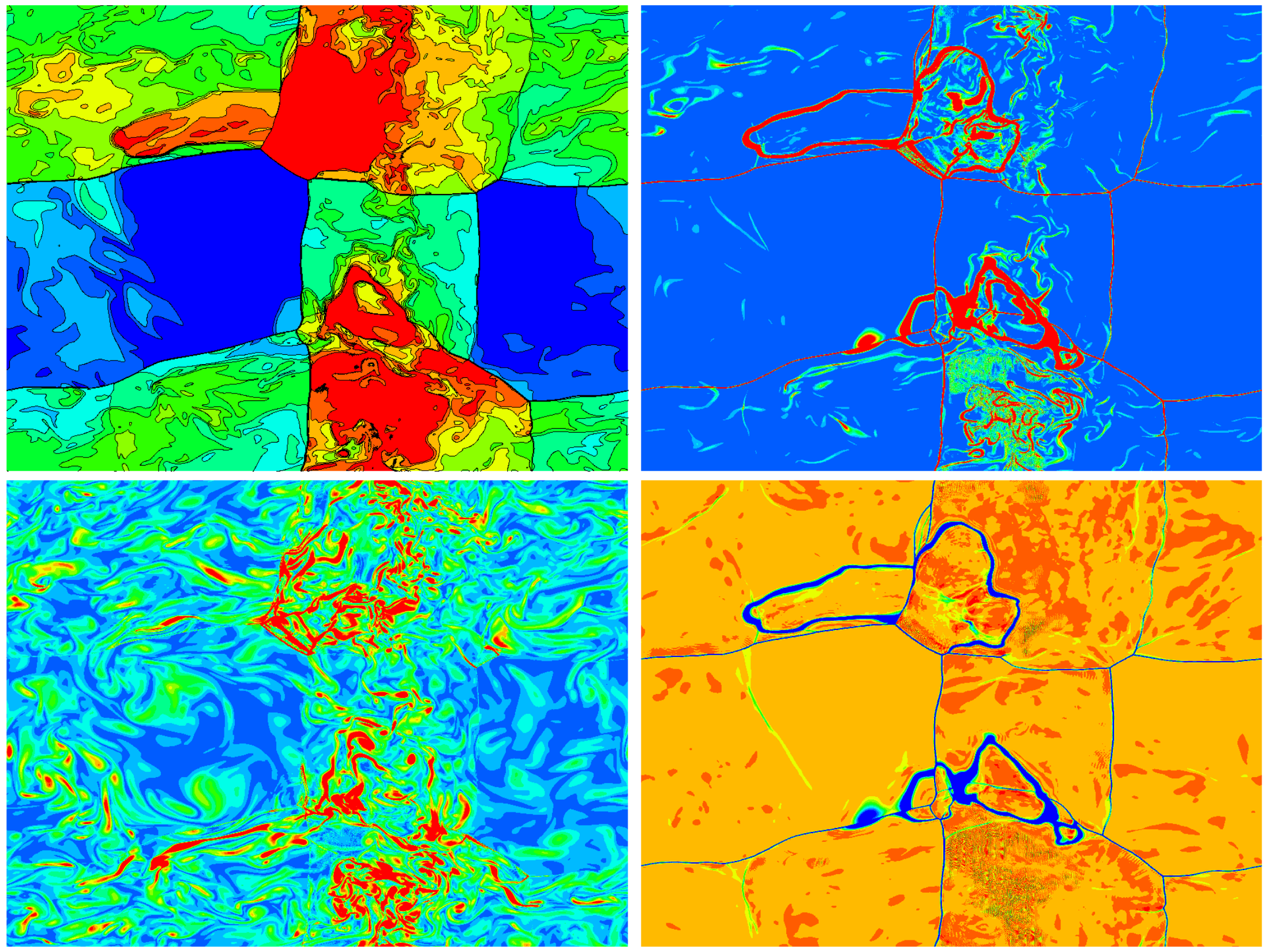}}
\caption{Two-dimensional contours of temperature (top left), temperature dissipation rate
(top right), vorticity magnitude (bottom left), and dilatation (bottom right) in CFT.}
\label{fig:fig4}
\end{figure}

We now take attention on the field structures. Figure~\ref{fig:fig3} provides the two-dimensional (2D) contours
of temperature (top left), temperature dissipation rate (top right), vorticity magnitude (bottom left) and
dilatation (bottom right) in SFT. Here the temperature field has the "ramp-cliff" structures, to some extent,
is like the scalar field in incompressible flows \citep{Shraiman94}. In particular, the small-scale cliffs
with high gradients of fluctuations divide the large-scale ramps with low gradients of fluctuations.
The temperature dissipation field consists of the smooth, low-dissipation sea and the sharp, high-dissipation
discontinuities under random distribution. For the dilatation field, the sharp discontinuities are exactly
the small-scale shocklets. Moreover, the intermittency of the vorticity field is reflected by the inhomogeneous
distribution of the small-scale strong vortices. The same four 2D contours of CFT are depicted in Figure~\ref{fig:fig4}.
There appear large-scale discontinuities in the temperature, temperature dissipation and dilatation fields, causing
by the large-scale shock waves. In front of shock waves, the temperature fluctuations are small, however, they
increase quickly once across the shock waves. It further shows that for the vorticity field, the small-scale
strong vortices preferentially concentrate in the high temperature regions, and thus, displays intensive intermittency.

\subsection{\emph{Probability Distribution Functions}}

\begin{figure}
\centerline{\includegraphics[width=8cm]{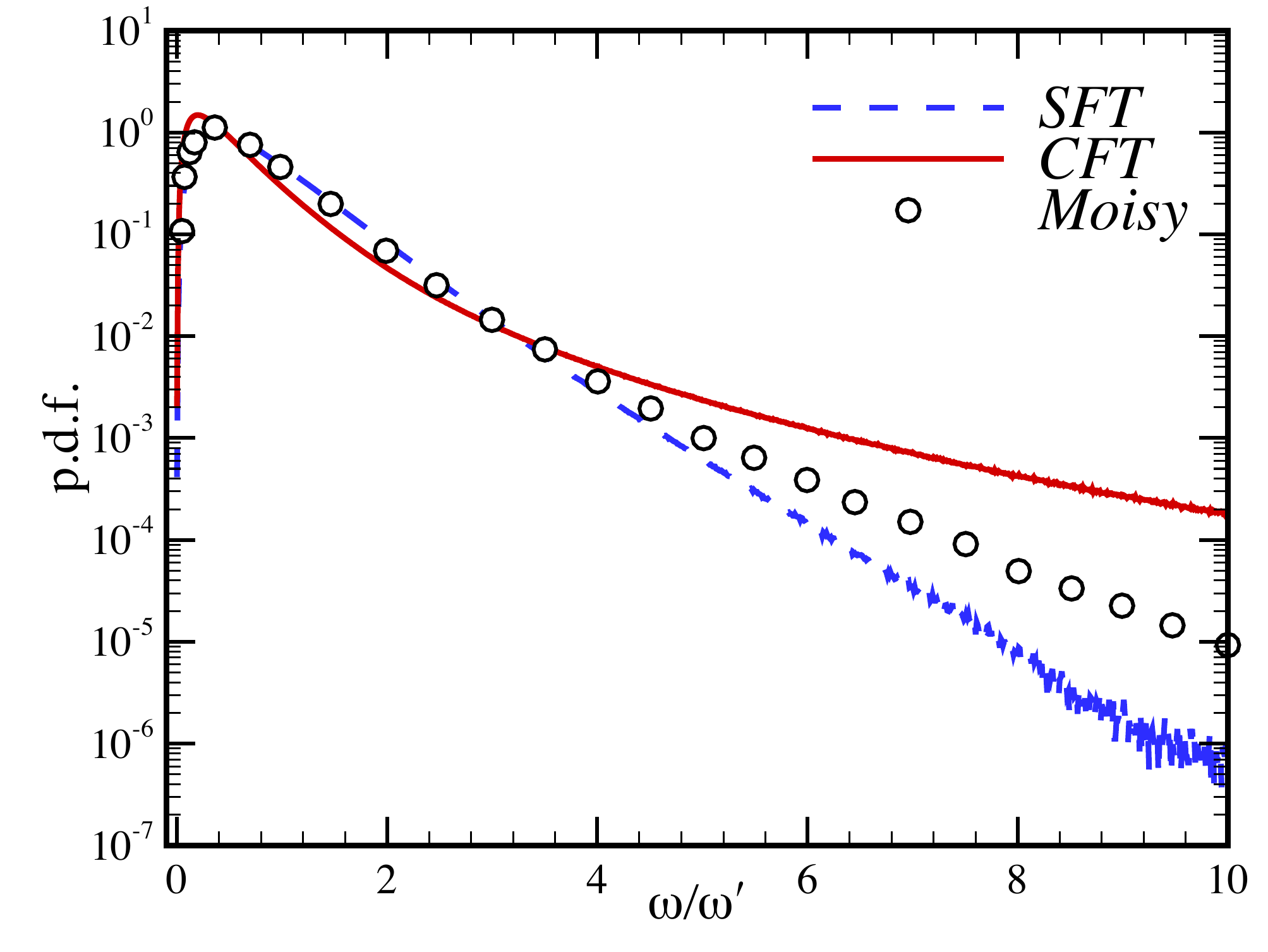}}
\caption{The p.d.f. of normalized vorticity magnitude, where the dashed and solid lines
are for SFT and CFT, respectively, and the circles are from incompressible turbulence \citep{Moisy04}}
\label{fig:fig5}
\end{figure}

In Figure~\ref{fig:fig5} we plot the p.d.f.s of the normalized vorticity magnitude. As expected, they
exhibit well-defined exponential tails, and the one in CFT is very long, showing strong vorticity
fluctuations in the condition that the compressive mode of velocity is stimulated. Here we also present
the incompressible p.d.f. provided by \citet{Moisy04}. At large amplitudes, the p.d.f. collapses between
those from SFT and CFT. This reveals that the declaration in \citet{Wang12b} that the intense vorticity
was suppressed in compressible turbulence works only for the solenoidal forced flow.

\begin{figure}
\centerline{\includegraphics[width=8cm]{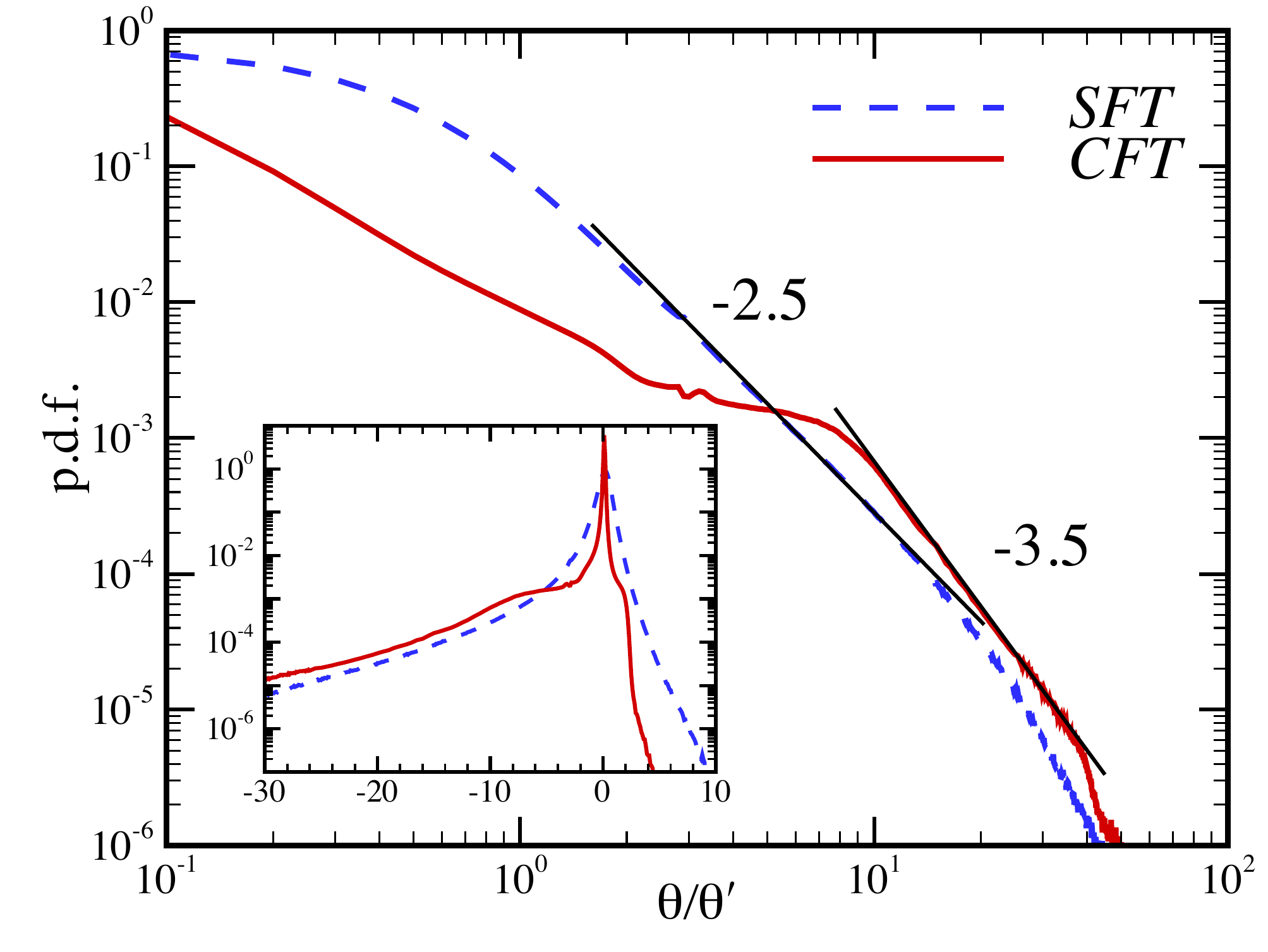}}
\caption{Log-log plot of the p.d.f. of normalized dilatation in compression region, where the
dashed and solid lines are for SFT and CFT, respectively. Inset: the p.d.f.s of normalized dilatation.}
\label{fig:fig6}
\end{figure}

The log-log plot of the p.d.f.s against the normalized negative dilatations is shown in Figure~\ref{fig:fig6}.
In strong compression region, the p.d.f.s have the power-law tails, which are qualitatively similar to the p.d.f.
of velocity derivative in Burgers turbulence \citep{Bec07}. According to the stochastic theory of Burgers equation,
it is the preshock leads to the large negative velocity derivative, and then the power-law tail
\citep{E97, E99, E00, Bec00, Bec01}. In our simulations the exponents of the power-law tails are around $-2.5$ in
SFT and $-3.5$ in CFT. We notice that the exponent value in the compressive forced flow is the same to that in the
Burgers turbulence. Indeed, previous study \citep{Wang12a} showed that the power-law region in compressible turbulence
is mainly caused by preshocks and weak shocklets rather than strong shock waves. In the inset we plot the p.d.f.s
throughout the entire dilatation range. Obviously, they exhibit strong skewness towards negative side.

To understand the mechanism of the power-law behavior, we write the Liouville equation for the p.d.f. of
dilatation, $Q(\theta)$, as follows
\begin{equation}
\frac{\partial}{\partial t}Q -\frac{\partial}{\partial\theta}\Big[\big(P+R-D+\theta^2\big)Q\Big] = \theta Q.
\end{equation}
Here $P(\theta)$, $R(\theta)$ and $D(\theta)$ are the ensemble averages of pressure,
anisotropic straining and viscous dissipation conditioned on dilatation, respectively
\begin{eqnarray}
&& P(\theta)=\big\langle\frac{1}{\gamma M^2}\frac{\partial}{\partial x_j}
(\frac{1}{\rho}\frac{\partial p}{\partial x_j})|\theta\big\rangle,
\label{pressure}\\
&& R(\theta)=\big\langle(\frac{\partial u_j}{\partial x_i}
\frac{\partial u_i}{\partial x_j}-\theta^2)|\theta\big\rangle,
\label{anisotropy}\\
&& D(\theta)=\frac{4\nu_0}{3Re}\big\langle\frac{\partial^2\theta}{\partial x_j^2}|\theta\big\rangle.
\label{viscosity}
\end{eqnarray}

\begin{figure}
\begin{center}
\subfigure{
\resizebox*{6.5cm}{!}{\rotatebox{0}{\includegraphics{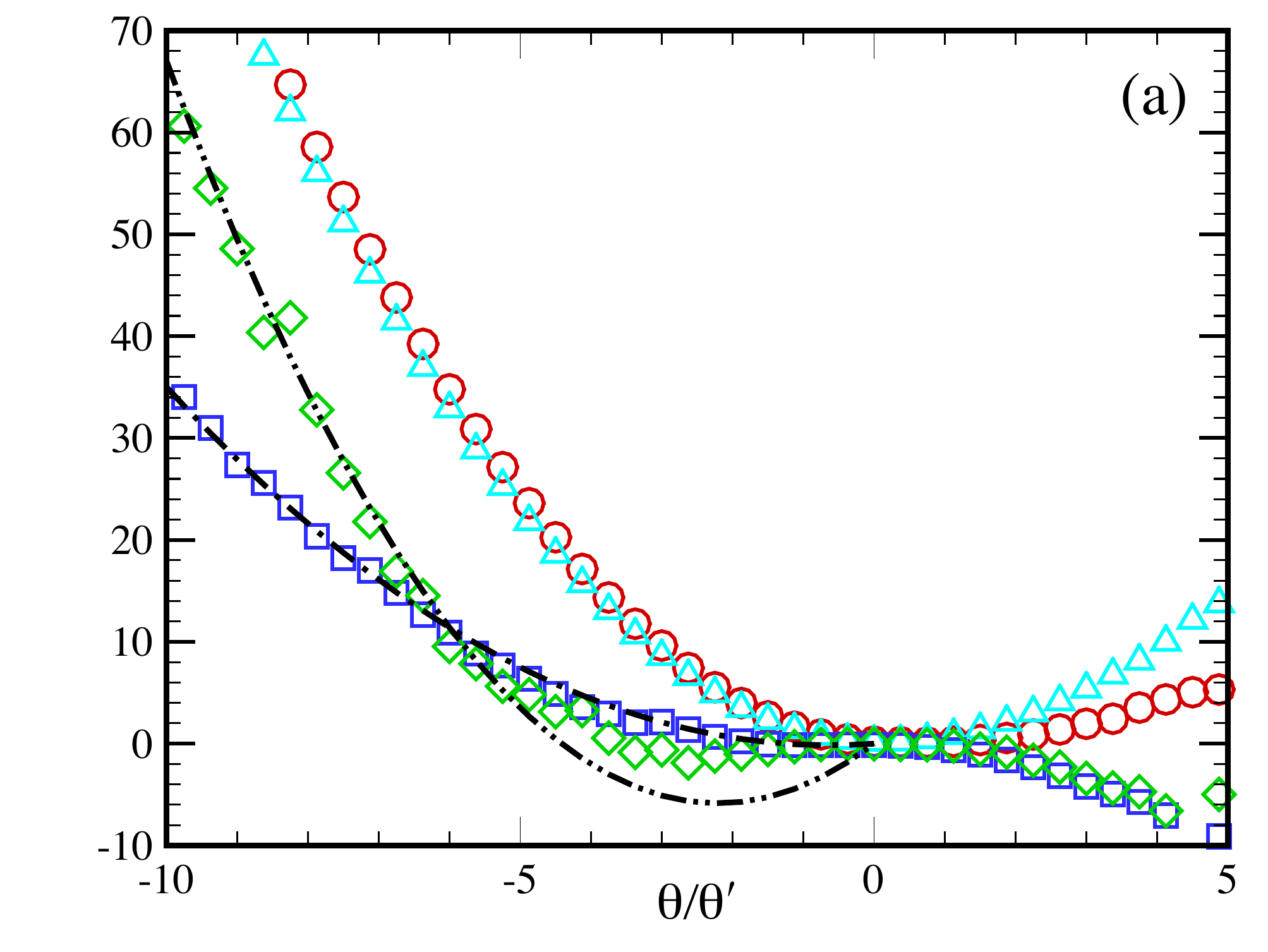}}}}%
\subfigure{
\resizebox*{6.5cm}{!}{\rotatebox{0}{\includegraphics{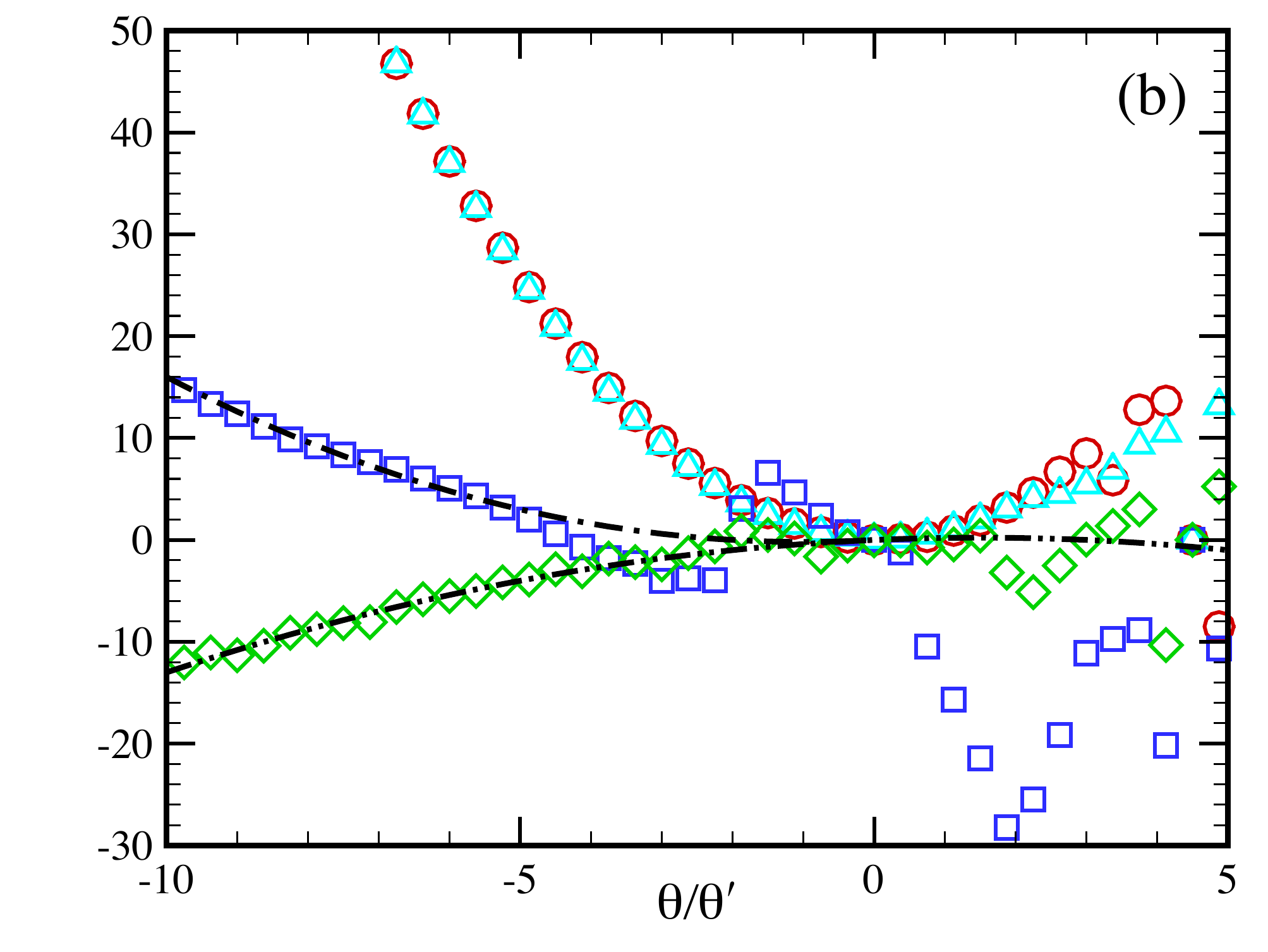}}}}%
\caption{$S(\theta/\theta')$ (circles), $S^c(\theta/\theta')$ (deltas), $P(\theta/\theta')$ (diamonds) and
$D(\theta/\theta')$ (circles) as functions of normalized dilatation. The lines are for the fitting by a
quadratic-parabolic formulation at various values of coefficients. (a) SFT, (b) CFT.}
\label{fig:fig7}
\end{center}
\end{figure}

In Figure~\ref{fig:fig7} we first plot the conditional averages of straining and its
compressive component,
$S(\theta/\theta')=\langle\partial u_j/\partial x_i\partial
u_i/\partial x_j/\theta'^2|\theta/\theta'\rangle$ and $S^c(\theta/\theta')=\langle\partial u^c_j/\partial x_i\partial
u^c_i/\partial x_j/\theta'^2|\theta/\theta'\rangle$, as functions of the normalized dilatation $\theta/\theta'$.
It shows that for in the compression regions for both flows, the compressive component dominates the overall straining.
Namely, the contribution from the solenoidal component is negligible. We then plot $P(\theta/\theta')$
and $D(\theta/\theta')$ against $\theta/\theta'$. Similar to \citet{Wang12b}, in SFT $P(\theta/\theta')$ and $D(\theta/\theta')$
are positive and can be well described by a quadratic-parabolic formulation of $c_1(\theta/\theta')^2+c_2(\theta/\theta')$.
The values of $c_1$ and $c_2$ are the same to those in \citet{Wang12b}. They are $c_{1p}=1.2$, $c_{2p}=5.3$, $c_{1d}=0.4$
and $c_{2d}=0.5$, where the subscripts $p$ and $d$ are for pressure and viscous dissipation, respectively. By contrast,
in CFT the values of $c_{1d}$ and $c_{2d}$ are $0.2$ and $0.4$, respectively. However, $P(\theta/\theta')$ is negative,
and its fitting coefficients are $c_{1p}=-0.1$ and $c_{2p}=0.3$.

For large negative dilatation, the stationary solution of Equation (3.7) is
\begin{equation}
Q(\theta) \propto \theta^{-q},
\end{equation}
where $q=2+1/(1+c_{1p}-c_{1d})$. Immediately, we obtain that $q=2.56$ for SFT and $q=3.43$ for CFT. which are very close
to the exponent values of $2.5$ and $3.5$ shown in Figure~\ref{fig:fig6}. It is worth to point out that in SFT the contributions
to the power-law exponent from the pressure and viscous dissipation are converse, while in CFT they become the same
because of the reversal of the contribution from pressure.

\begin{figure}
\begin{center}
\subfigure{
\resizebox*{6.5cm}{!}{\rotatebox{0}{\includegraphics{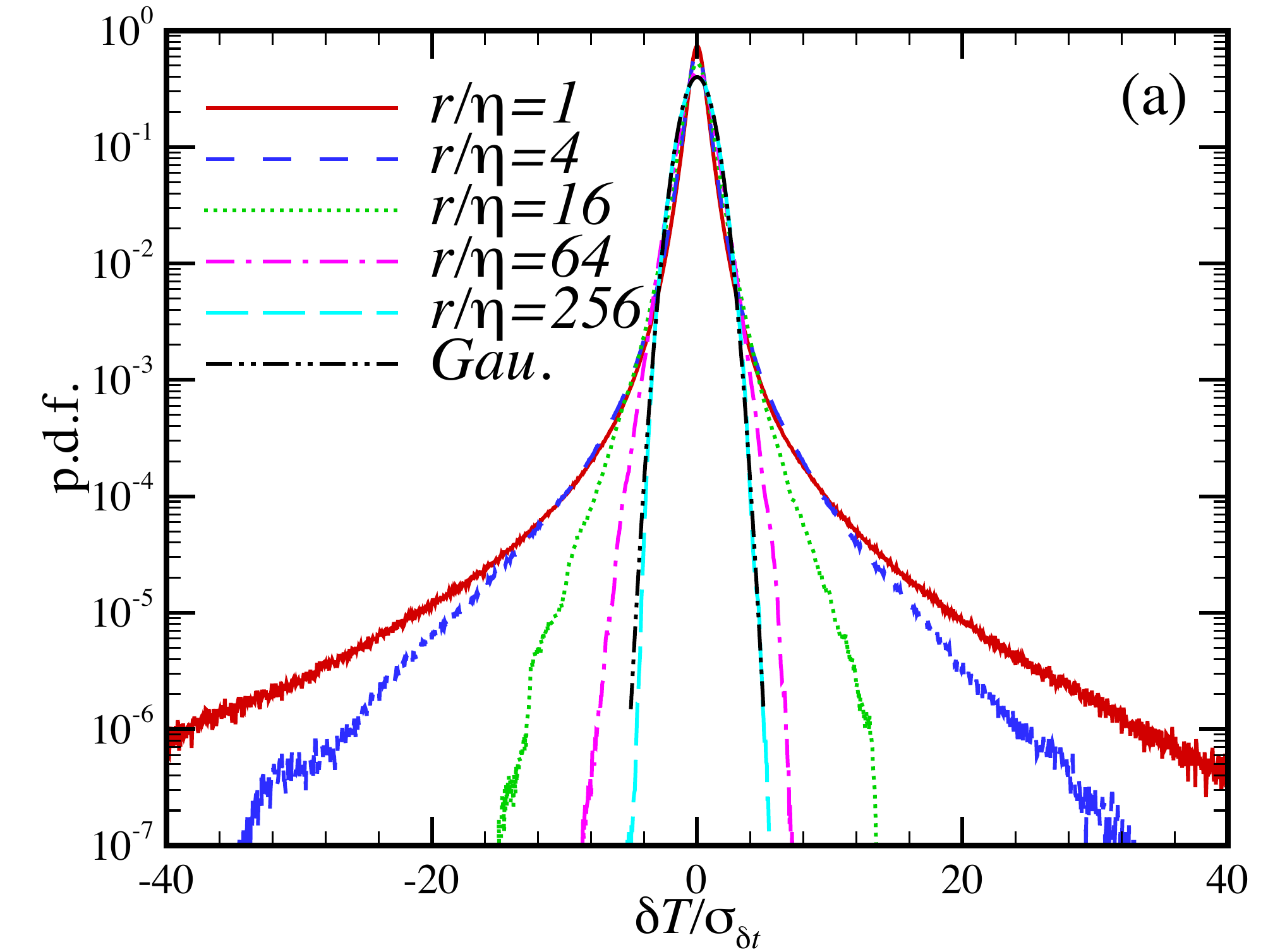}}}}%
\subfigure{
\resizebox*{6.5cm}{!}{\rotatebox{0}{\includegraphics{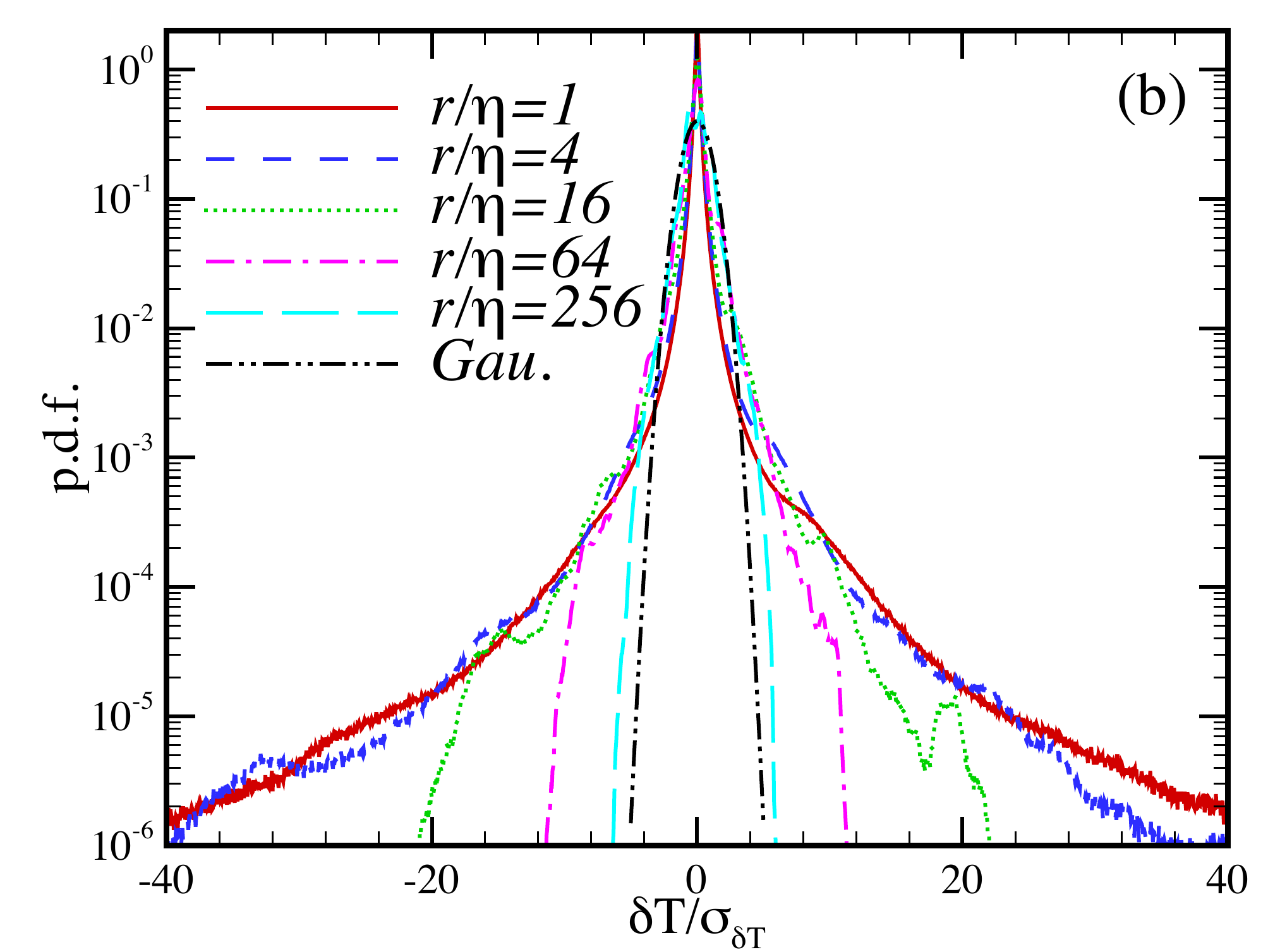}}}}%
\caption{The p.d.f.s of normalized temperature increment, where $\sigma_{\delta T}$ is
the standard deviation of temperature increment. (a) SFT, (b) CFT.}
\label{fig:fig8}
\end{center}
\end{figure}
\begin{figure}
\begin{center}
\subfigure{
\resizebox*{6.5cm}{!}{\rotatebox{0}{\includegraphics{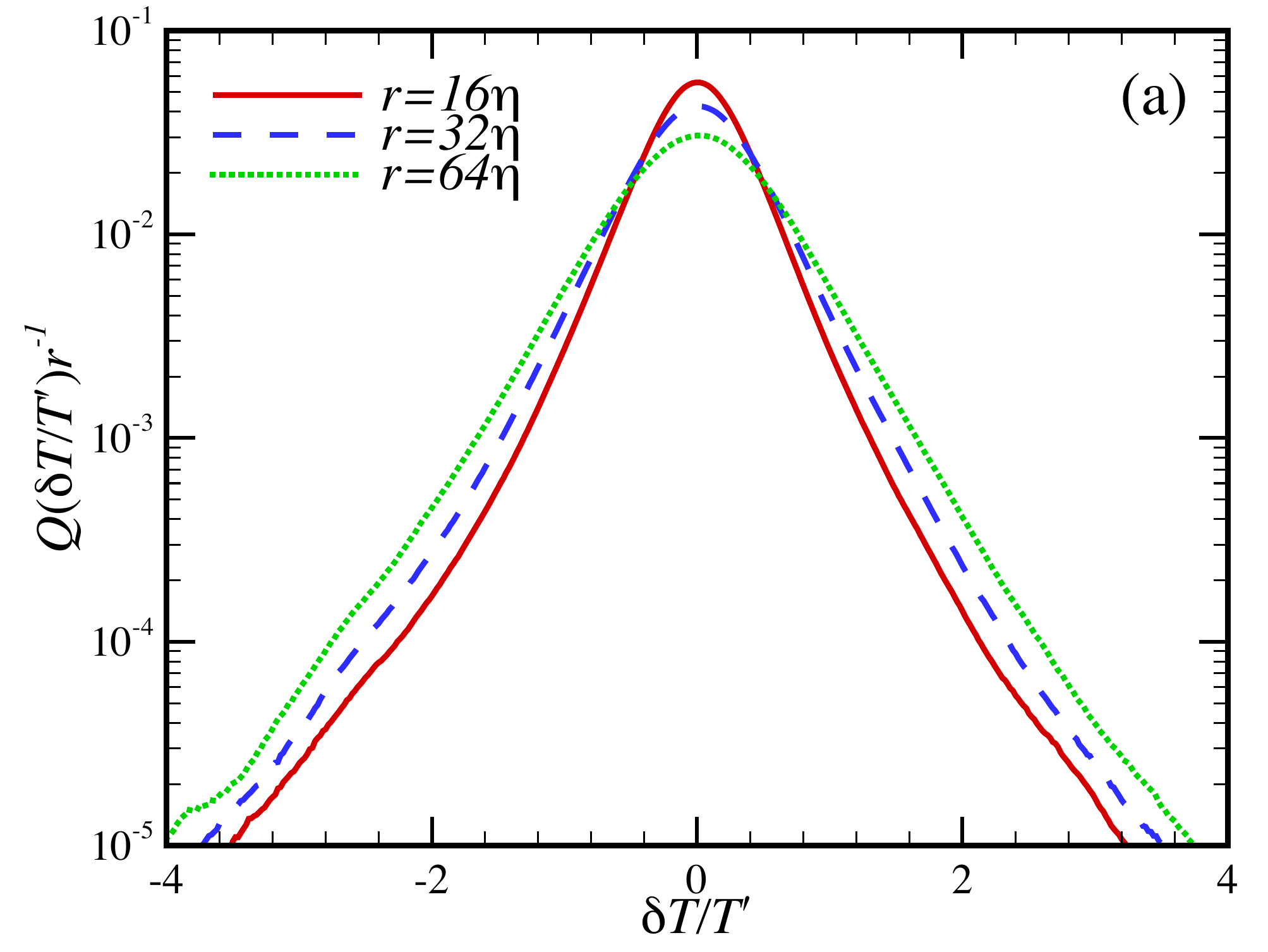}}}}%
\subfigure{
\resizebox*{6.5cm}{!}{\rotatebox{0}{\includegraphics{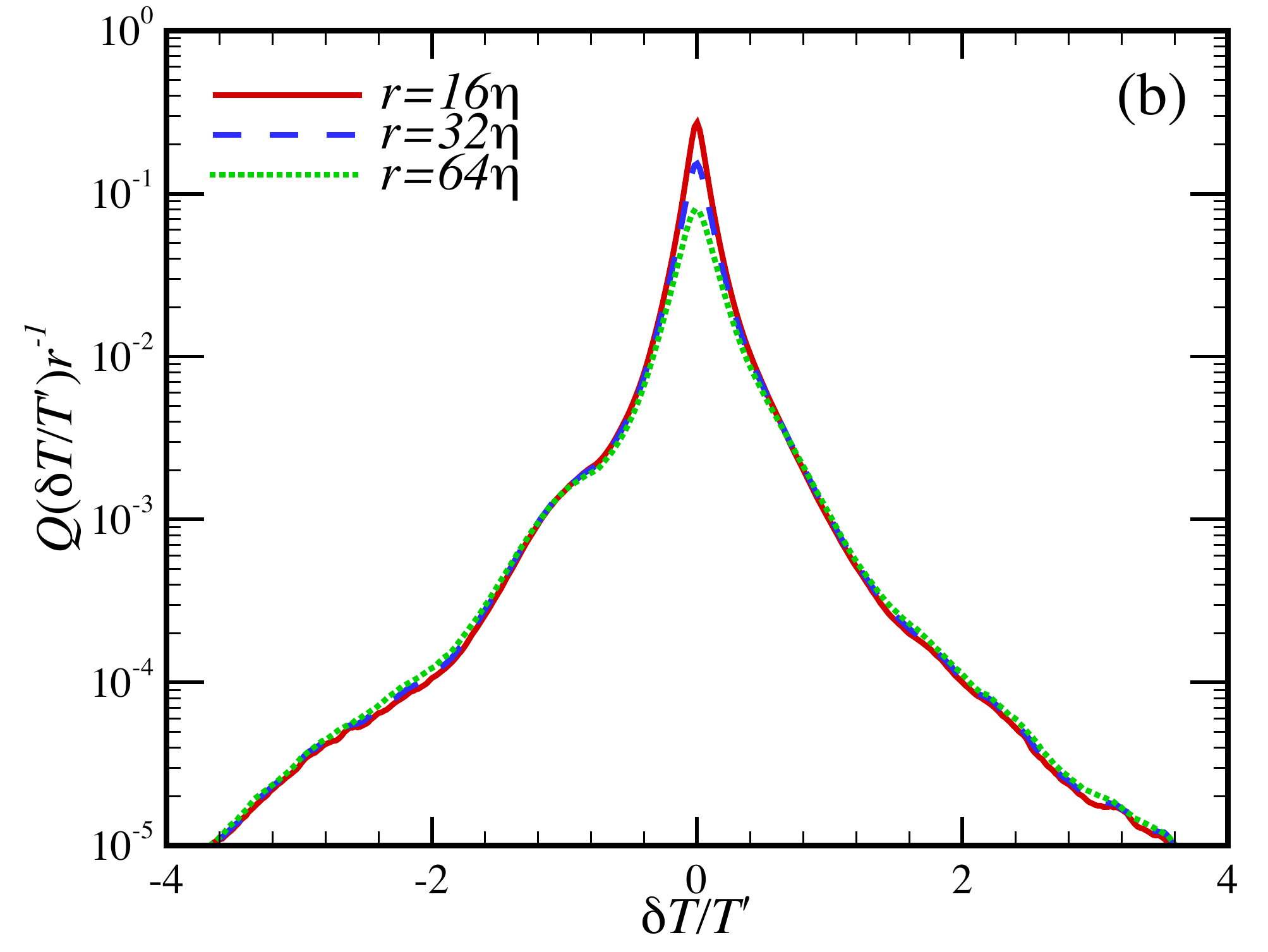}}}}%
\caption{The rescaled p.d.f.s of normalized temperature increment. (a) SFT, (b) CFT.}
\label{fig:fig9}
\end{center}
\end{figure}

In Figure~\ref{fig:fig8} we plot the two-point p.d.f.s of the normalized temperature increment at the
normalized separation distances of $r/\eta=1$, $4$, $16$, $64$ and $256$. Here $\delta T=T(\textbf{x}+r)-T(\textbf{x})$
is the temperature increment. It shows that the p.d.f. is basically symmetric for each scale, and gets
broader as scale increases. In particular, in SFT the p.d.f. at $r=256\eta$ is approximately Gaussian,
while that in CFT still keeps super-Gaussian. This implies that the presence of large-scale shock wave makes
the temperature field be intermittent, even though it lies on the integral scale. At large amplitudes,
the shape of p.d.f. is concave, similar to that of longitudinal velocity increment shown in \citet{Ni15a}.

The rescaled p.d.f., $Q(\delta T/T')/r$, for the inertial range of $r/\eta=16$, $32$ and $64$ are shown
in Figure~\ref{fig:fig9}. It is found that in CFT the p.d.f.s collapse to the same distribution.
According to multifractal theory \citep{Benzi08}, the scaling exponents for the statistical moments of temperature
increment in the inertial range of CFT will saturate at high order numbers. In other words,
$\langle|\delta T|^p\rangle \propto r^{z_{\infty}}$ for large $p$, and $z_{\infty}=1$.

\section{Isentropic Approximation on Thermodynamic Fields}

In compressible turbulence it is often valuable to reduce the number of thermodynamic variables, for simplifying theoretical
analysis and developing engineering model. A wide used approach, though not strictly exact because of the irreversible
dissipative nature of turbulence, is to assume that thermodynamic processes occur isentropically
\citep{Chandrasekhar51, Erlebacher90}. According to the state equation, the conservation of energy at constant entropy
indicates that the instantaneous density, temperature and pressure are related as follows
\begin{equation}
T/\langle T\rangle = \big(\rho/\langle\rho\rangle\big)^{\gamma-1} = \big(p/\langle p\rangle\big)^{1-1/\gamma}.
\end{equation}

\begin{figure}
\begin{center}
\subfigure{
\resizebox*{6.5cm}{!}{\rotatebox{0}{\includegraphics{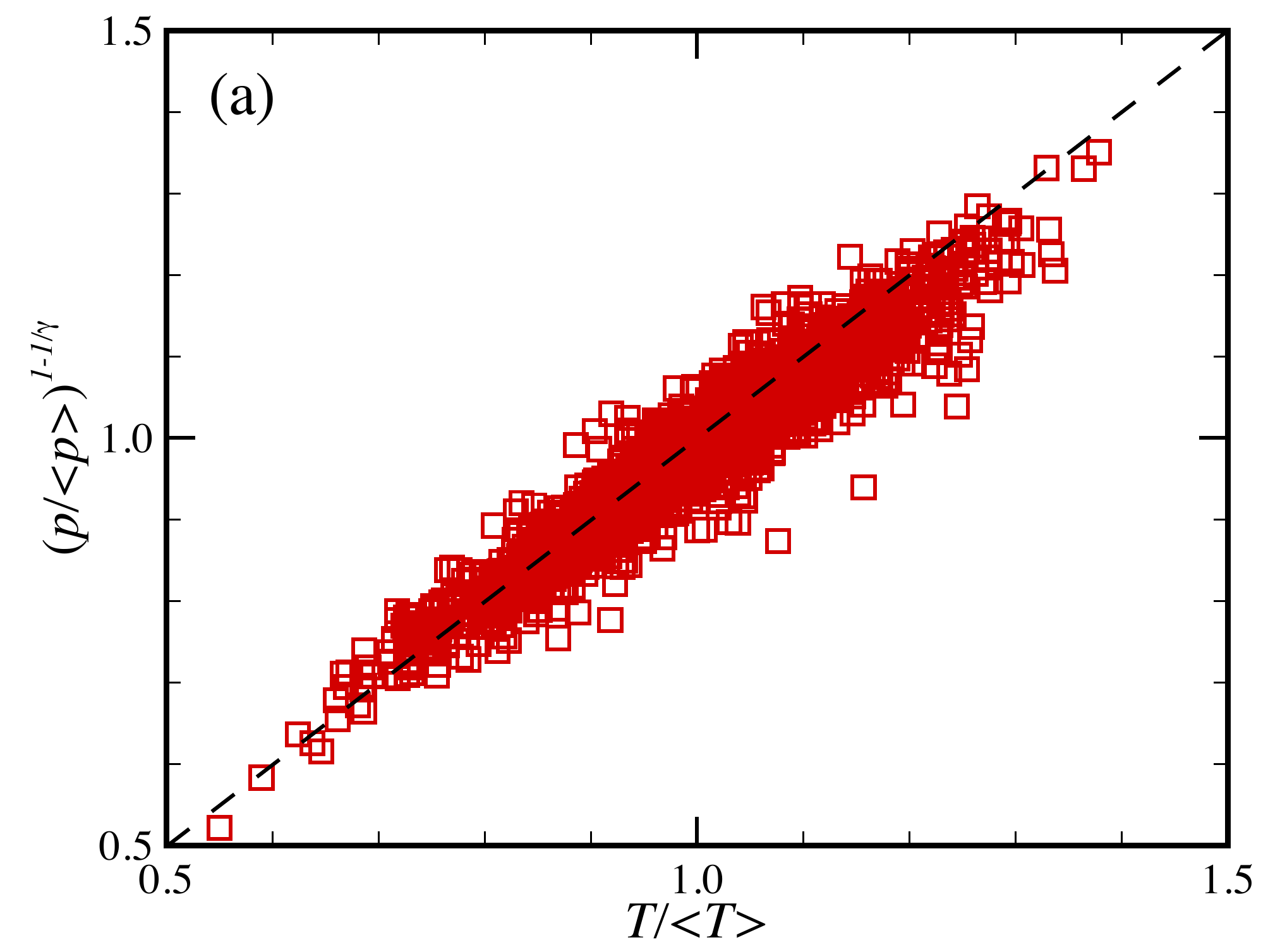}}}}%
\subfigure{
\resizebox*{6.5cm}{!}{\rotatebox{0}{\includegraphics{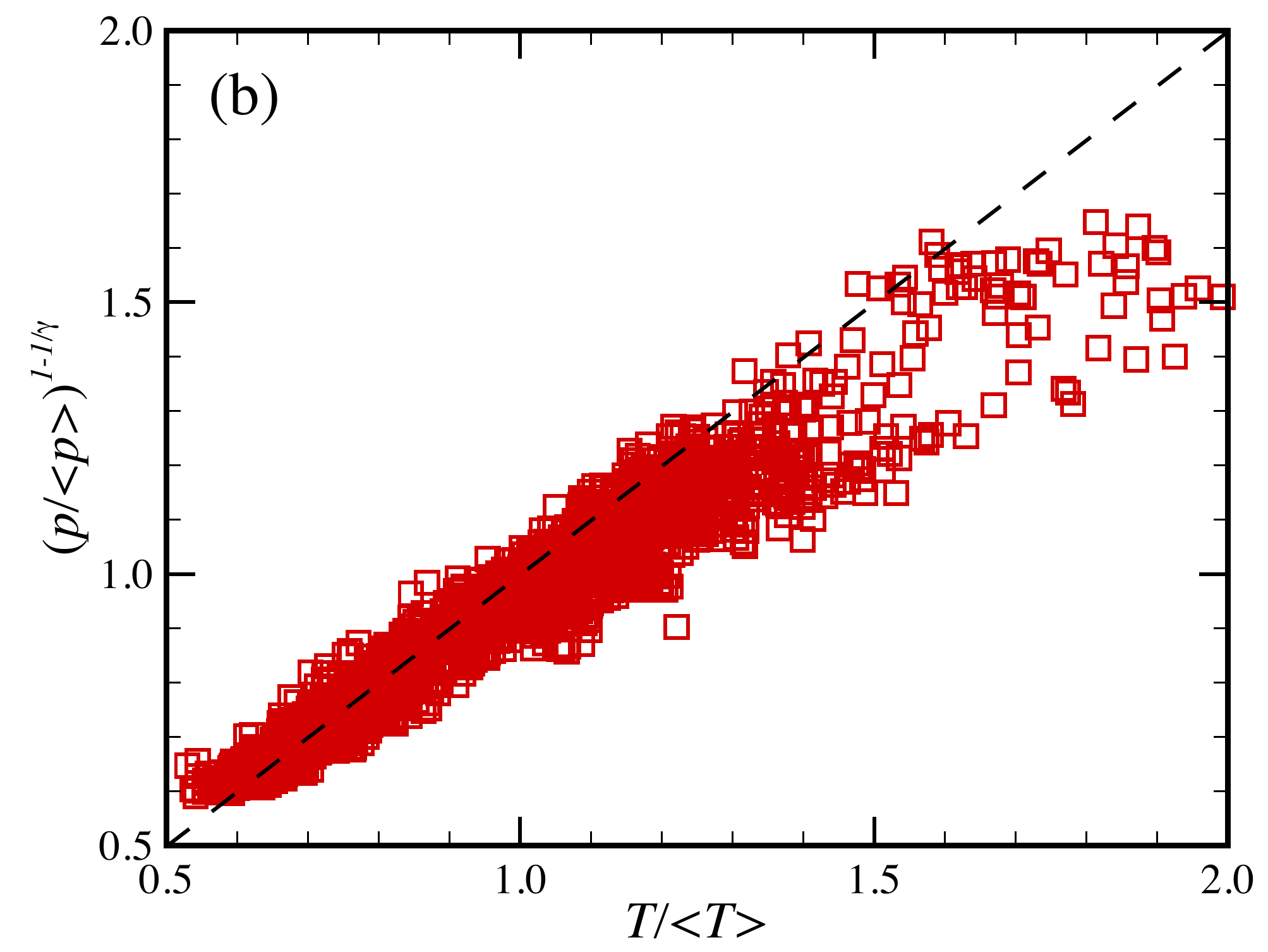}}}}%
\caption{$T/\langle T\rangle$ versus $(p/\langle p\rangle)^{1-1/\gamma}$, where $2197$ points are
used in the plot. The dashed line is for isentropic relation. (a) SFT, (b) CFT.}
\label{fig:fig10}
\end{center}
\end{figure}

In Figure~\ref{fig:fig10} we plot $T/\langle T\rangle$ against $(p/\langle p\rangle)^{1-1/\gamma}$.
A total of $2197$ points are presented, and the dashed line stands for the isentropic relation. In SFT most
of the points collapse onto the isentropic line. There are only about $1.6\%$ points failing to
satisfy the isentropic relation within a tolerance of $C_{tol}=10\%$, which is defined by
\begin{equation}
C_{tol}\equiv |1-(p/\langle p\rangle)^{1-1/\gamma}/(T/\langle T\rangle)|.
\end{equation}
Nevertheless, within the same tolerance there are about $18.9\%$ points failing to satisfy the
isentropic relation in CFT. The large-scale shock waves make the fluctuations of thermodynamic
variables intense and destroy the insentropity, especially at large amplitudes. We then define
the correlation coefficient between $T/\langle T\rangle$ and $(p/\langle p\rangle)^{1-1/\gamma}$ as follows
\begin{equation}
\mathcal{C}_{Tp}\equiv \frac{\Big\langle\big(T/\langle T\rangle-1\big)\big((p/\langle p\rangle)^{1-1/\gamma}-1\big)\Big\rangle}
{\Big\langle\big(T/\langle T\rangle-1\big)^2\Big\rangle^{1/2}\Big\langle\big((p/\langle p\rangle)^{1-1/\gamma}-1\big)^2\Big\rangle^{1/2}}.
\end{equation}
It shows that the values of $\mathcal{C}_{Tp}$ for both SFT and CFT are $0.96$.

\begin{figure}
\centerline{\includegraphics[width=8cm]{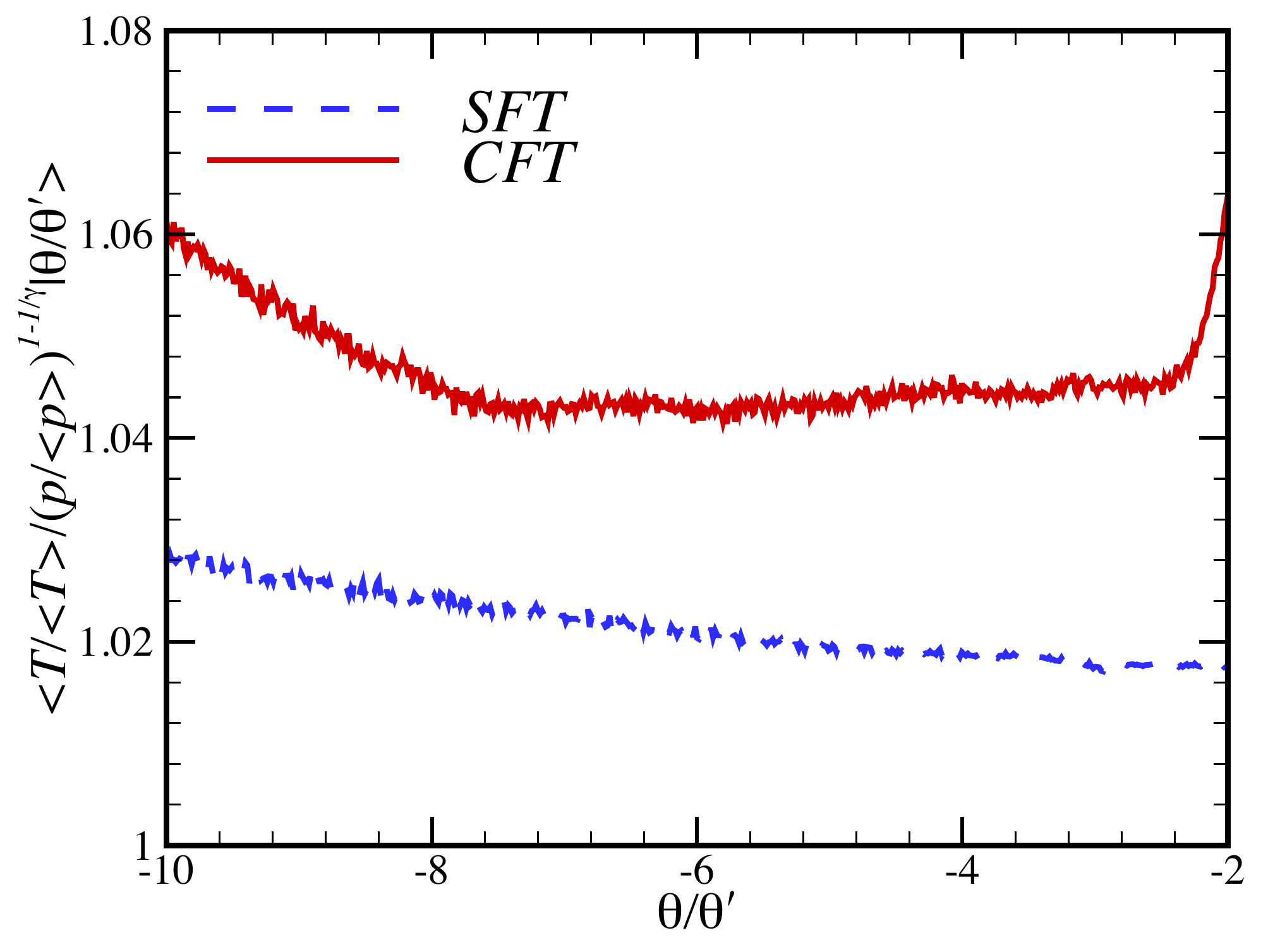}}
\caption{Conditional averages of ratio of $T/\langle T\rangle$ to $(p/\langle p\rangle)^{1-1/\gamma}$, where the dashed and
solid lines are for SFT and CFT, respectively.}
\label{fig:fig11}
\end{figure}

The average of the ratio of $T/\langle T\rangle$ to $(p/\langle p\rangle)^{1-1/\gamma}$ conditioned on dilatation,
as a function of the normalized dilatation, is shown in Figure~\ref{fig:fig11}. In the compression region of
$-10\leq \theta/\theta'\leq -2$, the average value in SFT is near unity, and slightly decreases as compression
decreases. This implies that the isentropic approximation of thermodynamic variables is valid in the small-scale shocklet
regions. By contrast, the average value in the same compression region of CFT is a bit higher, and is close
to $1.04$ in the range of $-8\leq \theta/\theta'\leq -2.4$. In a word, for the compression region of compressible turbulence,
one can use the isentropic approximation to facilitate the description of thermodynamic variables.

\begin{figure}
\centerline{\includegraphics[width=8cm]{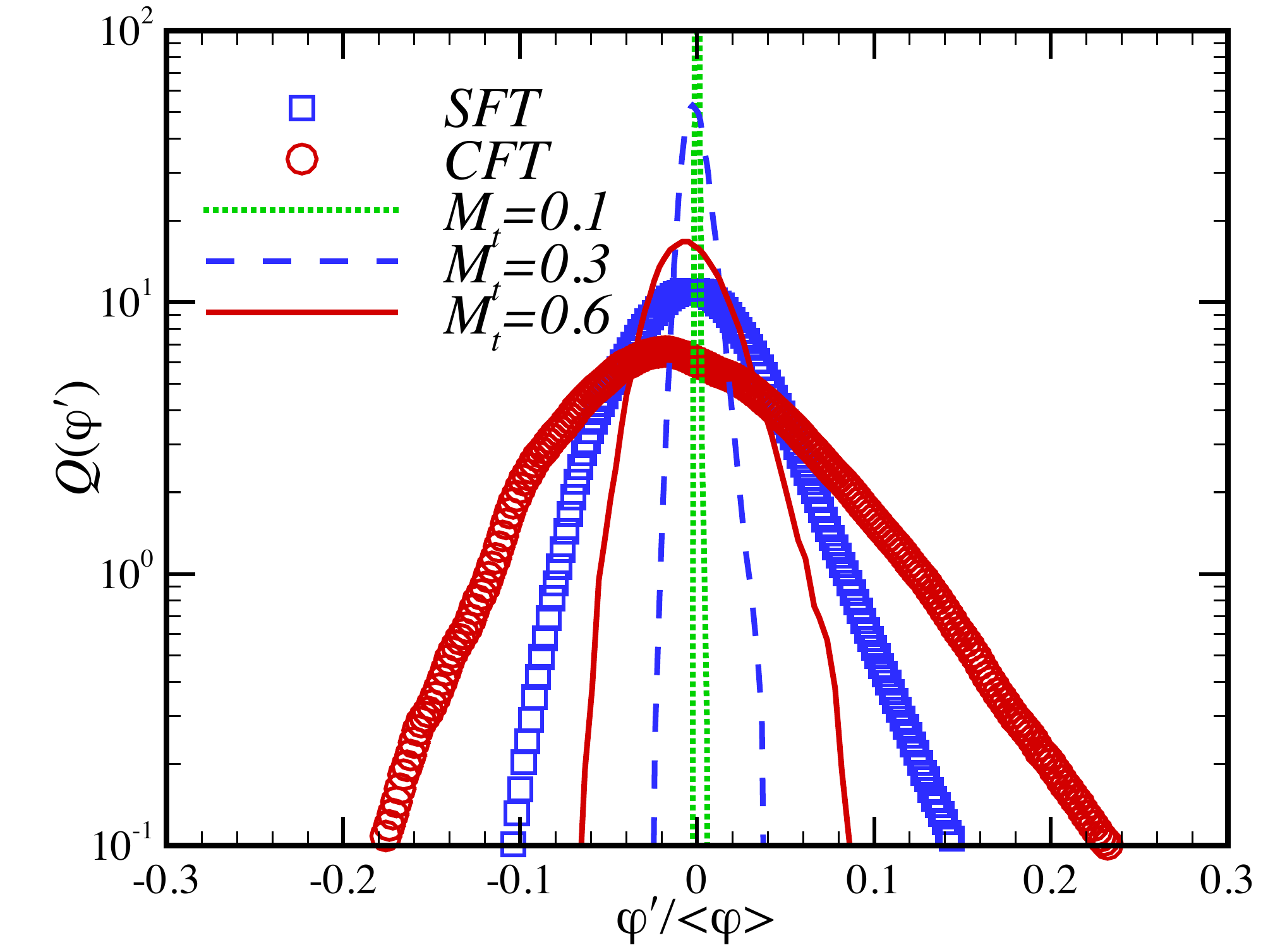}}
\caption{$Q(\varphi')$ in SFT (squares) and CFT (circles).The dotted, dashed and solid lines are for those
at $M_t=0.1$, $0.3$ and $0.6$ flows presented in \citet{Donzis13}.}
\label{fig:fig12}
\end{figure}

As one known, the thermodynamic fluctuations are related through Equation (4.1) on an instantaneous basis. We now
introduce a quantity of
\begin{equation}
\varphi\equiv\frac{p/\langle p\rangle}{\big(\rho/\langle\rho\rangle\big)^{\gamma}}
=\frac{p/\langle p\rangle}{\big(T/\langle T\rangle\big)^{\frac{\gamma}{\gamma-1}}}
\end{equation}
to measure the exactitude of isentropic approximation. Obviously, if the case is isentropic, the distribution of $\varphi'=\varphi-\langle\varphi\rangle$ will follow the Dirac delta function, $Q(\varphi')=\delta(\varphi')$.
In Figure~\ref{fig:fig12} we plot $Q(\varphi')$ from SFT and CFT. For comparison, those at $M_t=0.1$, $0.3$ and $0.6$
in \citet{Donzis13} are also presented. It shows that in the solenoidal forced flow with $M_t=0.1$, $Q(\varphi')$
is very close to $\delta(\varphi')$, and deviates from that as $M_t$ increases. At $M_t\approx 0.6$,
$Q(\varphi')$ of CFT is much wider than that of SFT. Furthermore, the value of the standard derivation,
$\langle(\varphi'/\langle\varphi\rangle)^2\rangle^{1/2}$, which quantifies the departure from isentropic assumption,
is $0.19$ in SFT and $0.27$ in CFT. Here we point out that the results from our simulations do not follow the
$0.1M_t^2$ scaling proposed by \citet{Donzis13}.

The above analysis shows that globally, the thermodynamic variables in compressible turbulence are not exactly
isentropic, because of $Q(\varphi')$ deviates from $\delta(\varphi')$. Nevertheless, by introducing an exponent
parameter $\gamma_c$, we can still use the following expression to connect pressure and temperature
\begin{equation}
p/\langle p\rangle\equiv \big(T/\langle T\rangle\big)^{\frac{\gamma_c}{\gamma_c-1}}.
\end{equation}
It gives that
\begin{equation}
\varphi=\big(T/\langle T\rangle\big)^{\frac{\gamma_c}{\gamma_c-1}-\frac{\gamma}{\gamma-1}}.
\end{equation}
We then use the constraint of probability, $Q(\varphi)d\varphi=Q(T)dT$, to evaluate to what degree
$\gamma_c$ can capture the fluctuations of thermodynamic variables. It yields that
\begin{equation}
Q\big(\varphi\big) = \frac{1}{|\frac{\gamma_c}{\gamma_c-1}-\frac{\gamma}{\gamma-1}|}\langle T\rangle
\big(T/\langle T\rangle\big)^{1+\frac{\gamma}{\gamma-1}-\frac{\gamma_c}{\gamma_c-1}}
Q\big(\langle T\rangle\varphi^{\frac{\gamma}{\gamma-1}-\frac{\gamma_c}{\gamma_c-1}}\big).
\end{equation}
By assuming that the temperature field follows Gaussian distribution, we obtain the expression of
$Q(\varphi)$
\begin{equation}
Q\big(\varphi\big) = \frac{1}{|\frac{\gamma}{\gamma-1}-\frac{\gamma_c}{\gamma_c-1}|}\frac{\varphi^{-\frac{1+\frac{\gamma}{\gamma-1}
+\frac{\gamma_c}{\gamma_c-1}}{\frac{\gamma}{\gamma-1}-\frac{\gamma_c}{\gamma_c-1}}}}{\sqrt{2\pi}\langle T^{*2}\rangle^{1/2}}
\exp\Big[-\frac{1}{2\langle T^{*2}\rangle}\big(\varphi^{\frac{\gamma}{\gamma-1}-\frac{\gamma_c}{\gamma_c-1}}-1\big)^2\Big],
\end{equation}
where $T^*=T^{''}/\langle T\rangle$ is the normalized temperature fluctuation, with $T^{''}$ representing
the temperature fluctuation.

\begin{figure}
\centerline{\includegraphics[width=8cm]{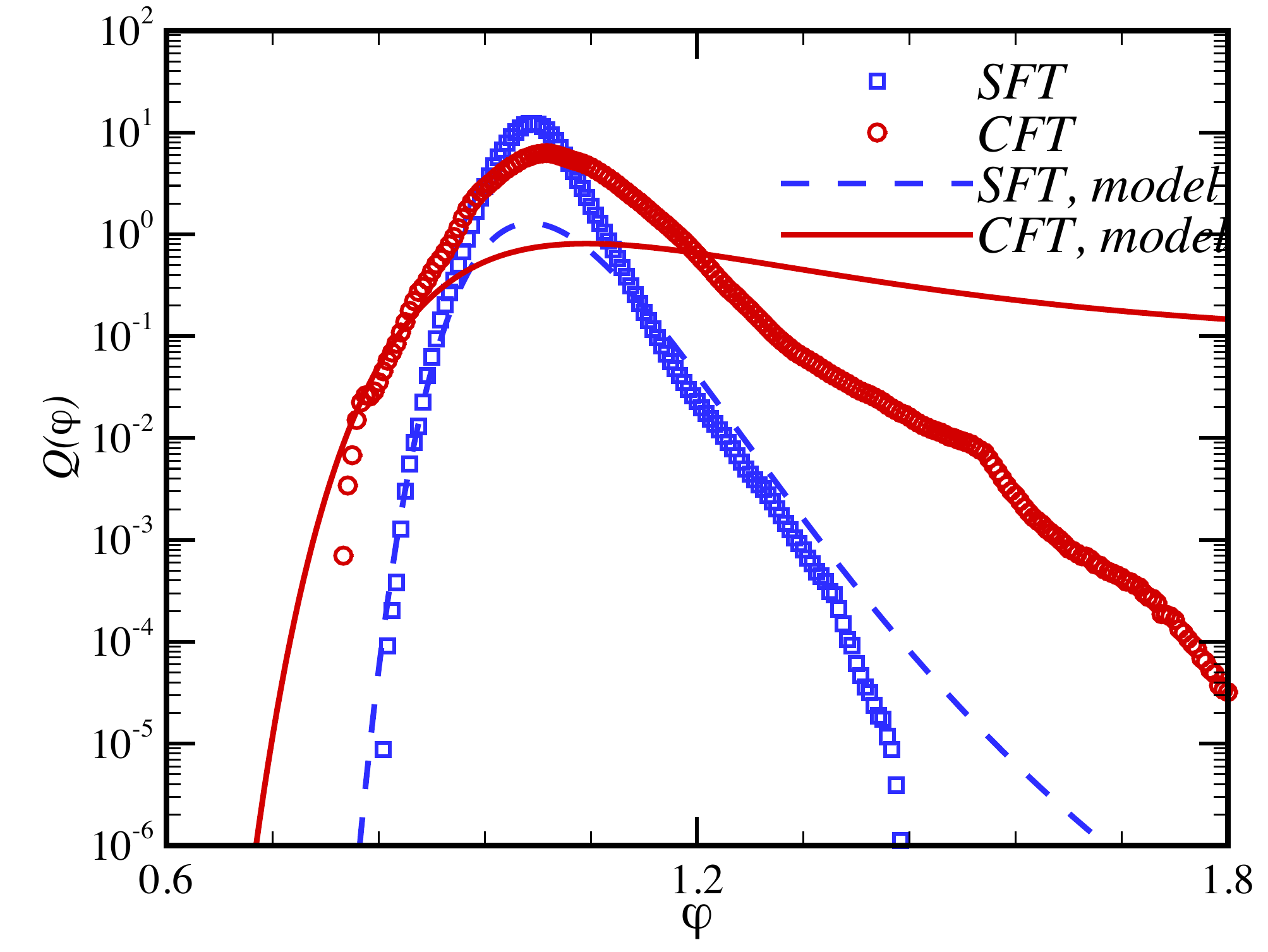}}
\caption{$Q(\varphi)$ in SFT (squares) and CFT (circles). The dashed and solid lines represent the model results for SFT and CFT, respectively.}
\label{fig:fig13}
\end{figure}

In Figure~\ref{fig:fig13} we compare $Q(\varphi)$ between theoretical model and numerical simulation. For a certain condition,
$\gamma_c$ can be obtained through the best fitting of simulation data. In our cases, the values of $\gamma_c$ in SFT and CFT are
around $1.20$ and $1.22$, respectively, smaller than the isentropic value of $1.4$. It displays that $Q(\varphi)$ from SFT
is basically well described by the model, except at the large positive amplitudes. By contrast, $Q(\varphi)$ from CFT
goes far away from the model, which indicates that the Gaussian assumption of temperature field in CFT works badly.

\section{Statistical Properties of Temperature Gradient}

\begin{figure}
\centerline{\includegraphics[width=8cm]{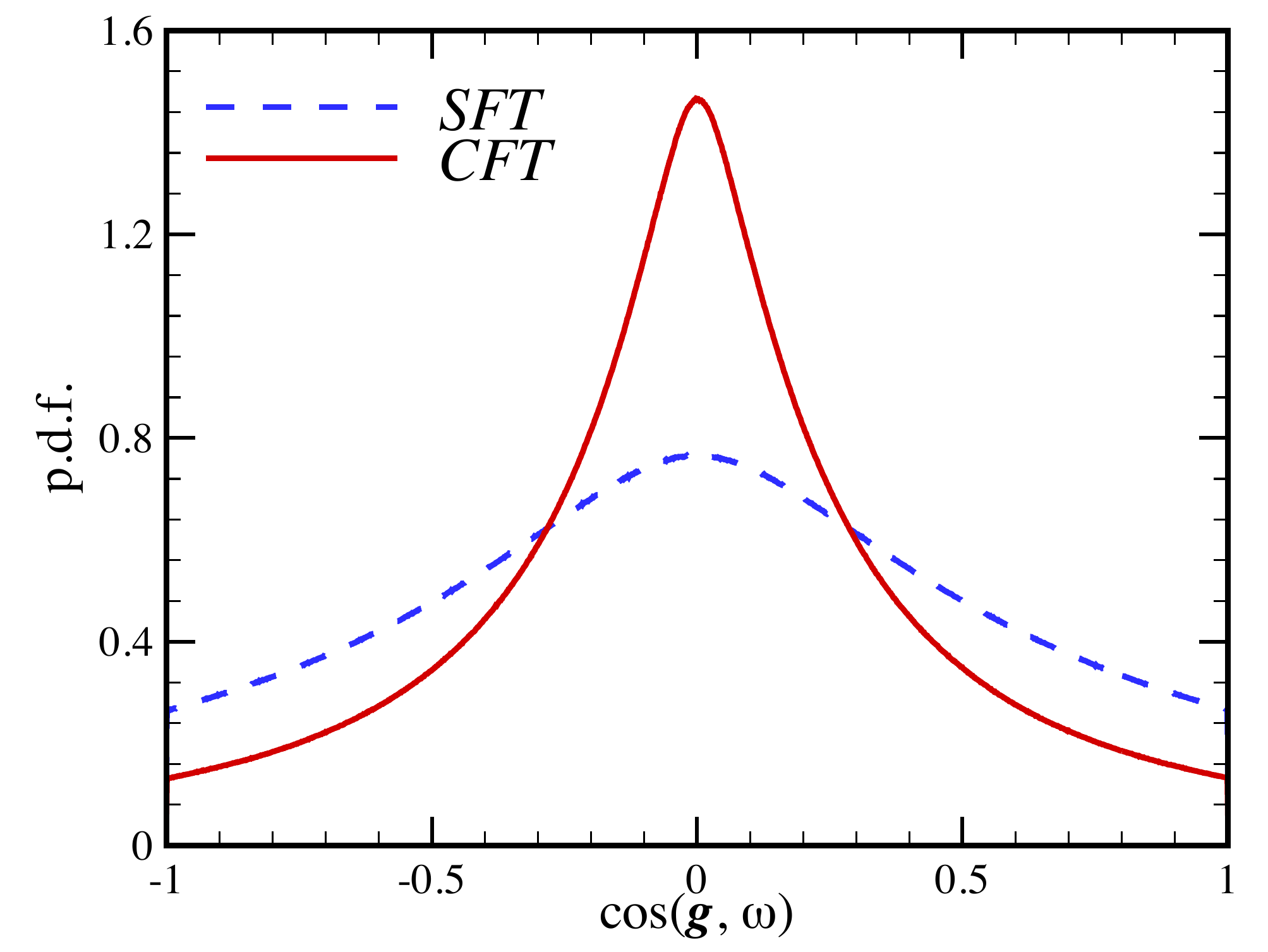}}
\caption{The p.d.f.s of cosine of angle between temperature gradient and vorticity, where
the dashed and solid lines are for SFT and CFT, respectively.}
\label{fig:fig14}
\end{figure}

Contrasting to that the properties of temperature in turbulence have received considerable attention, there are few
studies addressed the properties of temperature gradient. In this section we study the statistics of temperature gradient
on local flow structures. In Figure~\ref{fig:fig14} we plot the p.d.f. of the cosine of the angle between
temperature gradient
$g_i\equiv(\partial T/\partial x_i)/\sqrt{(\partial T/\partial x_1)^2+(\partial T/\partial x_2)^2+(\partial T/\partial x_3)^2}$
and vorticity. It shows that in highly compressible turbulence, there is a strong tendency for the temperature
gradient being orthogonal with the vorticity, especially for CFT. This feature is similar to the observation from
passive scalar transport in turbulence \citep{Ni15a}. Thus, we speculate that in some respects, temperature in compressible
turbulence may behave like passive scalar when the degree of compressibility increases.

\begin{figure}
\begin{center}
\subfigure{
\resizebox*{6.5cm}{!}{\rotatebox{0}{\includegraphics{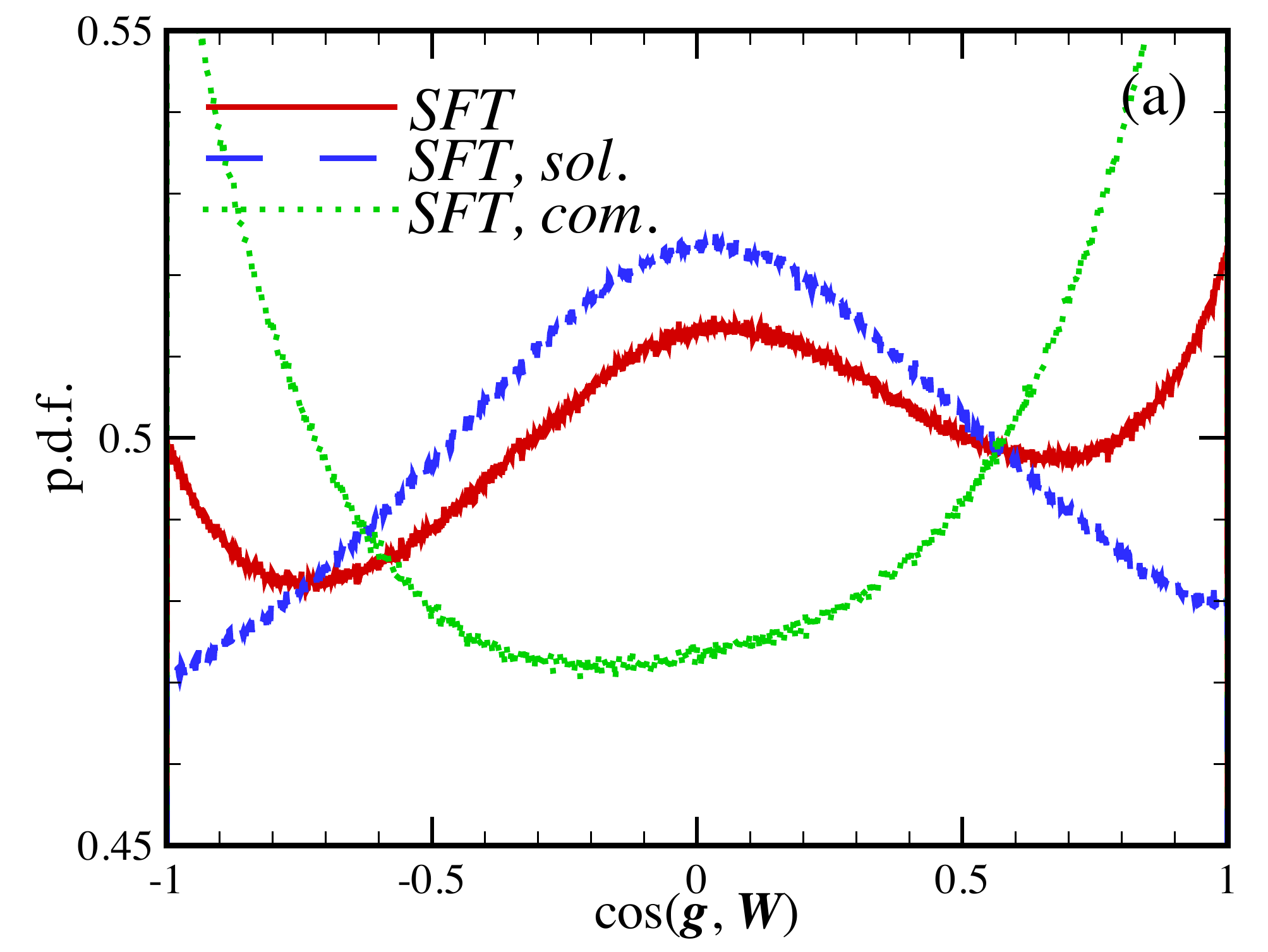}}}}%
\subfigure{
\resizebox*{6.5cm}{!}{\rotatebox{0}{\includegraphics{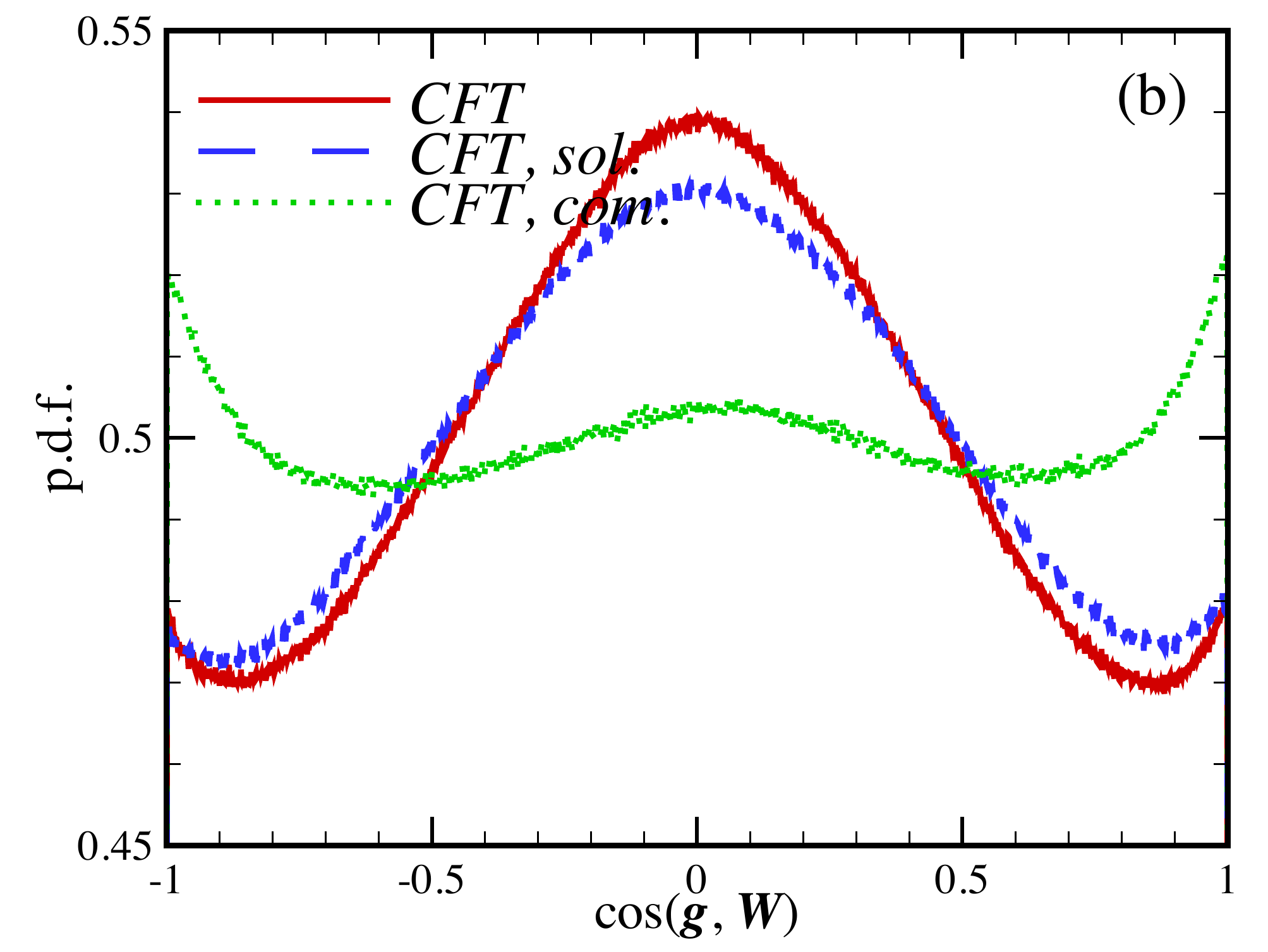}}}}%
\caption{The p.d.f.s of cosines of angles between temperature gradient and vortex stretching vector as well as
its components, where the solid, dashed and dotted lines are for the vortex stretching
vector and its solenoidal and compressive components, respectively. (a) SFT, (b) CFT.}
\label{fig:fig15}
\end{center}
\end{figure}

As a further investigation, we now consider the statistical correlation between the temperature gradient
and the anisotropic strain rate tensor $S^*_{ij}=S_{ij}-S_{kk}\delta_{ij}/3$ \citep{Erlebacher93,Pirozzoli04},
which is written through the enstrophy equation as follows
$$\Big(\frac{\partial}{\partial t}+u_j\frac{\partial}{\partial x_j}\Big)\frac{\omega^2}{2}=\omega_i\omega_jS_{ij}^s
+\omega_i\omega_jS_{ij}^{c*}-\frac{2}{3}\theta\omega^2$$
\begin{equation}
+\omega_i\frac{\varepsilon_{ijk}}{\gamma M^2}\frac{1}{\rho^2}
\frac{\partial\rho}{\partial x_j}\frac{\partial p}{\partial x_k}+\omega_i\frac{\varepsilon_{ijk}}{Re}\frac{\partial}{\partial x_j}
\Big(\frac{1}{\rho}\frac{\partial\sigma_{mk}}{\partial x_m}\Big).
\end{equation}
Here $S_{ij}=(\partial u_i/\partial x_j+\partial u_j/\partial x_j)/2$, and
$S_{ij}^s$ and $S_{ij}^{c*}$ are the solenoidal and compressive parts of $S_{ij}^*$, respectively. Further, the vortex stretching
vector $W_i\equiv\omega_jS_{ij}^*$ is decomposed as $W_i^s=\omega_jS_{ij}^s$ and $W_i^c=\omega_jS_{ij}^{c*}$.
In Figure~\ref{fig:fig15} we plot the p.d.f.s of the cosines of the angles between temperature gradient and vortex stretching
vector and its components. Globally, in SFT there are small positive alignments between the temperature gradient and the
vortex stretching vector and its components. In detail, the maximum p.d.f. for the solenoidal component appears in the
case where it is perpendicular to the temperature gradient, while that for the compressive component appears in the
case where it aligns with the temperature gradient. By contrast, in CFT the p.d.f.s for the vortex stretching vector
and its components are basically symmetric. It shows that the temperature gradient is preferentially perpendicular to the
solenoidal component, and preferentially aligns with the compressive component. These results reveal that in compressible
turbulence the increase in the degree of compressibility suppresses the anisotropy of strain rate tensor.

\begin{figure}
\begin{center}
\subfigure{
\resizebox*{6.5cm}{!}{\rotatebox{0}{\includegraphics{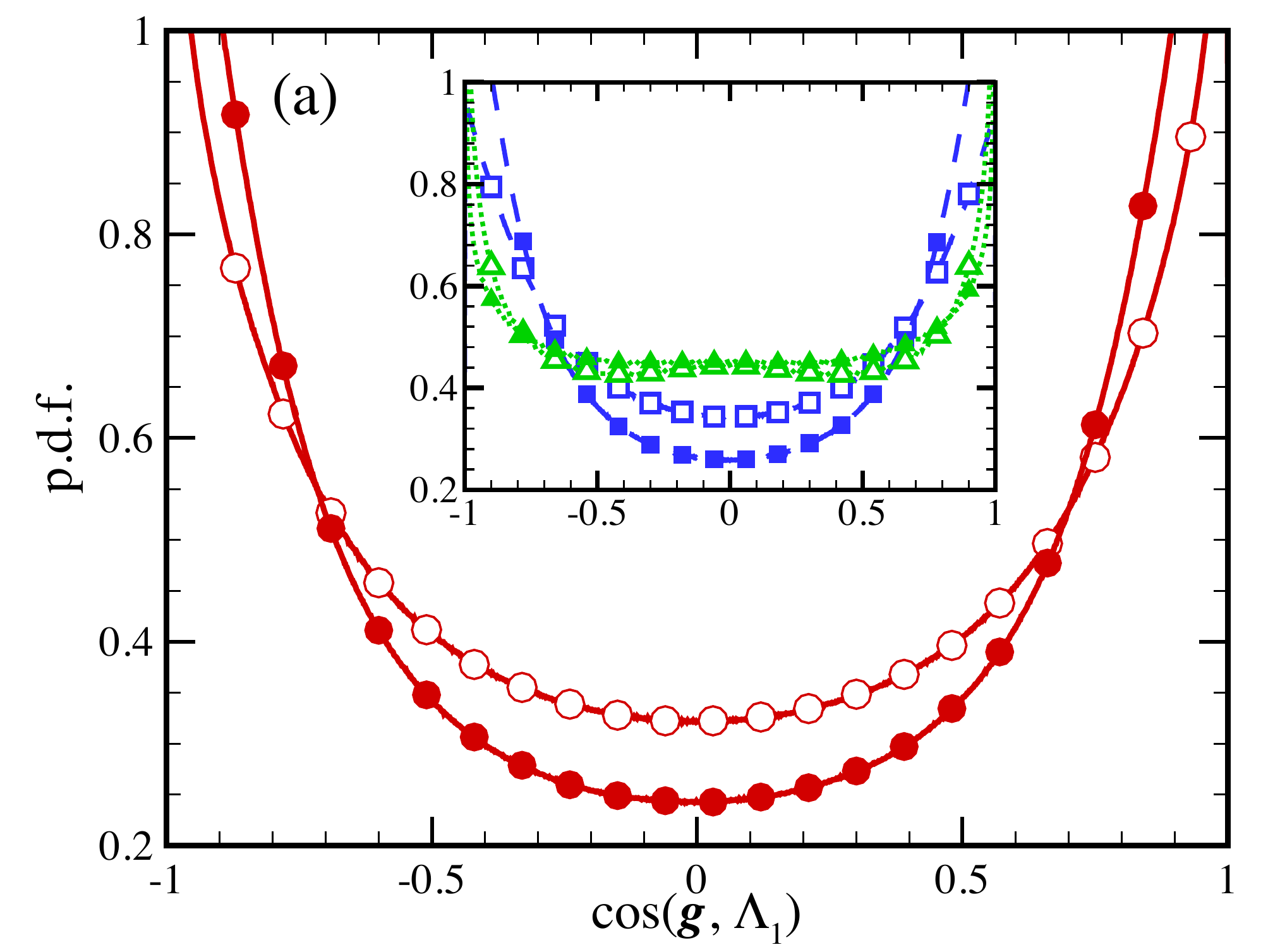}}}}%
\subfigure{
\resizebox*{6.5cm}{!}{\rotatebox{0}{\includegraphics{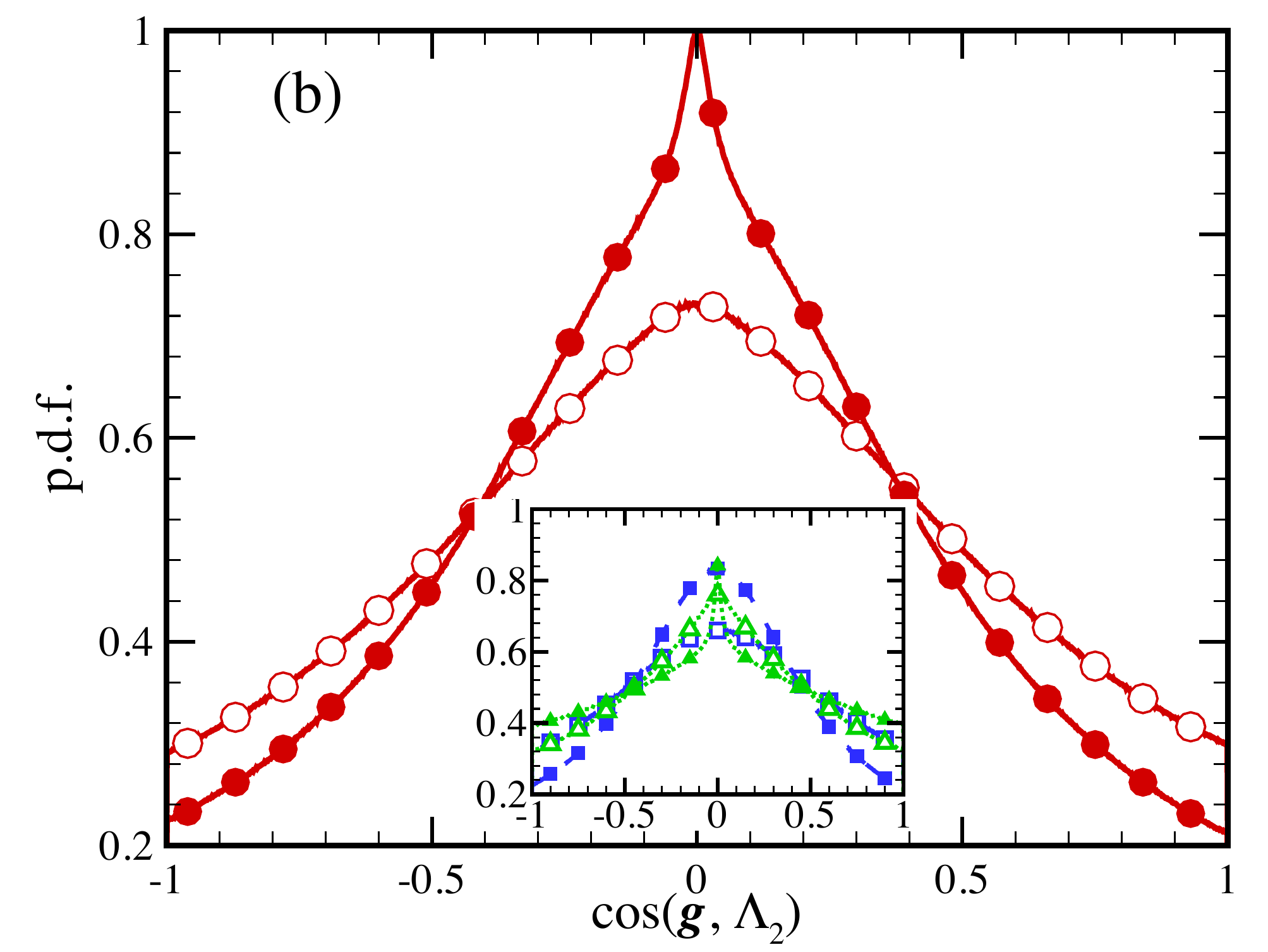}}}}%

\subfigure{
\resizebox*{6.5cm}{!}{\rotatebox{0}{\includegraphics{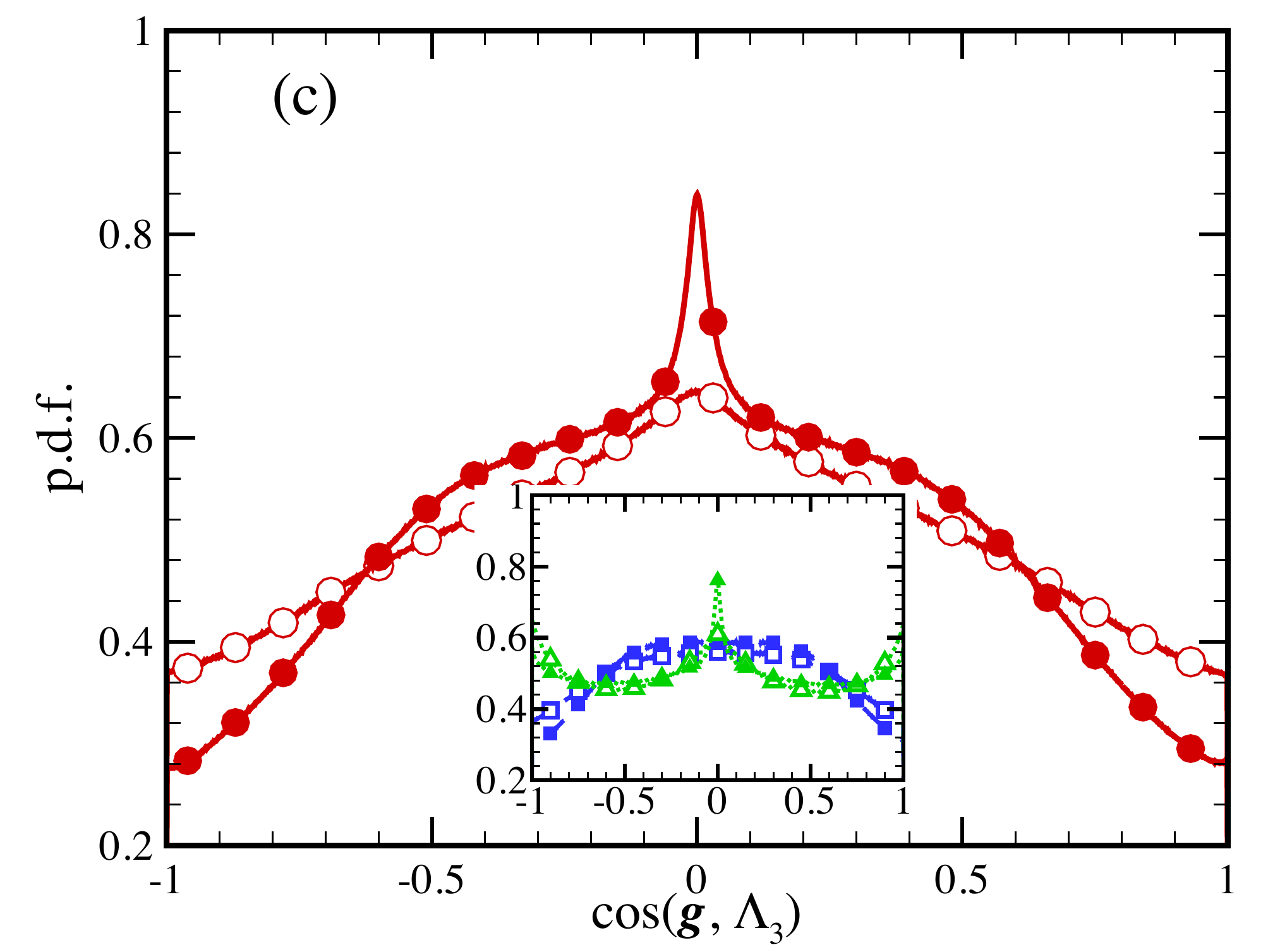}}}}%
\caption{The p.d.f.s of cosines of angles between temperature gradient and the eigenvector $\bm{\Lambda_i}$,
where $i=1$, $2$ and $3$. The lines with open and solid symbols are for SFT and CFT, respectively.
Inset: the p.d.f.s for the solenoidal (dashed line) and compressive (dotted line) components.}
\label{fig:fig16}
\end{center}
\end{figure}

Let us denote the three eigenvectors of the anisotropic strain rate tensor as $\bm{\Lambda}_1$, $\bm{\Lambda}_2$
and $\bm{\Lambda}_3$. The corresponding eigenvalues, arranged in ascending order, are $\lambda^*_1$, $\lambda^*_2$ and $\lambda^*_3$,
which satisfy the following condition
\begin{equation}
\lambda^*_1+\lambda^*_2+\lambda^*_3=0.
\end{equation}
Figure~\ref{fig:fig16} presents the alignment statistics between the temperature gradient and the strain rate eigenvectors.
The results are that: (1) there is a strong tendency for the temperature gradient to align with the first eigenvector corresponding
to the most negative eigenvalue; (2) there is also a clear tendency for the temperature gradient to be perpendicular to
the second eigenvector; and (3) the tendency for the temperature gradient to be perpendicular to the third eigenvector,
the one with the most positive eigenvalue, is also noticeable. The insets of Figure~\ref{fig:fig16} show that for each flow
the solenoidal component dominates the alignment statistics, and the contribution from the compressive component mainly occurs
at small angles.

\begin{figure}
\centerline{\includegraphics[width=8cm]{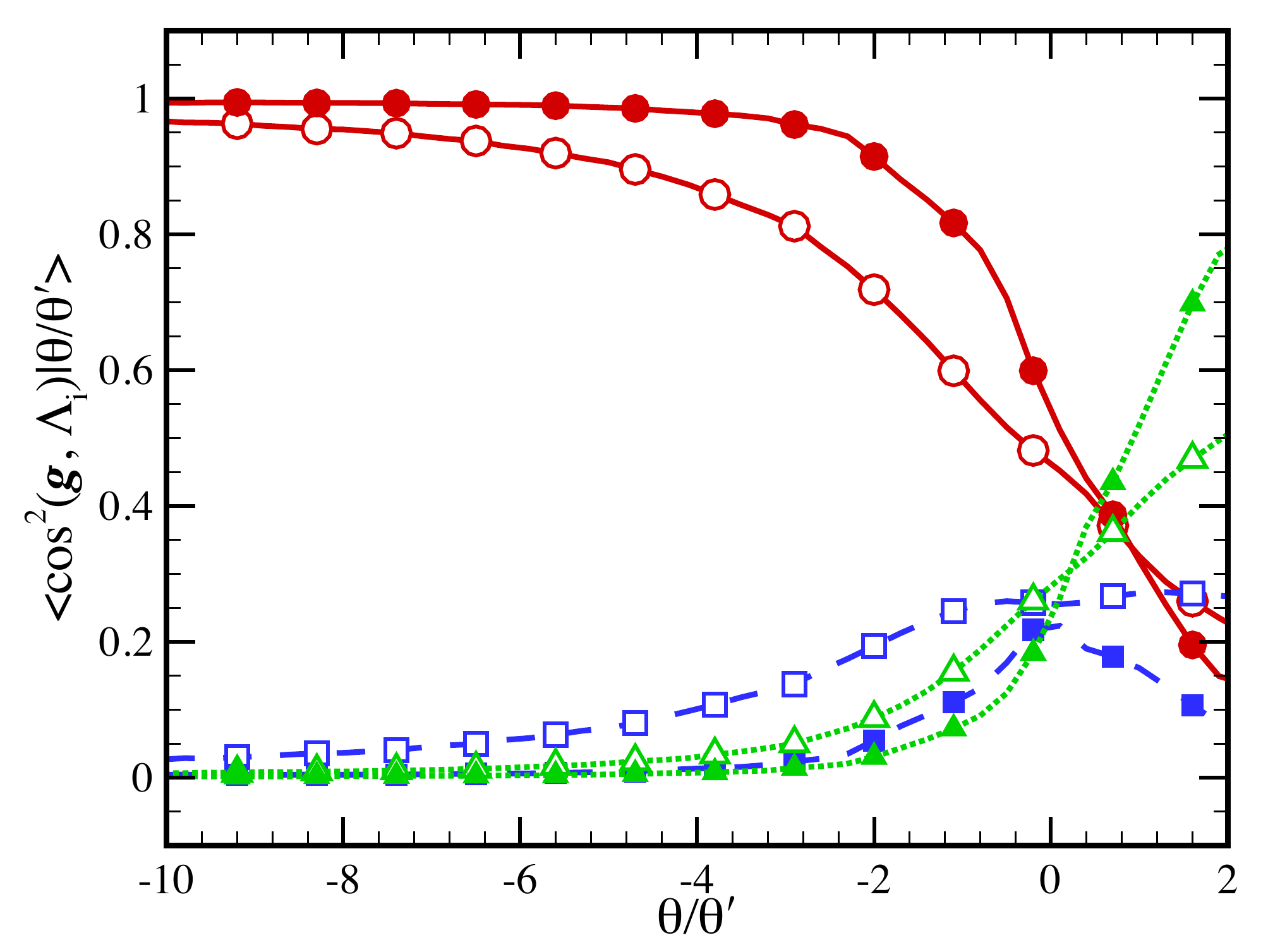}}
\caption{Conditional averages of squares of cosines of angles between temperature gradient and the
eigenvector $\bm{\Lambda_i}$, where the solid, dashed and dotted lines are for $i=1$, $2$ and $3$, respectively.
The open and solid symbols are for SFT and CFT, respectively.}
\label{fig:fig17}
\end{figure}

The conditional averages of the squares of $\cos(\bm{g}, \bm{\Lambda}_k)$ and are plotted in Figure~\ref{fig:fig17}.
As the compression increases, the conditional average for the first eigenvector approaches unity, while those for
the second and third eigenvectors approaches zero. This indicates that in strong compression region, the temperature
gradient always aligns with the most negative eigenvalue of the strain rate tensor. In fact, in the vicinity of
a shock, the temperature gradient and the first strain rate eigenvector are both perpendicular to the shock front.
Furthermore, in rarefaction region, there is a clear tendency for the temperature gradient to align with the third
eigenvector corresponding to the most positive eigenvalue.

Besides the angle statistics of temperature gradient, we now explore the effects on temperature gradient from
local flow structures. To facilitate the description of local flow structures in compressible turbulence,
we introduce the first, second and third invariants of the anisotropic velocity gradient tensor
$A_{ij}^*=A_{ij}-\theta\delta_{ij}/3$ as follows \citep{Wang12a},
\begin{eqnarray}
&& P^* = -\big(\xi_1^*+\xi_2^*+\xi_3^*\big) = 0,
\label{invariant1}\\
&& Q^* = \xi_1^*\xi_2^*+\xi_2^*\xi_3^*+\xi_3^*\xi_1^* = Q-\frac{1}{3}P^2,
\label{invariant2} \\
&& R^* = -\xi_1^*\xi_2^*\xi_3^* = R-\frac{1}{3}PQ+\frac{2}{27}P^3,
\label{invariant3}
\end{eqnarray}
where $\xi^*_i=\xi_i-\theta/3$ are the three eigenvalues of $A_{ij}^*$, and $\xi_i$ are the three eigenvalues of
$A_{ij}=\partial u_j/\partial x_i$. The details in $P$, $Q$ and $R$ can be found in \citet{Chong90}.

\begin{figure}
\begin{center}
\subfigure{
\resizebox*{6.5cm}{!}{\rotatebox{0}{\includegraphics{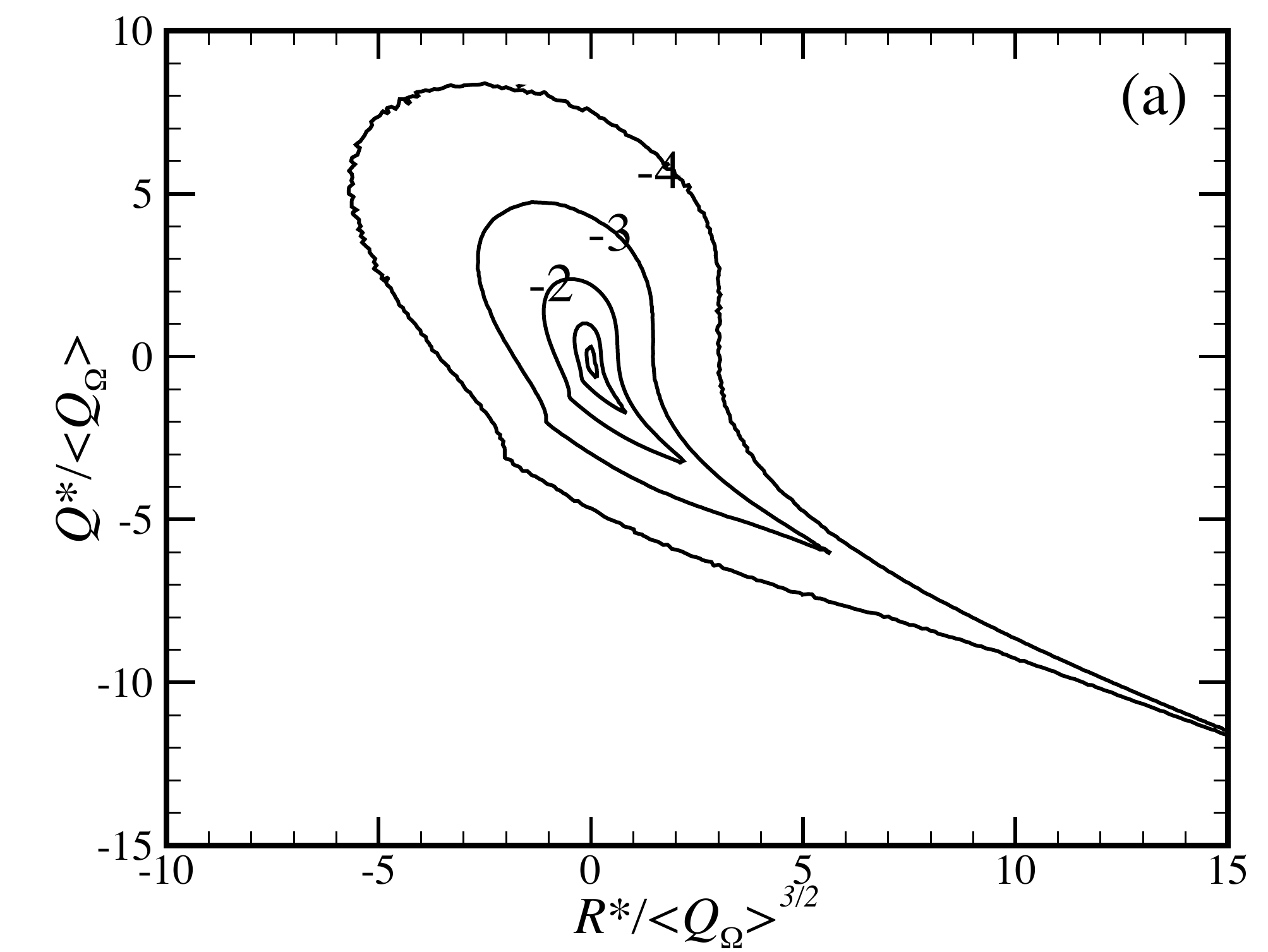}}}}%
\subfigure{
\resizebox*{6.5cm}{!}{\rotatebox{0}{\includegraphics{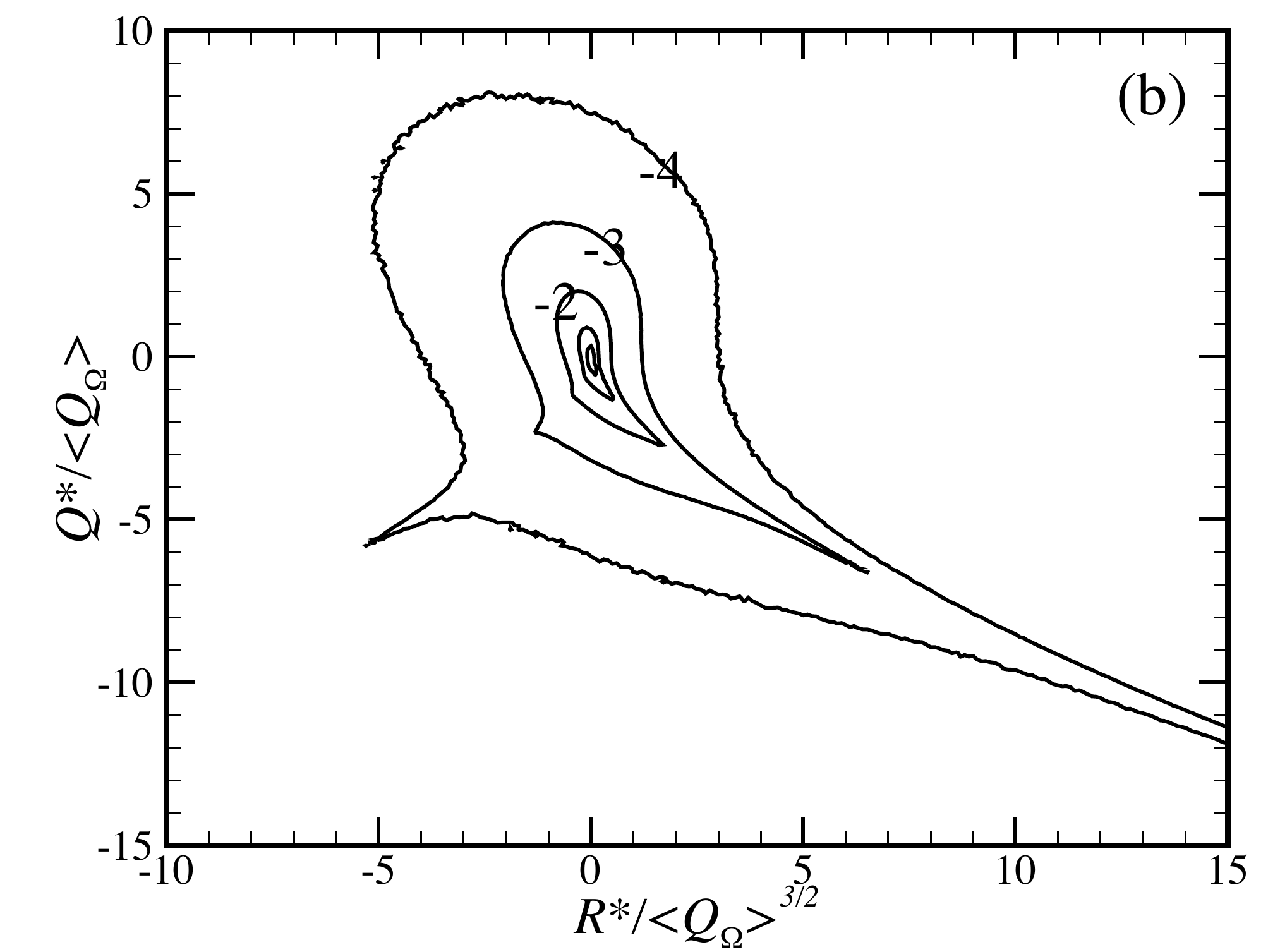}}}}%
\caption{Isocontour lines of logarithm of joint p.d.f. of $Q^*$ and $R^*$. Five contour lines at $0$, $-1$, $-2$, $-3$ and $-4$
are shown. (a) SFT, (b) CFT.}
\label{fig:fig18}
\end{center}
\end{figure}
\begin{figure}
\begin{center}
\subfigure{
\resizebox*{6.5cm}{!}{\rotatebox{0}{\includegraphics{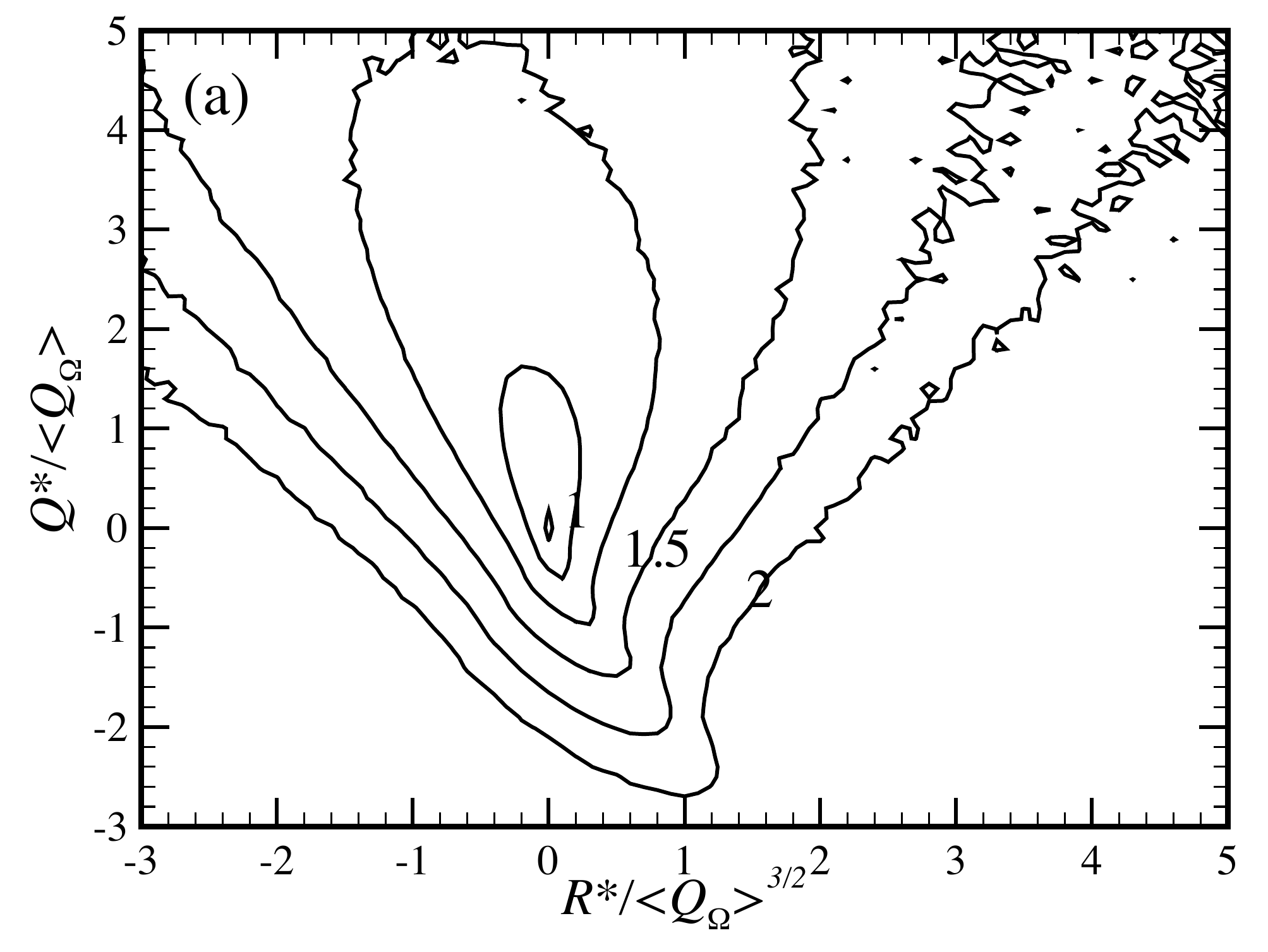}}}}%
\subfigure{
\resizebox*{6.5cm}{!}{\rotatebox{0}{\includegraphics{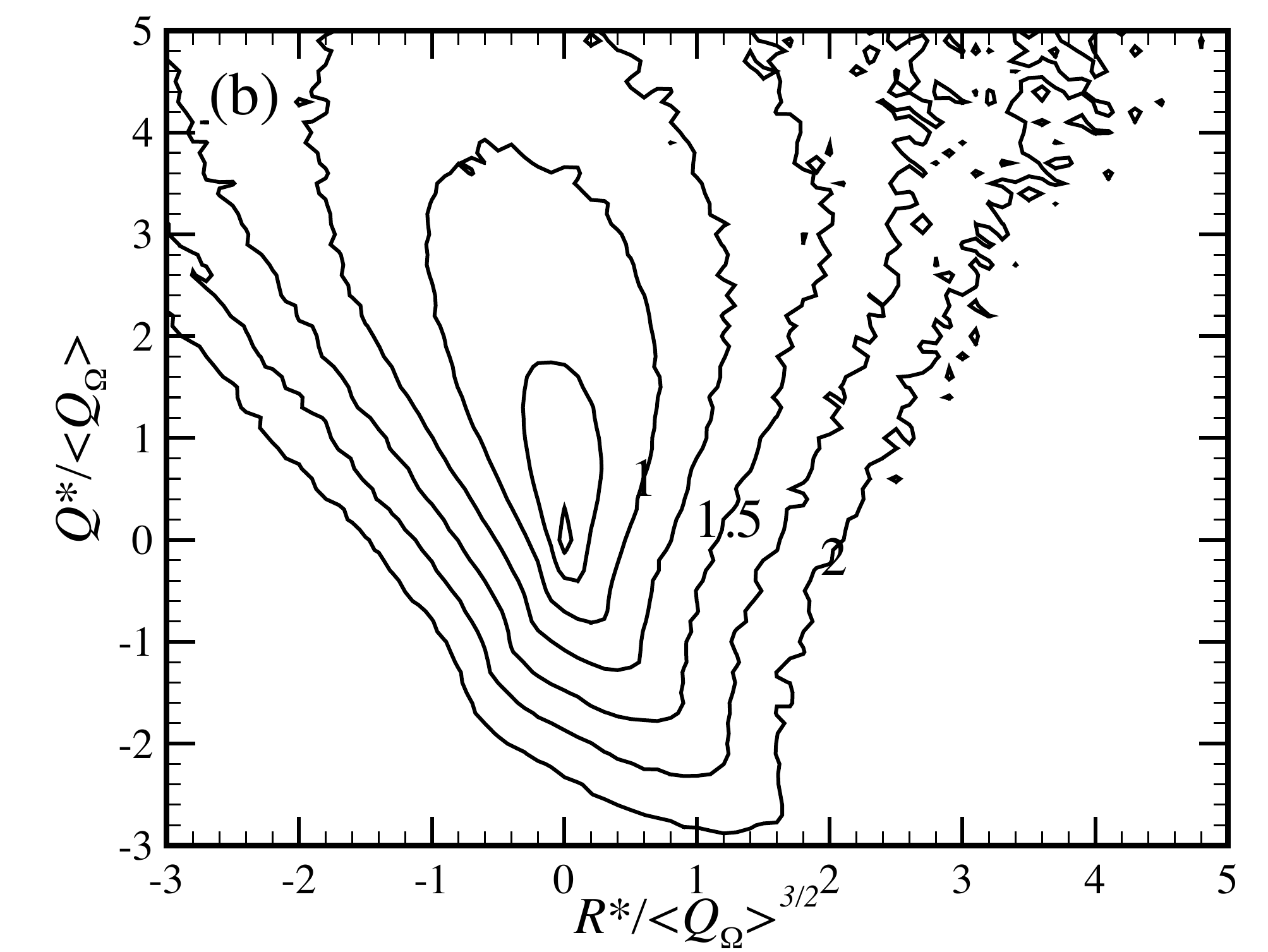}}}}%
\caption{Isocontour lines of temperature gradient magnitude conditioned on $Q^*$ and $R^*$. Nine lines at $0$, $0.25$, ...,
$2$ are shown. (a) SFT, (b) CFT.}
\label{fig:fig19}
\end{center}
\end{figure}

In Figure~\ref{fig:fig18} we plot the contour lines of the joint p.d.f. for the second and third invariants $(Q^*, R^*)$,
which are scaled by the second invariant of the rotation rate tensor $Q_{\Omega}=\Omega_{ij}\Omega_{ij}/2$. Similar to the
results from \citet{Pirozzoli04}, the joint p.d.f.s display the teardrop shape in both SFT and CFT. Compared with weakly and
moderately compressible turbulence \citep{Pirozzoli04},in our simulations the tails of the joint p.d.f.s in the fourth
quadrant are pronounced longer. It also shows that the joint p.d.f. from CFT has protrusive structures in the third quadrant.
Figure~\ref{fig:fig19} presents the contour lines of the magnitude averages of temperature gradient conditioned on $Q^*$ and $R^*$.
There also appears the teardrop in the conditional average, which shifts towards the positive part of $Q^*$ and
the negative part of $R^*$. This reveals that the temperature gradient is relative large in the region
where the second and third invariants of the anisotropic velocity gradient tensor are respectively positive and negative.

\section{Cascade of Temperature Field}

In incompressible flow, the cascade of temperature involves the generation of temperature fluctuations at large scales,
the stretching, contracting and folding of temperature by velocity, producing progressively smaller and smaller scales,
until the ultimate dissipation of temperature fluctuations at the smallest scale. In compressible turbulence,
the fact of the nonlinear interplay between velocity and temperature as well as that between solenoidal and
compressive modes of velocity greatly complicate the cascade of temperature. In this section, we carry out
investigation on this topic, especially in analyzing the important role of pressure-dilatation in temperature cascade.

\begin{figure}
\begin{center}
\subfigure{
\resizebox*{6.5cm}{!}{\rotatebox{0}{\includegraphics{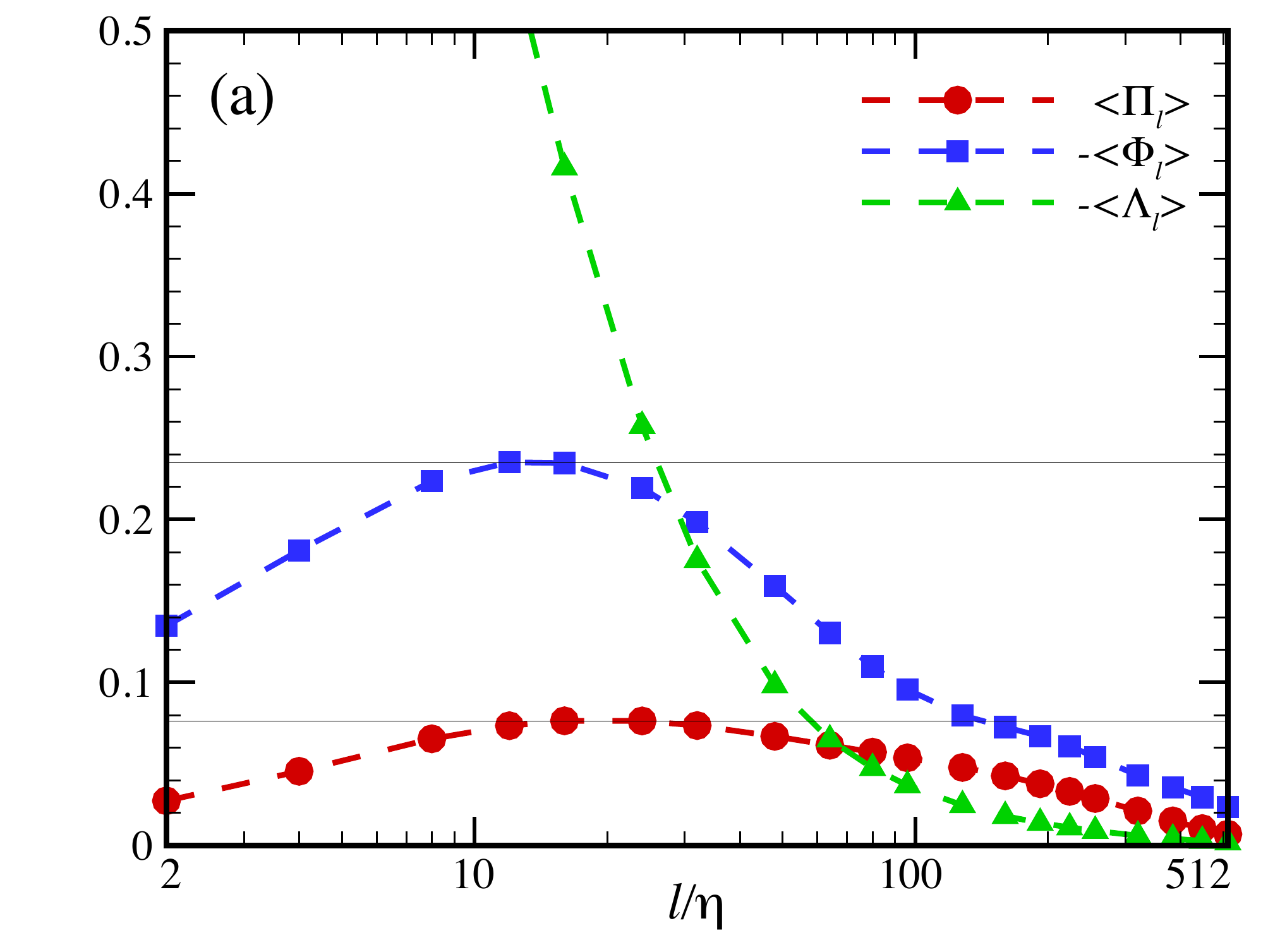}}}}%
\subfigure{
\resizebox*{6.5cm}{!}{\rotatebox{0}{\includegraphics{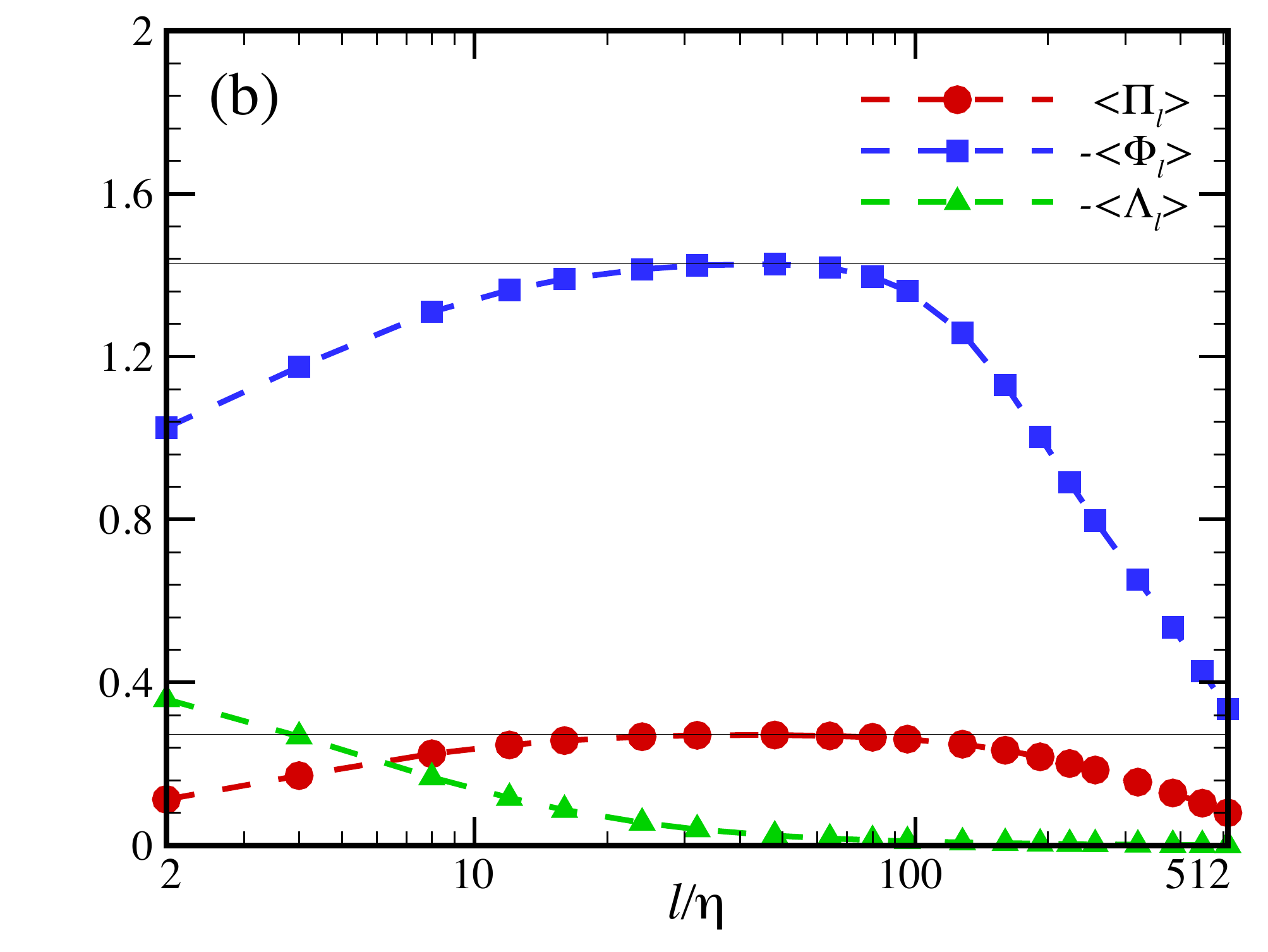}}}}%
\caption{Subgrid-scale temperature flux (circles), pressure-dilatation (squares), and viscous
dissipation (deltas), as functions of normalized scale $l/\eta$. (a) SFT, (b) CFT.}
\label{fig:fig20}
\end{center}
\end{figure}

A "coarse-graining" approach \citep{Aluie11,Aluie12} is employed to study the transport of temperature fluctuations
at different scales. According to the definition of a classically filtered field $\overline{a}_\emph{l}(\textbf{x})$
\begin{equation}
\overline{a}_\emph{l}(\textbf{x})\equiv \int d^3\textbf{r}G_\emph{l}(\textbf{r})a(\textbf{x}+\textbf{r}),
\end{equation}
the density-weighted filtered field is obtained
\begin{equation}
\widetilde{a}_\emph{l}(\textbf{x}) \equiv \frac{\overline{\rho a}_\emph{l}(\textbf{x})}{\overline{\rho}_\emph{l}(\textbf{x})}.
\end{equation}
Here $G_\emph{l}(\textbf{r})=G(\textbf{r}/\emph{l})/\emph{l}^3$ is a kernel, and $G(\textbf{r})$ is a window function.
By the large-scale continuity and temperature equations, it is straightforward to derive the governing equation for temperature
variance at large scales as follows \citep{Ni15a}
\begin{equation}
\frac{\partial}{\partial t}\Big(\frac{1}{2}\overline{\rho}_\emph{l}\widetilde{Te}_\emph{l}^2\Big)
+ \nabla\cdot\textbf{J}_\emph{l} = -\Pi_\emph{l} - \Phi_\emph{l} - \Lambda_\emph{l} - D_\emph{l} + \varepsilon_\emph{l}^{cool}.
\end{equation}
In Figure~\ref{fig:fig20} we plot the ensemble averages of SGS temperature flux, pressure-dilatation, and viscous dissipation,
as functions of the normalized scale $l/\eta$. In both SFT and CFT, the viscous dissipation primarily occurs at small scales,
and declines quickly as scale increases. Throughout scale ranges, the SGS temperature flux is positive, indicating that the
temperature fluctuations are always transported from large to small scales. The appearance of plateau in the SGS temperature
flux confirms the conservation in temperature cascade. In terms of the magnitude of pressure-dilatation, in SFT it first
increases and undergoes a flat region in the range of $8\leq l/\eta\leq 24$, then decreases and reaches zero at large scales.
By contrast, in CFT the flat region shifts to a larger scale range of $16\leq l/\eta\leq 80$, and when scale increases
it decreases and approaches a finite positive value. These observations reveal that in the transport of temperature
fluctuations, the pressure-dilatation mainly takes activities at moderate and large scales.

\begin{figure}
\centerline{\includegraphics[width=8cm]{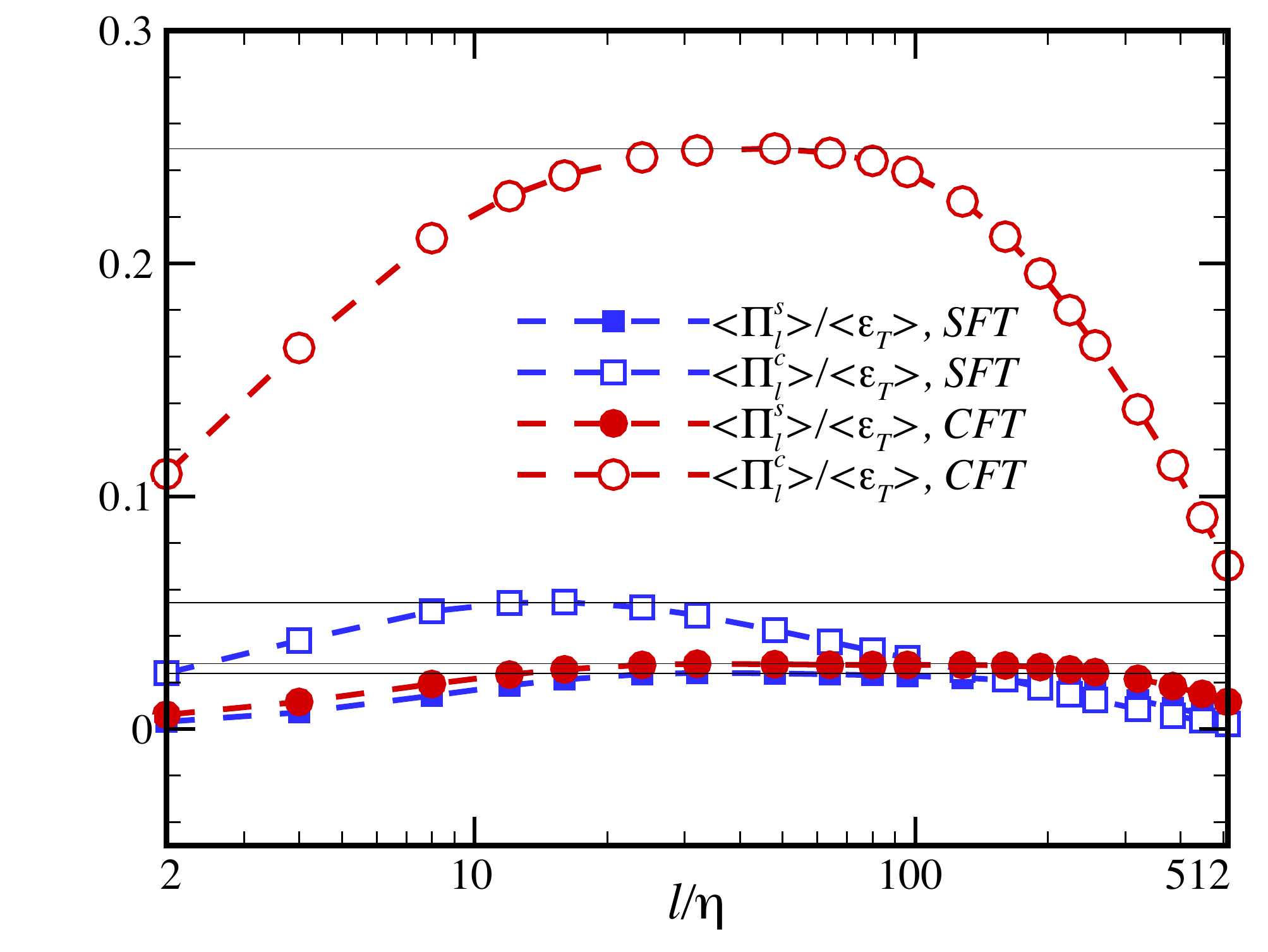}}
\caption{Normalized solenoidal (solid symbols) and compressive (open symbols) components of the subgrid-scale fluxes
of temperature, as functions of normalized scale $l/\eta$, where the symbols of squares and circles are for SFT
and CFT, respectively.}
\label{fig:fig21}
\end{figure}
\begin{figure}
\begin{center}
\subfigure{
\resizebox*{6.5cm}{!}{\rotatebox{0}{\includegraphics{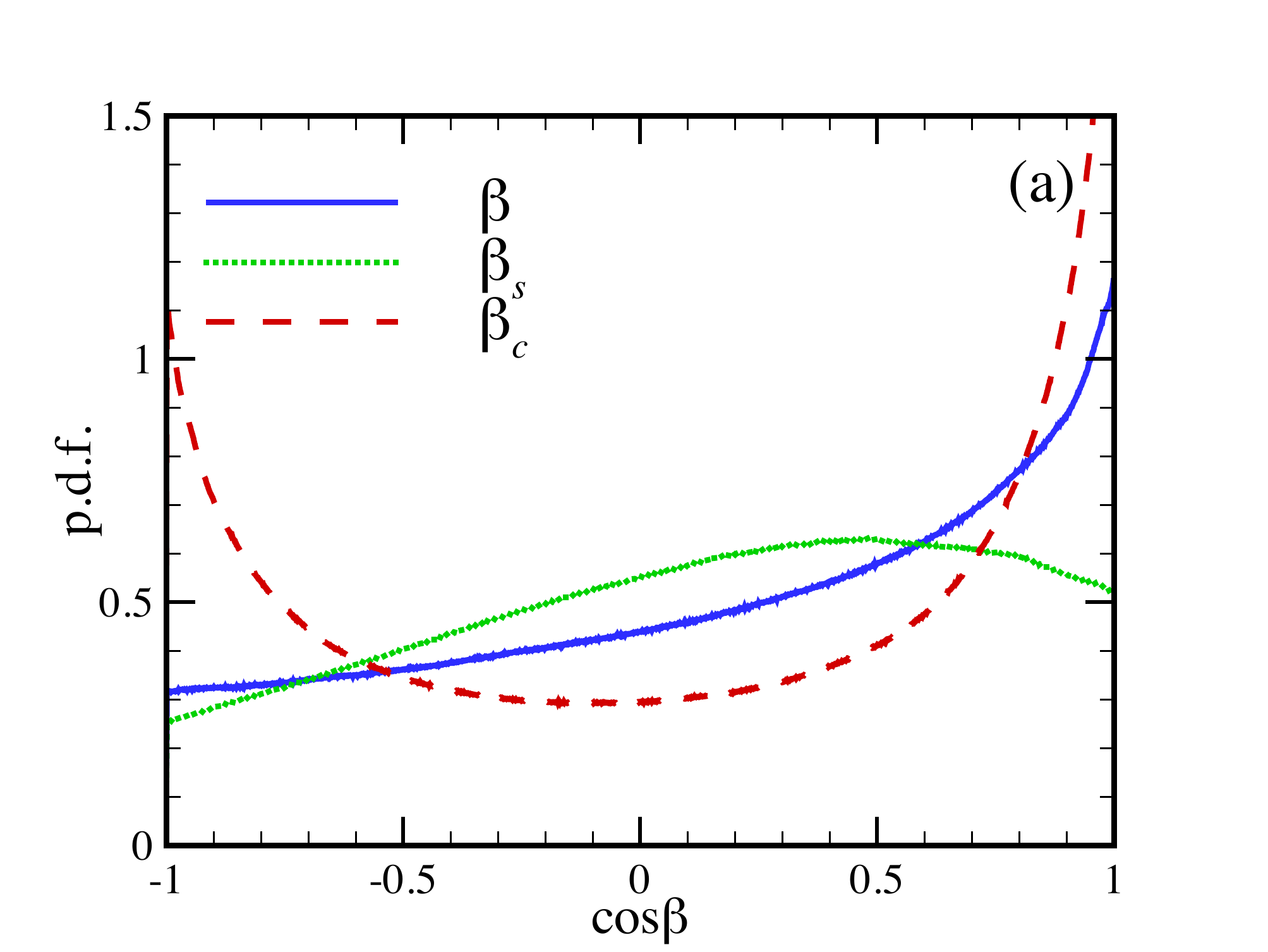}}}}%
\subfigure{
\resizebox*{6.5cm}{!}{\rotatebox{0}{\includegraphics{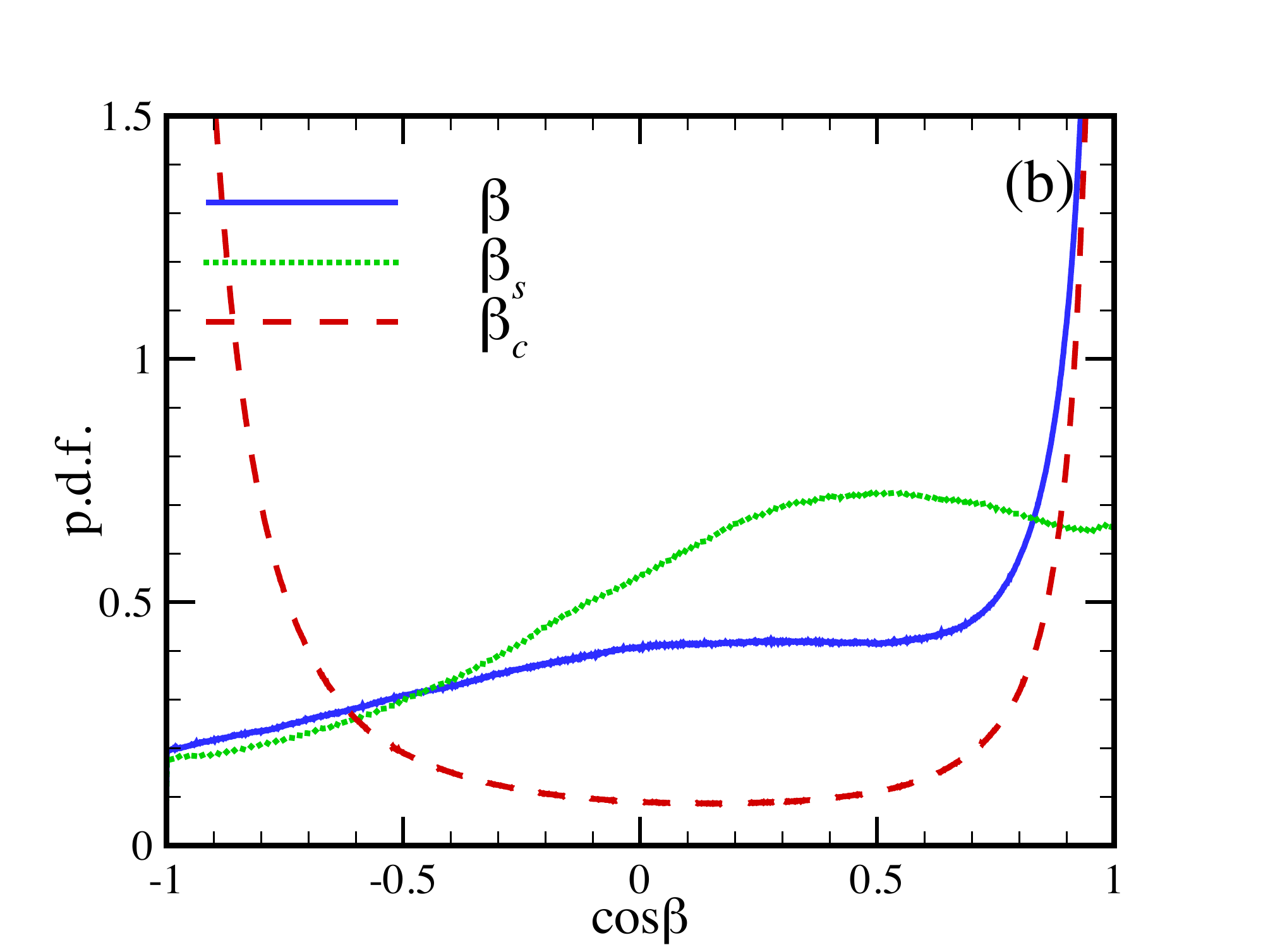}}}}%
\caption{The p.d.f.s of cosine of angle between $\widetilde{\tau}_j(u,T)$ and $\partial \widetilde{T}/\partial x_j$
(solid lines), and its solenoidal (dotted lines) and compressive (dashed lines) components. (a) SFT, (b) CFT.}
\label{fig:fig22}
\end{center}
\end{figure}

The solenoidal and compressive components of the SGS temperature flux normalized by the ensemble average of temperature
dissipation rate are depicted in Figure~\ref{fig:fig21}. Obviously, the two components are positive throughout scale ranges.
In SFT the solenoidal component is smaller than the compressive component in the range of $l/\eta\leq 96$,
and becomes comparable at larger scales. By contrast, in CFT the solenoidal component is always smaller than the
compressive component, revealing the overwhelming effect of compressibility.
To explore the angle statistics of temperature gradient in the cascade process, we rewrite the expression of the SGS
temperature flux as follows
\begin{equation}
\Pi_\emph{l} = -\widetilde{\tau_j}\big(u, T\big)\partial\widetilde{T}/\partial x_j,
\end{equation}
where the SGS temperature-velocity coupling $\widetilde{\tau_j}$ is defined as
\begin{equation}
\widetilde{\tau}_j\big(u, T\big) \equiv \overline{\rho}_\emph{l}\big[\big(\widetilde{Tu_j}\big)_\emph{l}
-\widetilde{T}_\emph{l}\widetilde{u_j}_\emph{l}\big].
\end{equation}
Then, we obtain the angle between $\widetilde{\tau_j}$ and $\partial\widetilde{T}/\partial x_j$
\begin{equation}
\cos\beta \equiv \frac{\tau_j\partial\widetilde{T}/\partial x_j}{|\tau_j||\partial\widetilde{T}/\partial x_j|}
\end{equation}
In Figure~\ref{fig:fig22} we plot the p.d.f.s of the cosine of the angle $\beta$ and its components $\beta_s$ and
$\beta_c$. For both SFT and CFT, the p.d.f.s of $\cos\beta$ are asymmetric and peak at $\beta=0$. This indicates
that the transfer of temperature flux preferentially occurs in the orientation where $\widetilde{\tau_j}$ is
anti-aligns with $\partial\widetilde{T}/\partial x_j$. The p.d.f.s of $\cos\beta_s$ are also asymmetric, but
peak in the range of $0<\beta_s<\pi/2$. On the contrary, the p.d.f.s of $\cos\beta_c$ are basically symmetric
and peak at $\beta_c=0$ and $\pi$.

\begin{figure}
\centerline{\includegraphics[width=8cm]{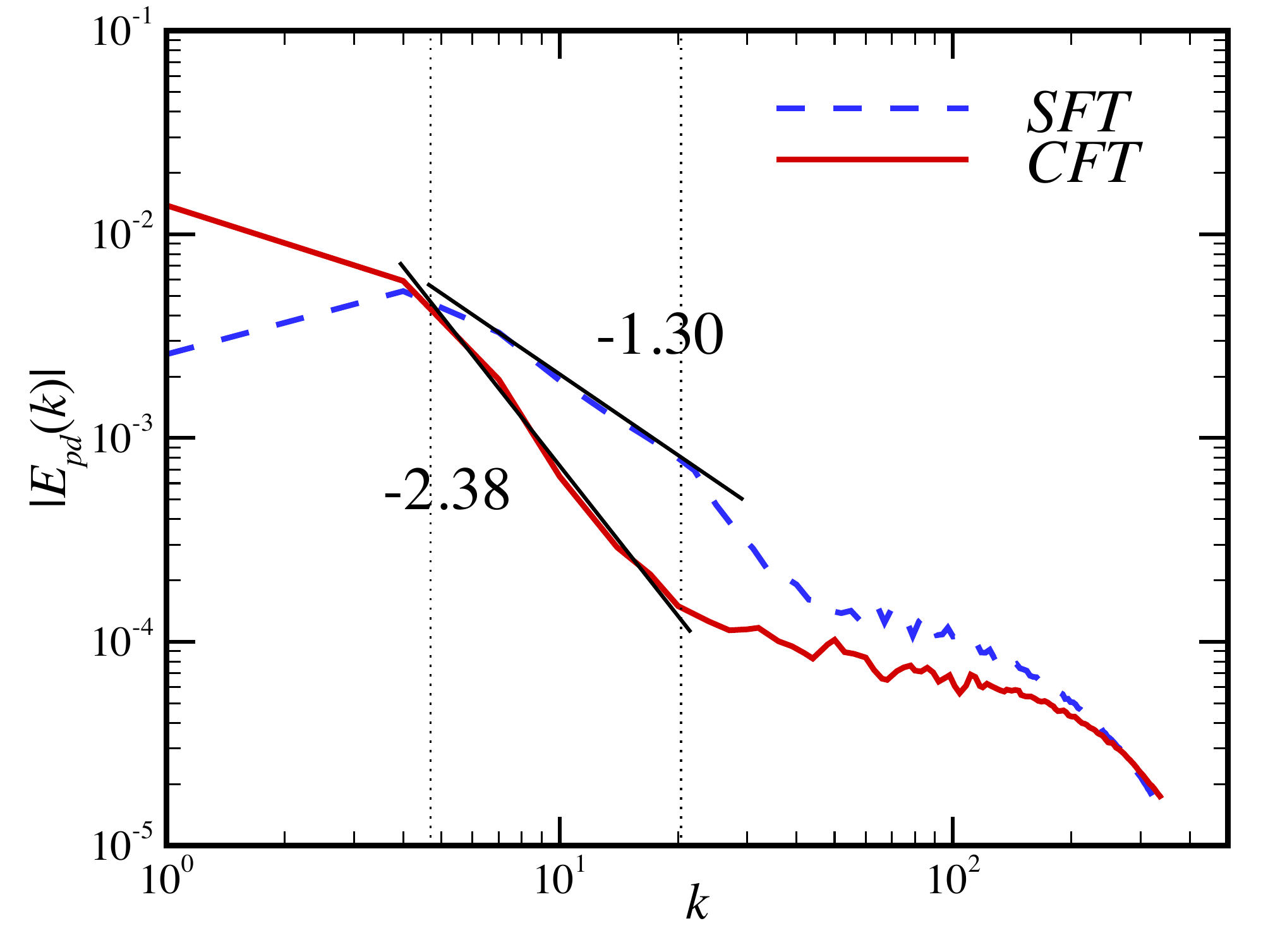}}
\caption{Cospectrum of pressure-dilatation, where the dashed and solid lines are for SFT and CFT, respectively.}
\label{fig:fig23}
\end{figure}

The pressure-dilatation $PD(l)=\overline{p}_\emph{l}\overline{\theta}_\emph{l}$ plays an important role in the conversion
between kinetic and internal energy, namely, if $\overline{\theta}_\emph{l}<0$, the energy is transported from the large-scale
kinetic to internal energy; if $\overline{\theta}_\emph{l}>0$, the process reverses. In Figure~\ref{fig:fig23} we plot
the cospectrum of pressure-dilatation, which is defined by
\begin{equation}
E_{pd}(k) \equiv \sum\limits_{k-1/2<|\textbf{k}|<k+1/2}-\widehat{p}(\textbf{k})\widehat{\theta}(\textbf{-k}).
\end{equation}
It shows that the slope value of the cospectrum is $-1.30$ for SFT and $-2.38$ for CFT, exhibiting that in
our simulations $E_{pd}(k)$ decay at rates faster than $k^{-1}$. This is in agreement with the following
criterion proposed by \citet{Aluie11}
\begin{equation}
|E_{pd}(k)| \leq Cu'p'(kL)^{-\varpi}, \quad \varpi>1,
\end{equation}
where $C$ is a nondimensional constant and $L$ is an integral scale. This criterion deduces that the
pressure-dilatation would converge and become independent of $l$ at small enough scales
\begin{equation}
\lim\limits_{l\rightarrow0}PD(l) = \lim\limits_{K\rightarrow\infty}\sum\limits_{0<k<K}E_{pd}(k)=\Theta
\end{equation}
It provides the picture that the pressure-dilatation mainly exchanges the kinetic and internal energy
over moderately large scales of limited extent. At smaller scales the kinetic and internal energy budgets
statistically decouple, giving rise to an inertial range in which the temperature undergoes a conservative
cascade. Furthermore, in CFT the faster decaying of $E_{pd}(k)$ undoubtedly connects with the larger
compressive component of the SGS temperature flux.

\begin{figure}
\centerline{\includegraphics[width=8cm]{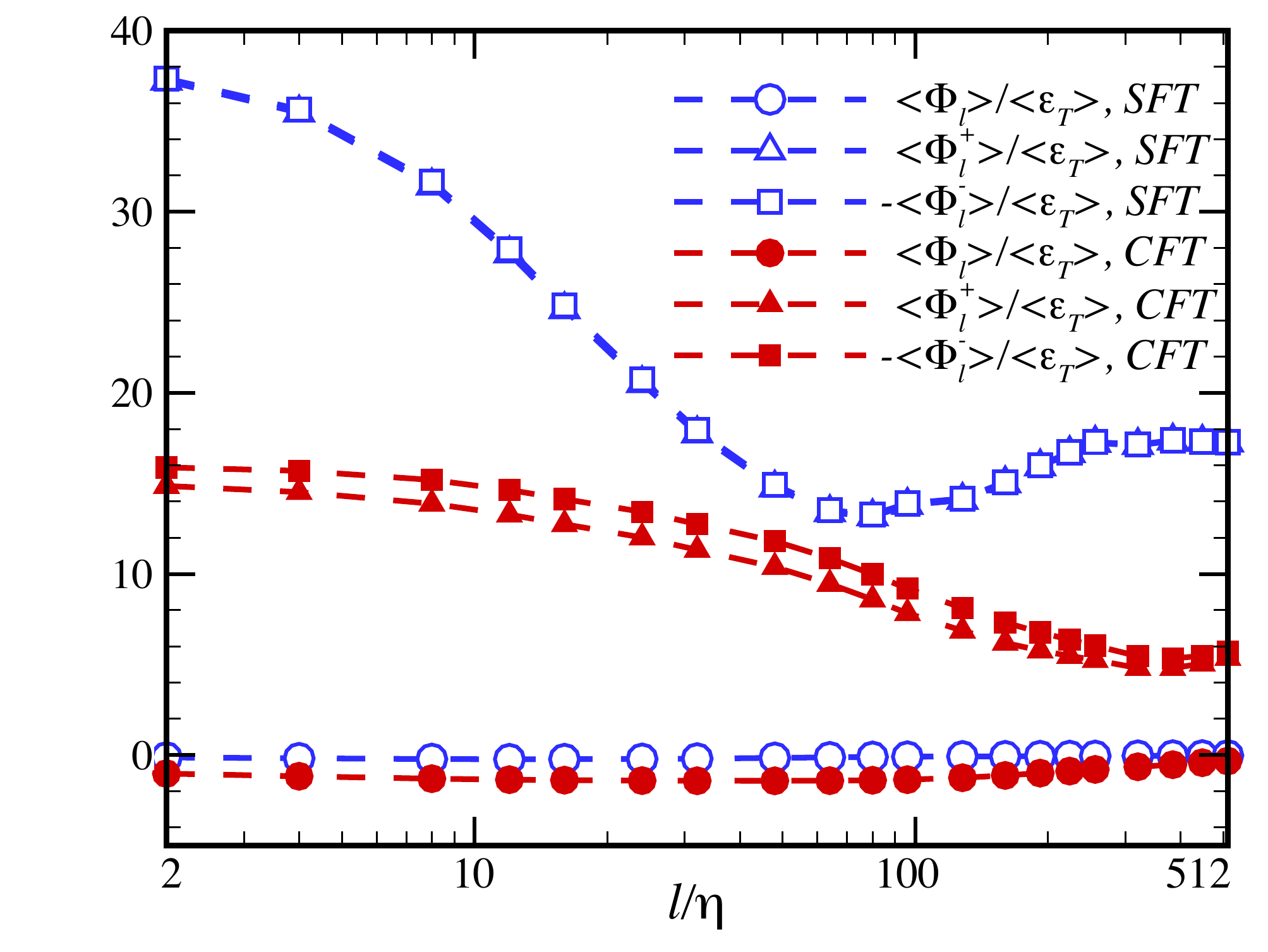}}
\caption{Pressure-dilatation (circles) and its positive (deltas) and negative (squares) components, where
the open and solid symbols are for SFT and CFT, respectively.}
\label{fig:fig24}
\end{figure}
\begin{figure}
\begin{center}
\subfigure{
\resizebox*{6.5cm}{!}{\rotatebox{0}{\includegraphics{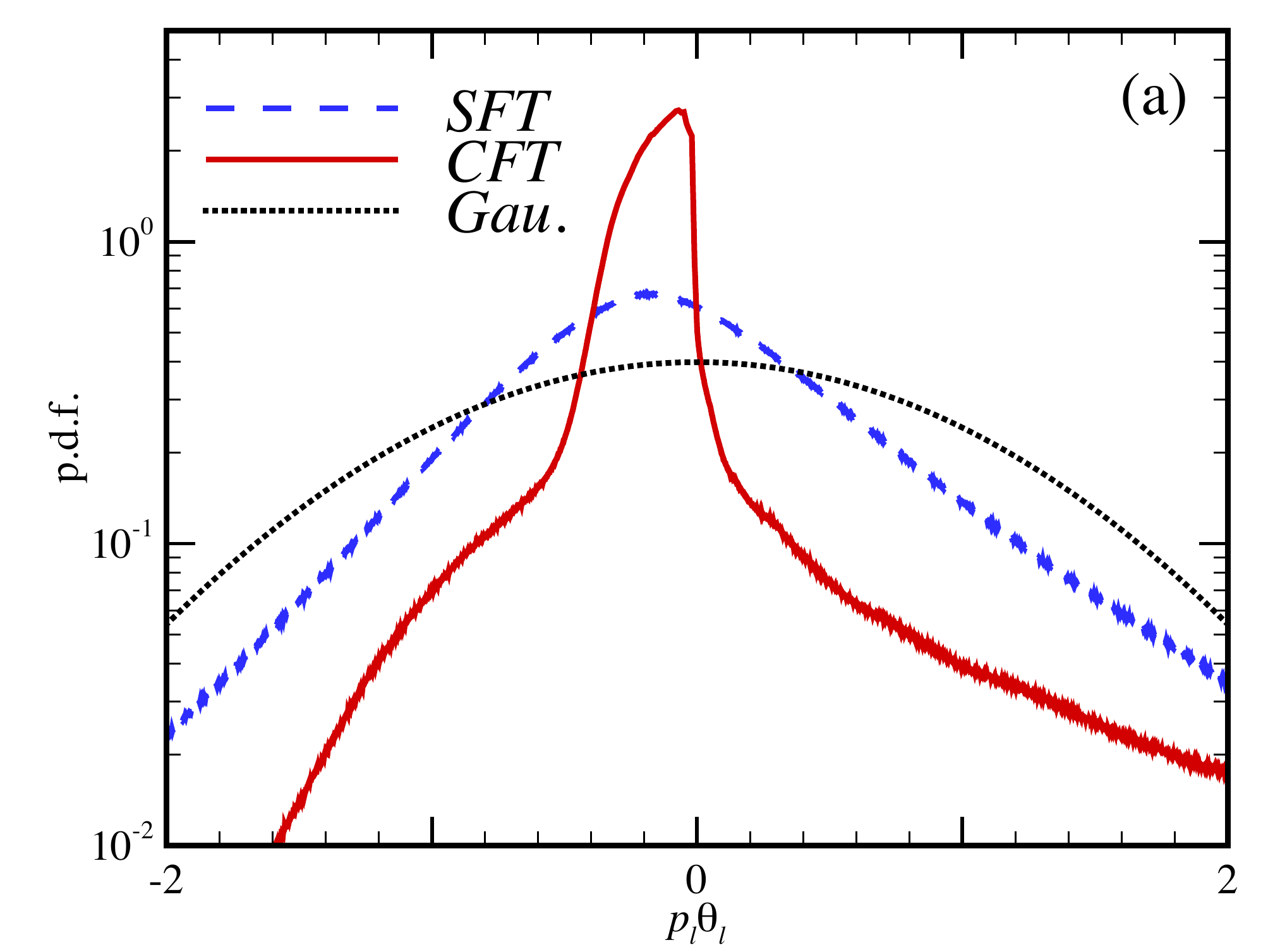}}}}%
\subfigure{
\resizebox*{6.5cm}{!}{\rotatebox{0}{\includegraphics{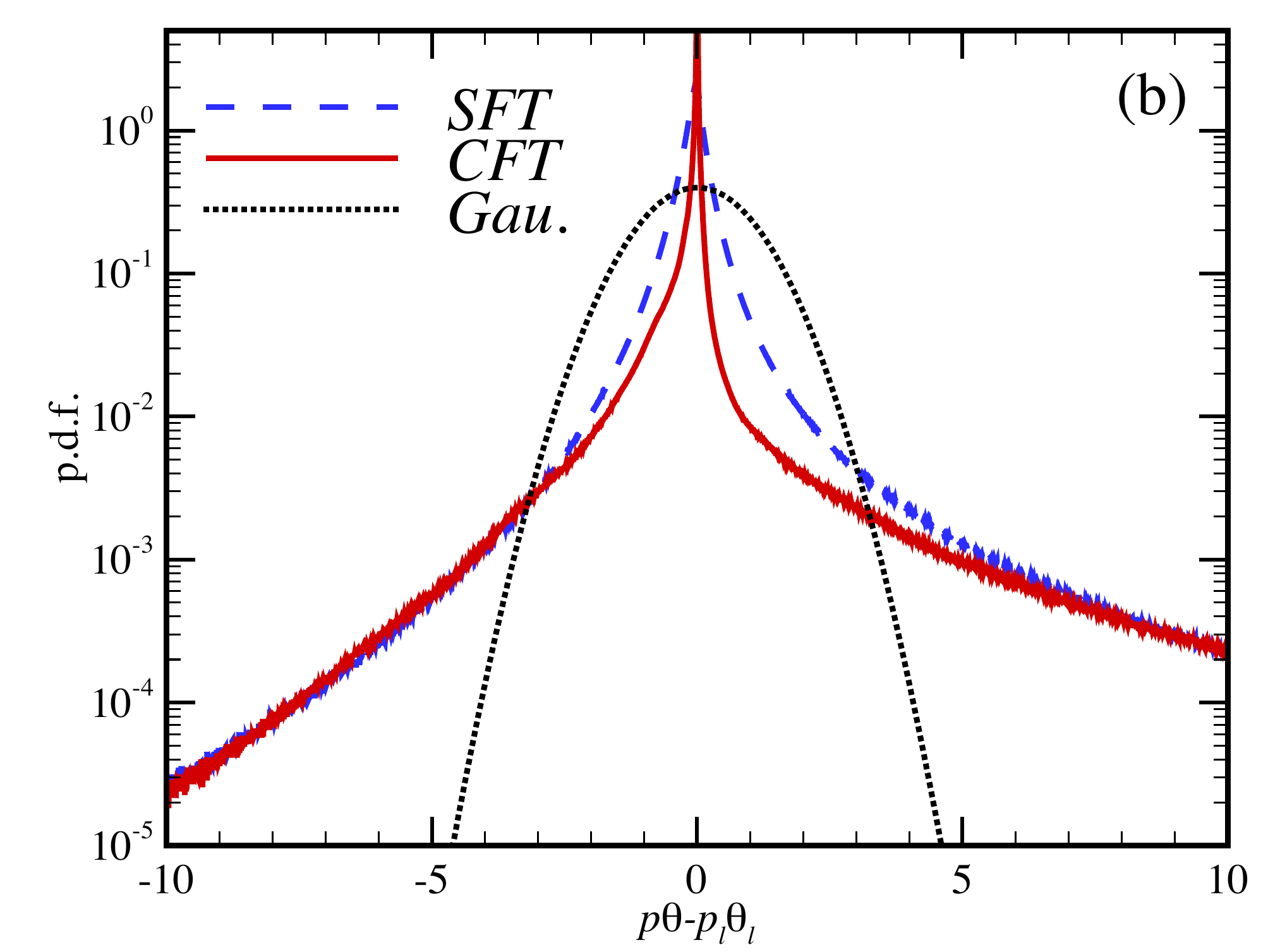}}}}%
\caption{The p.d.f.s of the pressure-dilatation for large scales (a) and the residual part from small scales (b)
at $l=32\eta$, where the dashed and solid lines are for SFT and CFT, respectively. The dotted lines are for Gaussian.}
\label{fig:fig25}
\end{center}
\end{figure}
\begin{figure}
\begin{center}
\subfigure{
\resizebox*{6.5cm}{!}{\rotatebox{-90}{\includegraphics{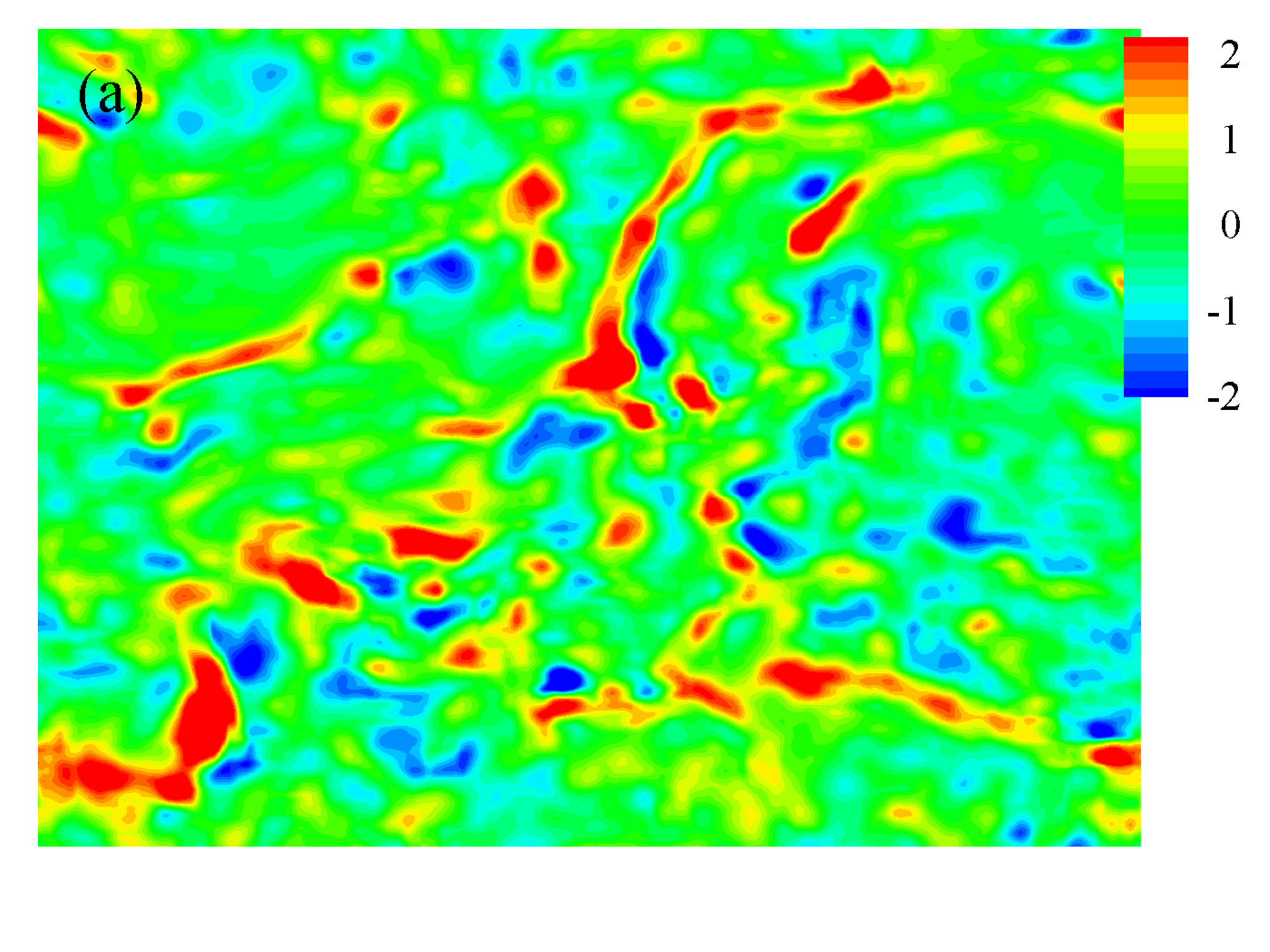}}}}%
\subfigure{
\resizebox*{6.5cm}{!}{\rotatebox{-90}{\includegraphics{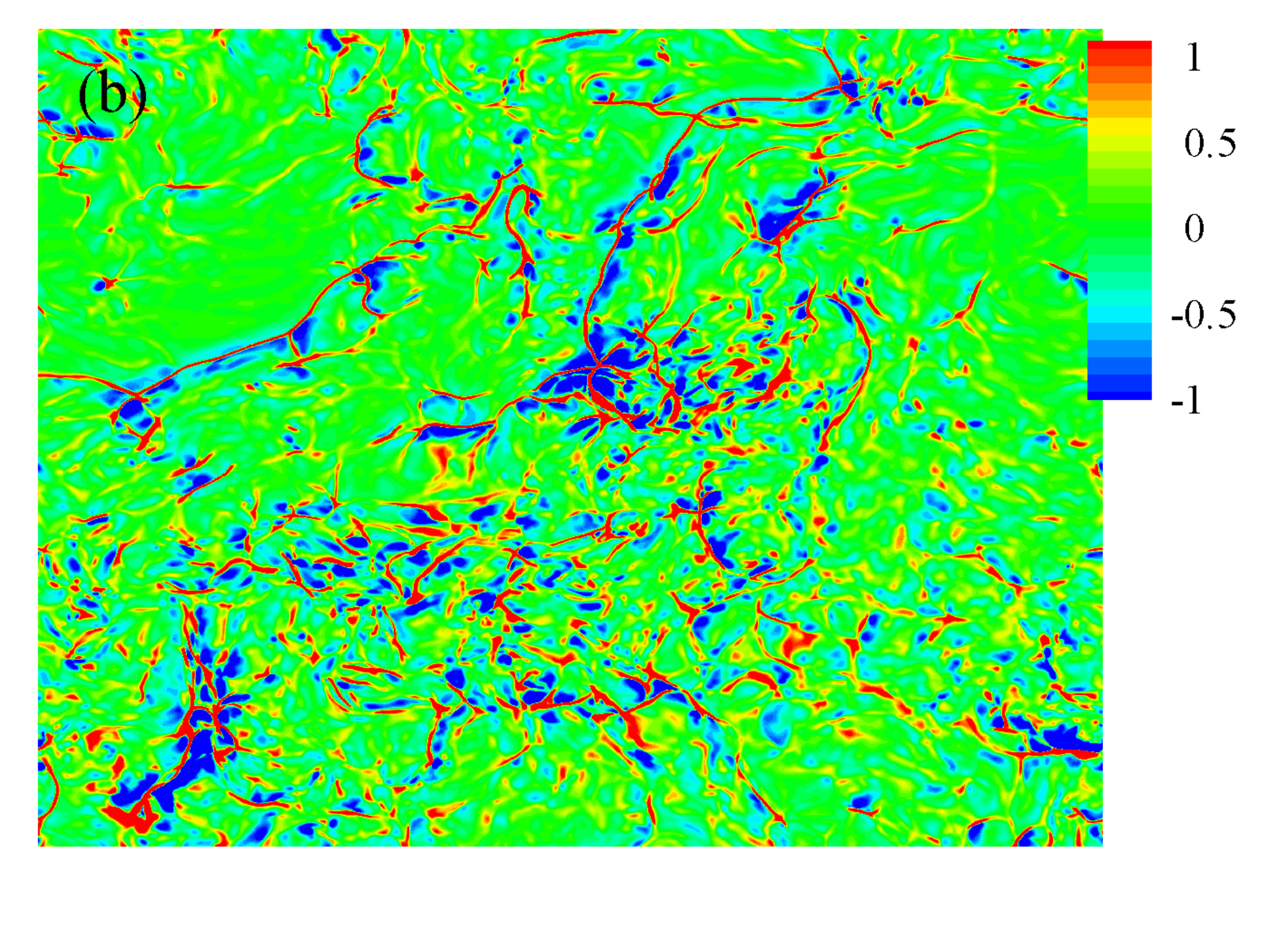}}}}%
\caption{Two-dimensional contours of the pressure-dilatation for large scales (a) and the residual part from small
scales (b) at $l=32\eta$, in SFT.}
\label{fig:fig26}
\end{center}
\end{figure}

The pressure-dilatation and its positive and negative components, as functions of the normalized scale
$l/\eta$, are depicted in Figure~\ref{fig:fig24}. Throughout scale ranges, $\langle\Phi^+_\emph{l}\rangle$
and $\langle\Phi^-_\emph{l}\rangle$ themselves are substantial, however, when adding together they almost cancel
each other and make the outcome small, especially at small scales. In other words, the high values
of pressure-dilatation generated in the vicinity of shock structures will basically vanish after taking
global averages. Further, we point out that the picture of the negligible contribution of pressure-dilatation at
small scales does not contradict to the motions of rarefaction and compression appearing at all scales, which
is the fundamental property of compressible turbulence. In CFT $\langle\Phi^-_\emph{l}\rangle$ is slightly
larger than that of $\langle\Phi^+_\emph{l}\rangle$, while in SFT the two components mostly overlay throughout
scale ranges.

\begin{figure}
\begin{center}
\subfigure{
\resizebox*{6.5cm}{!}{\rotatebox{-90}{\includegraphics{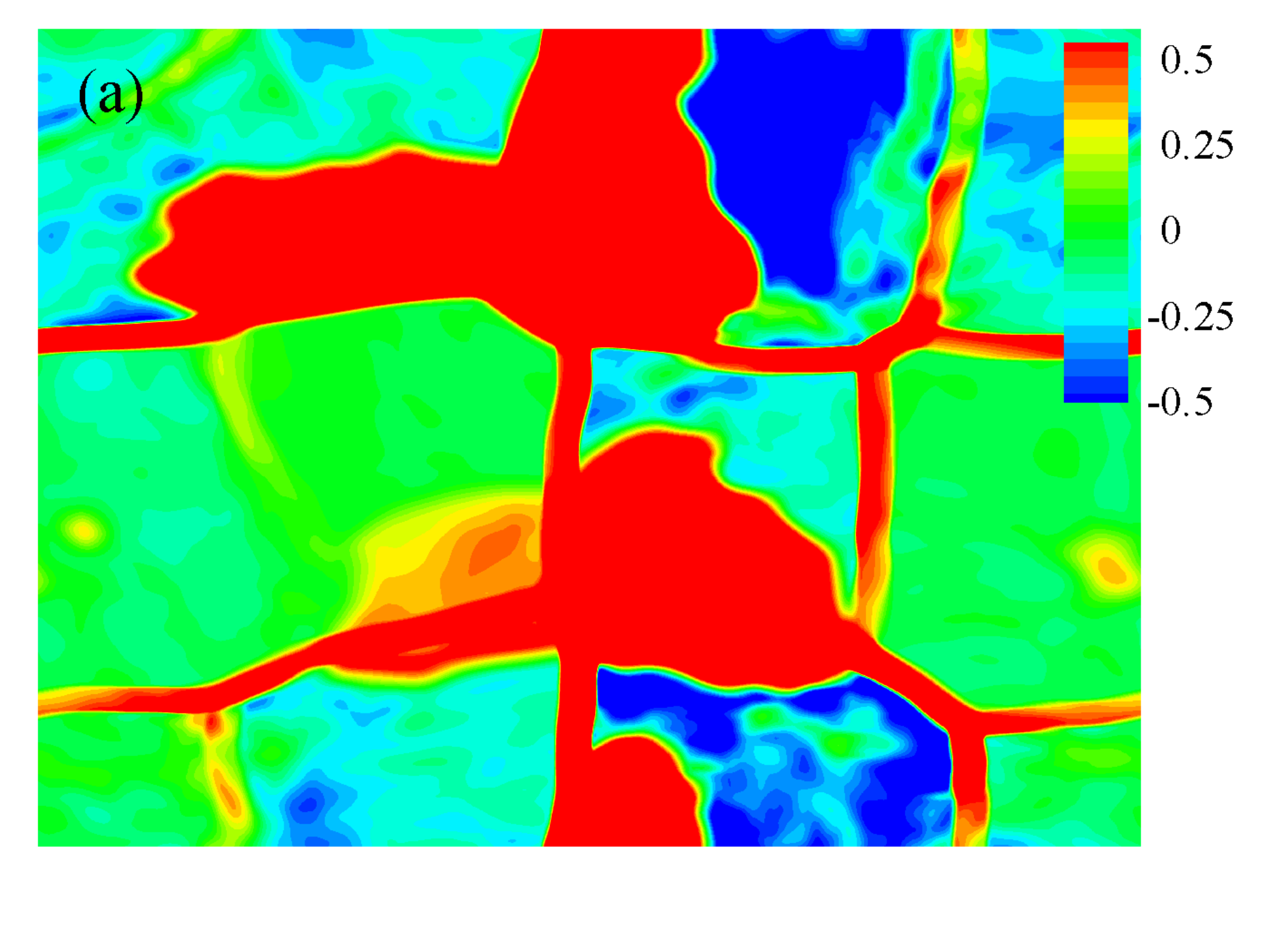}}}}%
\subfigure{
\resizebox*{6.5cm}{!}{\rotatebox{-90}{\includegraphics{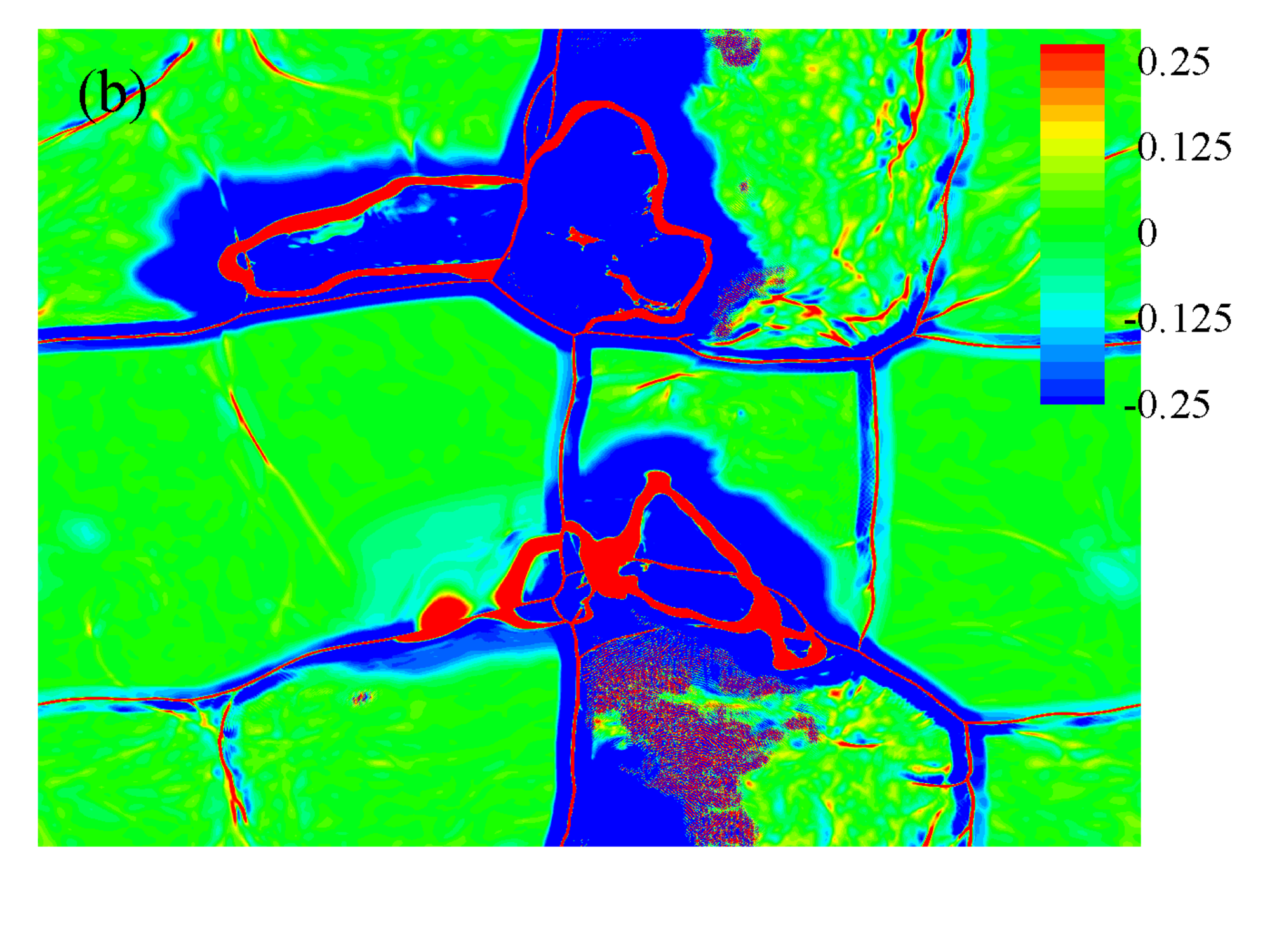}}}}%
\caption{Two-dimensional contours of the pressure-dilatation for large scales (a) and the residual part from small
scales (b) at $l=32\eta$, in CFT.}
\label{fig:fig27}
\end{center}
\end{figure}
\begin{figure}
\centerline{\includegraphics[width=8cm]{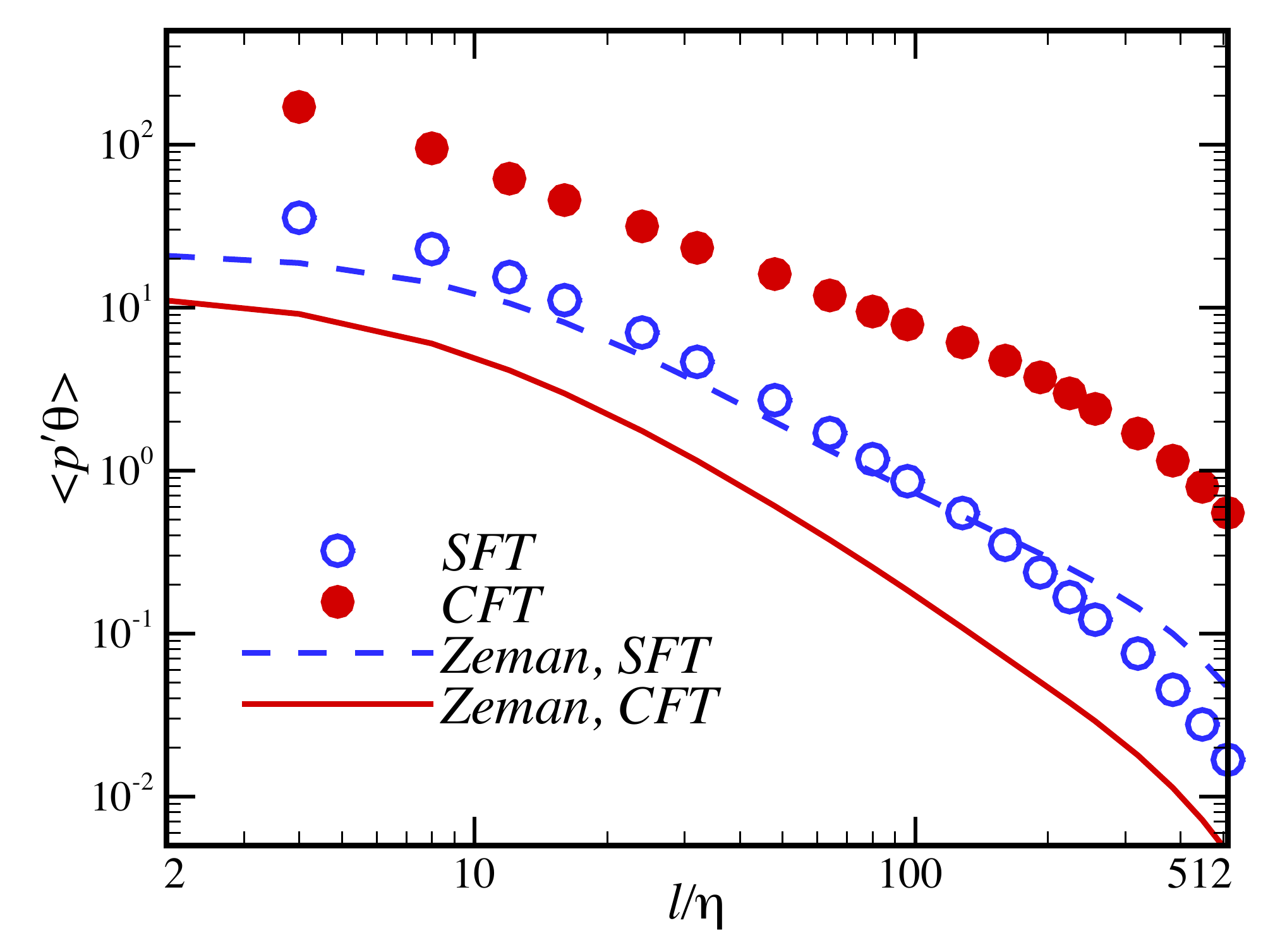}}
\caption{Fluctuation component of pressure-dilatation. The open and solid circles
are for SFT and CFT, respectively, and the dashed and solid lines represent the results
computed from the Zeman model.}
\label{fig:fig28}
\end{figure}

In Figure~\ref{fig:fig25} we plot the p.d.f.s of the pressure-dilatation for large scales
$\overline{p}_\emph{l}\overline{\theta}_\emph{l}$, and the residual part from small scales
$p\theta-\overline{p}_\emph{l}\overline{\theta}_\emph{l}$. The results are that the p.d.f. of $\overline{p}_\emph{l}\overline{\theta}_\emph{l}$ is sub-Gaussian and has small positive skewness.
Compared to those in CFT, the p.d.f. tails in SFT are broader and thus are less intermittent.
The super-Gaussian p.d.f.s of the residual parts display heavy tails, implying spatially rare but intense two-way
exchange between kinetic and internal energy. Moreover, these p.d.f.s are strongly positive skewed, indicating that
the conversion from kinetic to internal energy through compression is more intense than the inverse process through
rarefaction. In Figures~\ref{fig:fig26} and \ref{fig:fig27}, we plot the 2D contours of the pressure-dilatation for
large and small scales. The filter width is $l=32\eta$. Obviously, there exists appreciable differences in the
behavior of pressure-dilatation between the solenoidal and compressive forced flows. In SFT the pressure-dilatation
for large scales and the residual part from small scales distribute randomly, and the discontinuities shown in Figure~\ref{fig:fig26}(b) are the small-scale shocklets. By contrast, in CFT the two parts pressure-dilatation
concentrate in the rarefaction and compression regions, and the discontinuities shown in Figure~\ref{fig:fig27}(b)
are the large-scale shock waves. In Figure~\ref{fig:fig28} we depict the fluctuation component of pressure-dilatation,
as a function of the normalized scale $l/\eta$. For comparison, we also present the results computed from the
Zeman model \citep{Zeman91}. It shows that $\langle p'\theta\rangle$ from SFT basically collapses onto the line
representing the model, except a few derivations at small and large scales. Nevertheless, because of the
strong degree of compressibility, $\langle p'\theta\rangle$ from CFT deviates far away from the model.

\section{Summary and Conclusions}

In this paper, we performed a systematic investigation on the effects of shock topology on the statistics of
temperature in isotropic compressible turbulence. The simulations were solved numerically using
a hybrid method of a seventh-order WENO scheme for shock region and an eighth-order CCFD scheme for smooth region
outside shock. The small-scale shocklets and large-scale shock waves appeared in the compressible turbulence
driven by the large-scale, solenoidal and compressive forcings, respectively, where the stationary values of
the turbulent Mach number and Taylor microscale Reynolds number were ($M_t$, $Re_\lambda$) $=$ ($1.03$, $255$)
for SFT and ($M_t$, $Re_\lambda$) $=$ ($0.62$, $164$) for CFT. A variety of issues including the spectrum,
field structure, probability distribution function, isentropic assumption and cascade were reported. The results
revealed that there are appreciable differences in the statistics of temperature between SFT and CFT.

The kinetic energy spectra follow the $k^{-5/3}$ power law, and the Kolmogorov constant $C_K$ is $2.17$ in SFT
and $1.05$ in CFT. The Helmholtz decomposition on the velocity showed that in SFT the compressive component of
the kinetic kinetic energy spectrum in SFT obeys the $k^{-5/3}$ power law, while in CFT it defers to the $k^{-2}$
power law. The temperature has the Kolmogorov spectrum in SFT and the Burgers spectrum in CFT. The 2D contours
of temperature and temperature dissipation in SFT display the sufficient mixing of the large-scale ramps and the
small-scale cliffs. By contrast, the same 2D contours in CFT are dominated by the large-scale motions of
rarefaction and compression. The major contribution to the power-law region of large negative dilatation are
from preshocks and weak shocklets rather than strong shock waves. Our results showed that the power-law exponents
are $-2.5$ for SFT and $-3.5$ for CFT. Using a theoretical model to handle the conditional averages of pressure
and viscous, we obtained $-2.56$ for SFT and $-3.43$ for CFT, which are close to the numerical values.
The p.d.f. of temperature increment is concave and symmetric. When scale is comparable to the integral scale,
in SFT it becomes Gaussian but in CFT it is still super-Gaussian. Moreover, unlike that in SFT, the rescaled p.d.f.
in CFT collapses to the same distance, indicating that the scaling exponents for the statistical moments of
temperature increment should saturate at high order numbers.

We further studied the isentropic approximation in thermodynamic variables. Within a tolerance of $10\%$,
in SFT the pointwise values of temperature and pressure is only $1.6\%$ failing to satisfy the isentropic relation.
However, in CFT the deviation increases to $18.9\%$. For large negative dilatation, the ratio of
$T/\langle T\rangle$ to $(p/\langle p\rangle)^{1-1/\gamma}$ is near unity, meaning that the thermodynamic process
is approximately isentropic in the compression region. Then, the accuracy of isentropic approximation was measured
by introducing the quantity of $\varphi=p/\langle p\rangle/\big(\rho/\langle\rho\rangle\big)^{\gamma}$. It showed
that in SFT the p.d.f. of $\varphi$ is close to the Dirac delta function at $M_t=0.1$, and deviates from that
as $M_t$ increases. At $M_t\approx 0.6$, $\varphi$ of CFT displays a broader distribution than that of SFT.
Furthermore, it showed that the tails of the p.d.f. of $\varphi$, $Q(\varphi)$, from SFT are well described
by a theoretical model based on the Gaussian assumption of temperature distribution.

The description in the statistical properties of temperature gradient revealed that the
temperature tends to tangent to the vorticity in highly compressible turbulence. Compared to that in SFT, the
temperature gradient in CFT is preferentially perpendicular to the solenoidal component and preferentially
aligns with the compressive component of the anisotropic strain rate tensor. There are clear tendencies for
the temperature gradient to align with the first eigenvector corresponding to the most negative eigenvalue,
and to be perpendicular to the second and third eigenvectors. The conditional magnitude average
of temperature gradient has the teardrop shape, and is substantial in the region where the second and third
invariants of the anisotropic velocity gradient tensor are respectively positive and negative.

By employing a "coarse-graining" approach, we studied the cascade of temperature in compressible turbulence.
It was found that throughout scale ranges, the transport of temperature fluctuations is increased by the
viscous dissipation at small scales and the pressure-dilatation at moderately large scales, however, is decreased
by the SGS temperature flux, which preferentially occurs in the orientation where the temperature gradient
anti-align with the SGS velocity-temperature coupling. The appearance of plateau in the SGS temperature flux indicates
the conservation of temperature cascade from large to small scales. The slope values for the cospectrum of
pressure-dilatation is $-1.30$ in SFT and $-2.38$ in CFT, indicating that the pressure-dilatation converges
and be independent of scale at high enough wavenumbers. It provided the picture that the conversion between the
kinetic and internal energy by the pressure-dilatation mainly occurs over moderately large scales of limited extent.
The positive and negative components of pressure-dilatation are substantial at small scales. Once taking global
averages, they basically cancel each other and make the outcome small. This does not contradict to the fact that
the motions of rarefaction and compression happen at all scales in compressible turbulence.
The strongly positive skewness of the p.d.f. of pressure-dilatation from small scales implies that the conversion
from kinetic to internal energy through compression is more intense than the inverse process through rarefaction.
The 2D contours showed that in SFT the pressure-dilatation for large scales and the residual part from small scales
distribute randomly, while in CFT they are concentrate in the rarefaction and compression regions. Finally, we
observed that the fluctuation component of pressure-dilatation in SFT is well described by the Zeman model.

The current investigation reveals a variety of unique statistical properties of temperature in
compressible turbulence, relative to the features of temperature in incompressible flow. These new findings
can be largely understood through the effects of shock topology and the degree of compressibility.
We limit our study to numerical simulation, a further work addressed the theoretical models
for temperature in compressible turbulence will be performed in the near future.

\section{Acknowledgement}

We thank J. Wang for many useful discussions. This work was supported by the National Natural Science
Foundation of China (Grant 11221061) and the National Basic Research Program of China (973) (Grant 2009CB724101).
Q. N. acknowledges partial support by China Postdoctoral Science
Foundation Grant 2014M550557. Simulations were done on the TH-1A supercomputer in Tianjin, National
Supercomputer Center of China.

\bibliographystyle{jfm}

\begin{thebibliography}{0}
\expandafter\ifx\csname natexlab\endcsname\relax\def\natexlab#1{#1}\fi

\end{thebibliography}


\begin{thebibliography}{99}
\expandafter\ifx\csname
natexlab\endcsname\relax\def\natexlab#1{#1}\fi

\bibitem[Aluie(2011)]{Aluie11}
{\sc Aluie, H.} 2011 Compressible turbulence: the cascade and its locality.
{\em Phys. Rev. Lett.\/} {\bf 106}, 174502.

\bibitem[Aluie {\em et al.}(2012)]{Aluie12}
{\sc Aluie, H., Li, S. \& Li, H.} 2012
Conservative cascade of kinetic energy in compressible turbulence.
{\em Astrophy. J.\/} {\bf 751}, L29.

\bibitem[Andreopoulos {\em et al.}(2000)]{Andreopoulos00}
{\sc Andreopoulos, Y., Agui, J.~H. \& Briassulis, G.} 2000
Shock wave-turbulence interactions.
{\em Annu. Rev. Fluid Mech.\/} {\bf 32}, 309--345.

\bibitem[Balsara \& Shu(2000)]{Balsara00}
{\sc Balsara, D.~S. \& Shu, C.~W.} 2000 Monotonicity preserving
weighted essentially non-oscillatory schemes with increasingly high
order of accuracy. {\em J.~Comp. Phys.\/} {\bf 160}, 405--452.

\bibitem[Bec {\em et al.}(2000)]{Bec00}
{\sc Bec, J., Frisch, U. \& Khanin, K.} 2000
Kicked Burgers turbulence. {\em J.~Fluid Mech.\/} {\bf 416}, 239--267.

\bibitem[Bec(2001)]{Bec01}
{\sc Bec, J.} 2001
Universality of velocity gradients in forced Burgers turbulence. {\em Phys. Rev. Lett.\/} {\bf 87}, 104501.

\bibitem[Bec \& Khanin(2007)]{Bec07}
{\sc Bec, J. \& Khanin, K.} 2007 Burgers turbulence. {\em Phys. Rep.\/} {\bf 447}, 1--66.

\bibitem[Belmonte \& Libchaber(1996)]{Belmonte96}
{\sc Belmonte, A. \& Libchaber, A.} 1996 Thermal signature of plumes in turbulent convection: The
skewness of the derivative. {\em Phys. Rev. E\/} {\bf 53}, 4893--4898.

\bibitem[Benzi {\em et al.}(2008)]{Benzi08}
{\sc Benzi, R., Biferale, L., Fisher, R.~T., Kadanoff, L.~P., Lamb, D.~Q. \& Toschi, F.} 2008
Intermittency and university in fully developed inviscid and weakly compressible turbulent
flows. {\em Phy. Rev. Lett.\/} {\bf 100}, 234503.

\bibitem[Canuto(1997)]{Canuto97}
{\sc Canuto, V.~M.} 1997 Compressible turbulence. {\em Astrophys. J.\/} {\bf 482}, 827--851.

\bibitem[Cattaneo {\em et al.}(2003)]{Cattaneo03}
{\sc Cattaneo, F., Emonet, T. \& Weiss, N.} 2003
On the interaction between convection and magnetic fields. {\em Astrophys. J.\/} {\bf 588}, 1183--1198.

\bibitem[Celani {\em et al.}(2001)]{Celani01}
{\sc Celani, A., Lanotte, A., Mazzino A. \& Vergassola, M.} 2001
Fronts in passive scalar turbulence. {\em Phys. Fluids\/} {\bf 13}, 1768--1783.

\bibitem[Chandrasekhar(1951)]{Chandrasekhar51}
{\sc Chandrasekhar, S.} 1951 The fluctuations of density in isotropic turbulence.
{\em Proc. R. Soc. Lond. A\/} {\bf 210}, 18--25.

\bibitem[Chong {\em et al.}(1990)]{Chong90}
{\sc Chong M. S., Perry, A. E., \& Cantwell, B. J.} 1990
A general classification of three-dimensional flow fields. {\em Phys. Fluids A\/} {\bf 2}, 765--777.

\bibitem[Corrsin(1951)]{Corrsin51}
{\sc Corrsin, S.} 1951 On the spectrum of isotropic temperature fluctuations
in an isotropic turbulence. {\em J.~Appl. Phys.\/} {\bf 22}, 469--473.

\bibitem[Donzis \& Jagannathan(2013)]{Donzis13}
{\sc Donzis, D.~A. \& Jagannathan, S.} 2013 Fluctuations of thermodynamic variables in stationary compressible
turbulence. {\em J.~ Fluid Mech.\/} {\bf 733}, 221--244.

\bibitem[E {\em et al.}(1997)]{E97}
{\sc E, W., Khanin, K., Mazel, A. \& Sinai, Ya.} 1997
Probability distribution functions for the random forced Burgers equation. {\em Phys. Rev. Lett.\/} {\bf 78}, 1904--1907.

\bibitem[E \& Eijnden(1999)]{E99}
{\sc E, W. \& Eijnden, E.~V.} 1999 Asymptotic theory for the probability
density function in Burgers turbulence. {\em Phys. Rev. Lett.\/} {\bf 83}, 2572--2575.

\bibitem[E \& Eijnden(2000)]{E00}
{\sc E, W. \& Eijnden, E.~V.} 2000 Statistical theory for the stochastic
Burgers equation in the inviscid limit. {\em Comm. Pure Appl. Math.\/} {\bf 53}, 852--877.

\bibitem[Erlebacher {\em et al.}(1990)]{Erlebacher90}
{\sc Erlebacher, G., Hussaini, M.~Y., Kreiss, H.~O., \& Sarkar, S.} 1990
The analysis and simulation of compressible turbulence. {\em Theor. Comput. Fluid Dyn.\/} {\bf 2}, 73--95.

\bibitem[Erlebacher {\em et al.}(1993)]{Erlebacher93}
{\sc Erlebacher, G., \& Sarkar, S.} 1993
Statistical analysis of the rate of strain tensor in compressible homogeneous turbulence. {\em Phys. Fluids A\/} {\bf 5},
3240--3254.

\bibitem[Hill(1976)]{Hill76}
{\sc Hill, J.~C.} 1976 Homogeneous turbulent mixing with chemical reaction. {\em Annu. Rev. Fluid Mech.\/} {\bf 8}, 135--161.

\bibitem[Kritsuk {\em et al.}(2007)]{Kritsuk07}
{\sc Kritsuk, A.~G., Norman, M.~L., Padoan P. \& Wagner, R.}
2007 The statistics of supersonic isothermal turbulence.
{\em Astrophys. J.\/} {\bf 665}, 416--431.

\bibitem[Lele(1992)]{Lele92}
{\sc Lele, S.~K.} 1992 Compact finite difference schemes with
spectral-like resolution. {\em J.~Comp. Phys.\/} {\bf 103}, 16--42.

\bibitem[Mitra {\em et al.}(2005)]{Mitra05}
{\sc Mitra, D., Bec, J., Pandit R. \& Frisch, U.}
2005 Is multiscaling an artifact in the stochastically forced Burgers equation?
{\em Phys. Rev. Lett.\/} {\bf 94}, 194501.

\bibitem[Moisy \& Jimenez(2004)]{Moisy04}
{\sc Moisy, F. \& Jimenez, J.} 2004 Geometry and clustering of intense structures in isotropic turbulence.
{\em J.~Fluid Mech.\/} {\bf 513}, 111--113.

\bibitem[Ni \& Chen(2012)]{Ni12}
{\sc Ni, Q. \& Chen, S.} 2012 Statistics of active and
passive scalars in one-dimensional compressible turbulence. {\em Phys.
Rev. E\/} {\bf 86}, 066307.

\bibitem[Ni {\em et al.}(2013)]{Ni2013}
{\sc Ni, Q., Shi Y. \& Chen, S.} 2013 Statistics of one-dimensional compressible
turbulence with random large-scale force. {\em Phys. Fluids\/} {\bf 25}, 075106.

\bibitem[Ni {\em et al.}(2015a)]{Ni15a}
{\sc Ni, Q., Shi Y. \& Chen, S.} 2015 A numerical investigation on active and passive scalars in isotropic compressible
turbulence. {\em J.~Fluid Mech.\/}, submitted; arXiv:1505.02685.

\bibitem[Ni(2015b)]{Ni15b}
{\sc Ni, Q.} 2015 Compressible turbulent mixing: Effects of Schmidt number. {\em Phys. Rev. E\/} {\bf 91}, 053020.

\bibitem[Ni(2015c)]{Ni15c}
{\sc Ni, Q.} 2015 Effects of compressibility on passive scalar transport in isotropic
turbulence. {\em Phys. Rev. E\/}, to be submitted.

\bibitem[Pirozzoli \& Siggia(2004)]{Pirozzoli04}
{\sc Pirozzoli, S. \& Grasso, F.} 2004 Direct numerical simulations of isotropic compressible turbulence: influence of compressibility on
dynamics and structures. {\em Phys. Fluids\/} {\bf 16}, 4386-4407.

\bibitem[Samtaney {\em et al.}(2001)]{Samtaney01}
{\sc Samtaney, R., Pullin, D.~I. \& Kosovic, B.} 2001 Direct numerical simulation of decaying compressible turbulence and
shocklet statistics. {\em Phys. Fluids\/} {\bf 13}, 1415--1430.

\bibitem[Shraiman \& Siggia(1994)]{Shraiman94}
{\sc Shraiman, B.~I. \& Siggia, E.~D.} 1994 Lagrangian path integrals and fluctuations
in random flow. {\em Phys. Rev. E\/} {\bf 49}, 2912--2927.

\bibitem[Shraiman \& Siggia(2000)]{Shraiman00}
{\sc Shraiman, B.~I. \& Siggia, E.~D.} 2000 Scalar turbulence.
{\em Nature\/} {\bf 405}, 639.

\bibitem[Sreenivasan(1991)]{Sreenivasan91}
{\sc Sreenivasan, K. R.} 1991
On local isotropy of passive scalars in turbulent shear flows. {\em Proc. R. Soc. Lond. A\/} {\bf 434}, 165--182.

\bibitem[Stevens(2005)]{Stevens05}
{\sc Stevens, B.} 2005
Atmospheric moist convection. {\em Annu. Rev. Earth Planet Sci.\/} {\bf 33}, 605--643.

\bibitem[Sutherland(1992)]{Sutherland92}
{\sc Sutherland, W.} 1992 The viscosity of gases and molecular force. {\em Philos. Mag. Suppl.\/} {\bf 5},
507--531.

\bibitem[Wang {\em et al.}(1996)]{Wang96}
{\sc Wang, L.-P., Chen, S., Brasseur, J.~G. \& Wyngaard, J.~C.} 1996
Examination of hypotheses in the Kolmogorov refined turbulence
theory through high-resolution simulations. Part 1. Velocity field.
{\em J.~Fluid Mech.\/} {\bf 309}, 113--156.

\bibitem[Wang {\em et al.}(2010)]{Wang10}
{\sc Wang, J., Wang, L.-P., Xiao, Z., Shi, Y. \& Chen, S.} 2010 A
hybrid approach for direct numerical simulation of isotropic
compressible turbulence. {\em J.~Comp. Phys. \/} {\bf 229},
5257--5279.

\bibitem[Wang {\em et al.}(2011)]{Wang11}
{\sc Wang, J., Shi, Y., Wang, L.-P., Xiao, Z., He, X. \& Chen, S.} 2011 Effect of shocklets on
the velocity gradients in highly compressible isotropic turbulence. {\em Phys. Fluids\/} {\bf 23},
125103.

\bibitem[Wang {\em et al.}(2012a)]{Wang12a}
{\sc Wang, J., Shi, Y., Wang, L.-P., Xiao, Z., He, X. \& Chen, S.} 2012 Effect
of compressibility on the small-scale structures in isotropic turbulence.
{\em J.~Fluid Mech.\/} {\bf 713}, 588--631.

\bibitem[Wang {\em et al.}(2012b)]{Wang12b}
{\sc Wang, J., Shi, Y., Wang, L.-P., Xiao, Z., He, X. \& Chen, S.} 2012 Scaling
and statistics in three-dimensional compressible turbulence.
{\em Phys. Rev. Lett.\/} {\bf 108}, 214505.

\bibitem[Wang {\em et al.}(2013a)]{Wang13a}
{\sc Wang, J., Yang, Y., Shi, Y., Xiao, Z., He, X. \& Chen, S.} 2013 Cascade of kinetic energy in
three-dimensional compressible turbulence.
{\em Phys. Rev. Lett.\/} {\bf 110}, 214505.

\bibitem[Wang {\em et al.}(2013b)]{Wang13b}
{\sc Wang, J., Yang, Y., Shi, Y., Xiao, Z., He, X. \& Chen, S.} 2013 Statistics and structures of pressure
and density in compressible isotropic turbulence.
{\em J. Turb.\/} {\bf 14}, 21.

\bibitem[Warhaft(2000)]{Warhaft00}
{\sc Warhaft, Z.} 2000 Passive scalars in turbulent flows.
{\em Annu. Rev. Fluid Mech.\/} {\bf 32}, 203--240.

\bibitem[Zeman(1991)]{Zeman91}
{\sc Zeman, O.} 1991 On the decay of compressible isotropic turbulence.
{\em Phys. Fluids A\/} {\bf 3}, 951--955.

\bibitem[Zhou \& Xia(2008)]{Zhou08}
{\sc Zhou, Q. \& Xia, K.~Q.} 2008 Comparative experimental study of local mixing of active and passive
scalars in turbulent thermal convection. {\em Phys. Rev. E\/} {\bf 77}, 056312.

\bibitem[Zhou \& Xia(2002)]{Zhou02}
{\sc Zhou, S.~Q. \& Xia, K.~Q.} 2002 Plume statistics in thermal turbulence:
mixing of an active scalar. {\em Phys. Rev. Lett.\/} {\bf 89}, 184502.

\bibitem[Zhou(2013)]{Zhou13}
{\sc Zhou, Q.} 2013 Temporal evolution and scaling of mixing in two-dimensional Rayleigh-Taylor turbulence.
{\em Phys. Fluids\/} {\bf 25}, 085107.

\end{thebibliography}

\end{document}